\documentclass[twocolumn,times,tighten]{aastex61}
\usepackage[bookmarks=false]{hyperref}
\usepackage{natbib}
\usepackage{epsfig}
\usepackage{wrapfig}
\usepackage{graphicx}
\usepackage{spverbatim}
\usepackage{amsmath,amstext}

\newcommand\arcpt{${{\lower3pt\hbox{$^{\prime\prime}$}}\atop{\raise4pt\hbox{.}}}$}

\submitjournal{Astronomical Journal}

\shorttitle{LACEwING}
\shortauthors{Riedel et al.}

\begin{document}

\title{LACEwING: A New Moving Group Analysis Code}

\author{Adric~R.~Riedel}

\affil {Department of Astronomy, California Institute of Technology, Pasadena, CA, 91125, USA}
\affil {Department of Engineering Science and Physics, The College of Staten Island, Staten Island, NY, 10314, USA} 
\affil {Department of Physics and Astronomy, Hunter College, New York, NY, 10065, USA}
\affil {Department of Astrophysics, American Museum of Natural History, New York, NY, 10024, USA}

\email{arr@astro.caltech.edu}

\author{Sarah~C.~Blunt}

\affil {Department of Physics, Brown University, Providence, RI, 02912, USA}
\affil {Department of Astrophysics, American Museum of Natural History, New York, NY, 10024, USA}

\author{Erini~L.~Lambrides}

\affil {Department of Physics and Astronomy, Johns Hopkins University, Baltimore MD 21218}

\author {Emily~L.~Rice}

\affil {Department of Engineering Science and Physics, The College of Staten Island, Staten Island, NY, 10314, USA} 
\affil {Department of Astrophysics, American Museum of Natural History, New York, NY, 10024, USA}
\affil {Physics Ph.D. program, The Graduate Center, City University of New York, New York, NY 10016, USA}

\author {Kelle~L.~Cruz}

\affil {Department of Physics and Astronomy, Hunter College, New York, NY, 10065, USA}
\affil {Department of Astrophysics, American Museum of Natural History, New York, NY, 10024, USA}
\affil {Physics Ph.D. Program, The Graduate Center, City University of New York, New York, NY 10016, USA}

\author{Jacqueline~K.~Faherty}

\affil {Department of Terrestrial Magnetism, Carnegie Institute of Science, Washington DC, 20015, USA}
\affil {Department of Astrophysics, American Museum of Natural History, New York, NY, 10024, USA}

\begin{abstract}

We present a new nearby young moving group (NYMG) kinematic membership analysis code, LocAting Constituent mEmbers In Nearby Groups (LACEwING), a new Catalog of Suspected Nearby Young Stars, a new list of bona-fide members of moving groups, and a kinematic traceback code. LACEwING is a convergence-style algorithm with carefully vetted membership statistics based on a large numerical simulation of the Solar Neighborhood. Given spatial and kinematic information on stars, LACEwING calculates membership probabilities in 13 NYMGs and \replaced{3}{three} open clusters within 100 pc. In addition to describing the inputs, methods, and products of the code, we provide comparisons of LACEwING to other popular kinematic moving group membership identification codes. As a proof of concept, we use LACEwING to reconsider the membership of 930 stellar systems in the Solar Neighborhood (within 100 pc) that have reported measurable lithium equivalent widths. We quantify the evidence in support of a population of young stars not attached to any NYMGs, which is a possible sign of new as-yet-undiscovered groups or of a field population of young stars.


\end{abstract}

\section{Introduction}
\label{sec:intro}

Young stars were traditionally thought to exist in star forming regions and open clusters, the closest of which are the Scorpius-Centaurus complex and Taurus-Auriga, both over \replaced{150}{100} pc away\explain{I note the LCC region of Scorpius-Centaurus is also closer}. In the last 30 years (starting with studies like \citealt{Rucinski1983} and \citealt{de-la-Reza1989}) a number of stars have been discovered within that distance that are relatively young (5-500 Myr). This population of stars has been extensively studied and has immense scientific value as the nearest examples of the later stages of star formation. Nearby young moving groups (NYMGs) are older than star forming regions, but they are significantly closer and therefore their members are easier to study. As groups, the NYMGs are spread out over large (often overlapping) volumes of space and large areas of the sky, which makes defining groups and identifying interlopers challenging.

Currently, young stars within 100 pc of the Sun are thought to exist in three open clusters - Hyades, Coma Ber, and $\eta$ Cha - and roughly ten gravitationally unbound NYMGs \replaced{(\citealt{Zuckerman2004,Torres2008,Malo2013}, Table \ref{tab:grouptable})}{(Table \ref{tab:grouptable}; \citealt{Zuckerman2004,Torres2008,Malo2013})}. These moving groups (occasionally called ``loose associations'') are distinct from open clusters: they have no strong nuclei and are incredibly sparse, with a few dozen stars spread over thousands of cubic parsecs of space. They are also distinct from the streams and \replaced{kinematic overdensities discovered pre-{\it Hipparcos} like}{pre-{\it Hipparcos} kinematic overdensities like} the Local Association/Pleiades Moving Group  \citep{Jeffries1995,Montes2001a}, Hyades Supercluster \citep{Eggen1985}, and IC 2391 Supercluster \citep{Eggen1991}, which have been identified as heterogeneous assemblages of stars \citep{Famaey2008}. A few of the NYMGs appear to be related to open clusters: AB Dor to the Pleiades, Argus to IC 2391, and $\epsilon$ Cha to $\eta$ Cha, suggesting a common or at least related origin. The groups have ages between $\sim$5 Myr old ($\epsilon$ Cha) and \added{600--}800 Myr old (Hyades).

\begin{deluxetable*}{llrrrrrr}
\setlength{\tabcolsep}{0.04in}
\tablewidth{0pt} 
\tabletypesize{\scriptsize}
\tablecaption{Nearby Young Moving Groups and Open Clusters\label{tab:grouptable}}
\tablehead{
 \colhead{Name} &
 \colhead{Abbreviation} &
 \multicolumn{2}{c}{Members} &
 \multicolumn{2}{c}{Min Age~~~~~~~~~~~~~~~~~~~~~~~} & 
 \multicolumn{2}{c}{Max Age~~~~~~~~~~~~~~~~~~~~~~~} \\
 \colhead{} &
 \colhead{} &
 \colhead{All} &
 \colhead{OBAFGK} &
 \colhead{(Myr)} &
 \colhead{Reference} &
 \colhead{(Myr)} &
 \colhead{Reference}\\
 \colhead{(1)} &
 \colhead{(2)} &
 \colhead{(3)} &
 \colhead{(4)} &
 \colhead{(5)} &
 \colhead{(6)} &
 \colhead{(7)} &
 \colhead{(8)} }

\startdata
$\epsilon$ Cham\ae leontis & $\epsilon$ Cha&  35 & 17 &   5 & \citet{Murphy2013} & 8 & \citet{Torres2008} \\
$\eta$ Cham\ae leontis\tablenotemark{a}& $\eta$ Cha    &  21 &  6 &   6 & \citet{Torres2008} & 11 & \citet{Bell2015} \\
TW Hydrae               & TW Hya        &  38 &  7 &   3 & \citet{Weinberger2013} & 15 & \citet{Weinberger2013} \\
$\beta$ Pictoris        & $\beta$ Pic   &  94 & 34 &  10 & \citet{Torres2008} & 24 & \citet{Bell2015} \\
32 Orionis              & 32 Ori        &  16 & 12 &  15 & E.E. Mamajek (private communication) & 65 & \citet{David2015} \\
Octans                  & Octans        &  46 & 22 &  20 & \citet{Torres2008} & 40 & \citet{Murphy2015} \\
Tucana-Horologium       & Tuc-Hor       & 209 & 63 &  30 & \citet{Torres2008} & 45 & \citet{Kraus2014a} \\
Columba                 & Columba       &  82 & 52 &  30 & \citet{Torres2008} & 42 & \citet{Bell2015} \\
Carina                  & Carina        &  32 & 22 &  30 & \citet{Torres2008} & 45 & \citet{Bell2015} \\
Argus                   & Argus         &  90 & 38 &  35 & \citet{Barrado-y-Navascues1999b} & 50 & \citet{Barrado-y-Navascues1999b} \\
AB Doradus              & AB Dor        & 146 & 86 &  50 & \citet{Torres2008} & 150 & \citet{Bell2015} \\
Carina-Near             & Car-Near      &  13 & 10 & 150 & \citet{Zuckerman2006} & 250 & \citet{Zuckerman2006} \\
Coma Berenices\tablenotemark{a}& Coma Ber& 195&104 & 400 & \citet{Casewell2006} & & \\
Ursa Major              & Ursa Major    &  62 & 55 & 300 & \citet{Soderblom1993} & 500 & \citet{King2003} \\
$\chi^{01}$ Fornax       & $\chi^{01}$ For&  14 & 14 & 500 & \citet{Pohnl2010} & & \\
Hyades\tablenotemark{a} & Hyades        & 724 &260 & 600 & \citet{Zuckerman2004} & 800 & \citet{Brandt2015} \\
\hline
\enddata
\tablecomments{More details on these groups can be found in Section \ref{sec:NYMGs}.}
\tablenotetext{a}{Open Cluster}

\end{deluxetable*}

The fundamental assumption about these NYMGs and open clusters is that they are the products of single bursts of star formation. This means that every constituent member should be roughly the same age (with attendant constraints on activity, radius, and rotational velocity), have the same chemical composition, have been in the same location at the time of formation, and have formed under the same conditions. Although the moving groups are not gravitationally bound, they are young enough that their space motion should still trace the Galactic orbits of their natal gas clouds. Due to their proximity and lack of gas, NYMG members allow easy and uncomplicated analysis of their photometric and spectroscopic properties.

The existence of these groups has been beneficial to the study of extremely low mass objects - planets \citep{Baines2012,Delorme2013}, brown dwarfs \citep{Faherty2016}, and very low mass stars \citep{Mathieu2007} - whose formation and evolutionary sequence and properties are still largely unknown. Using the assumption of a common origin, the age, metallicity, and formation environment deduced from the high-mass members can be applied to very low mass objects.

The methods for identifying young stars vary with their mass and age. They include measurements of coronal activity, as seen in X-rays \citep{Schmitt1995,Micela1999,Feigelson2002,Torres2008} and UV activity \citep{Shkolnik2012,Rodriguez2013}, chromospheric activity, as seen in H\deleted{-}$\alpha$ \citep{West2008} and optical calcium \citep{Hillenbrand2013}, measuring the lithium equivalent width (e.g. \citealt{Mentuch2008,Malo2014a}), equivalent widths of gravity-sensitive spectral features (e.g. \citealt{Lyo2004a,Schlieder2012b}), rotational velocity measurements (e.g. \citealt{Mamajek2008}), emission line core widths (e.g. \citealt{Shkolnik2009}), chemical abundances (e.g. \citealt{DOrazi2012,Tabernero2012,De-Silva2013}), and isochrone fitting (e.g. \citealt{Torres2008,Malo2013}). Most of those techniques can establish or at least constrain ages for ranges of stellar temperatures and masses, but they cannot generally be used to identify memberships in a particular group. Conversely, kinematic memberships themselves do not generally convey any proof of youth \citep{Lopez-Santiago2009}, but they alone can group stars so that collective properties can be determined. Given that the spatial distributions of many NYMGs are overlapping and distributed across the sky (at least three -- AB Dor, $\beta$ Pic, Ursa Major -- are effectively all sky), space velocities are often the only practical way to identify memberships. This makes identifying members of NYMGs a different task from identifying members of more distant clusters, which are more localized on the sky. 

A variety of codes are publicly available to accomplish the task of identifying NYMG members kinematically: BANYAN \citep{Malo2013}, which implements Bayesian methods to choose between membership in seven moving groups and a field/old option; BANYAN II \citep{Gagne2014a}, a modification of BANYAN with updated kinematic models and algorithms, and a convergence algorithm \citep{Rodriguez2013} that uses the convergence points of the NYMGs to determine probable membership in six NYMGs. There are also other prominent but less widely available codes, including ones used in \citet{Montes2001a} and subsequent papers; \citet{Torres2008} and subsequent papers; \citet{Kraus2014a}; \citet{Lepine2009} and follow-up papers \citep{Schlieder2010,Schlieder2012a,Schlieder2012c};  \citet{Shkolnik2012}; \citet{Klutsch2014}; \citet{Riedel2014}; and \citet{Binks2015} \explain{Citations moved into chronological order, and different codes delimited by semicolons}. All require varying amounts of the six kinematic elements to identify objects by their kinematics, but none include all the moving groups and open clusters currently believed to exist within 100 parsecs of the Sun. Throughout the paper, we will follow this nomenclature for the six kinematic elements: right ascension $\alpha \equiv$ RA, declination $\delta \equiv$ DEC, parallax $\pi$ (or distance \replaced{$Dist.$}{$Dist$}), proper motion in right ascension \replaced{$\mu_{\alpha}\equiv\mu_{{RA}\cos{DEC}}$}{$\mu_{\alpha}\equiv\mu_{\textrm{RA}}\cos{\textrm{DEC}}$}, proper motion in declination $\mu_{\delta} \equiv \mu_{\textrm{DEC}}$, systemic radial velocity $\gamma \equiv$ RV.

In this paper we present a new kinematic moving group code, LocAting Constituent mEmbers In Nearby Groups (LACEwING), which has been presented in international conferences \citep{Riedel2016a} and publications \citep{Faherty2016,Riedel2016b}. LACEwING, given stellar kinematic properties, calculates membership probabilities for 16 groups, comprising the 13 moving groups and \replaced{3}{three} open clusters within 100 parsecs mentioned earlier and in Table \ref{tab:grouptable}. The following discussion, and the moving group code described herein, present three interrelated products:
\begin{itemize}
\item The LACEwING kinematic moving group identification code (Section \ref{sec:lacewing}), comprised of both the code itself (Section \ref{sec:design}) and its calibration (Section \ref{sec:calibration}).
\item An epicyclic traceback code, TRACEwING (Section \ref{sec:tracewing}) which was used to identify interlopers in samples of NYMG members.
\item A Catalog (Section \ref{sec:catalog}) of data on 5350 known and suspected nearby young stars, from which a sample of 400 high confidence members of NYMGs (later reduced to 297 systems, Section \ref{sec:bonafides}) and a sample of 930 lithium-detected objects (Section \ref{sec:lithium}) were taken. That new bona-fide sample is used to form the kinematic models and calibration of the particular implementation of LACEwING presented here.
\end{itemize}
As a proof of concept, we use LACEwING to calculate the membership probabilities for both the bonafide (Section \ref{sec:bonafides}) and lithium (Section \ref{sec:lithium}) samples. We then compare LACEwING's recovery of known members in the lithium sample to BANYAN, BANYAN II, and the \citet{Rodriguez2013} convergent point analysis (Section \ref{sec:results}), and outline our conclusions about the moving group members themselves (Section \ref{sec:NYMGs}). Conclusions are summarized in Section \ref{sec:conclusions}.

\section{The LACEwING Moving Group Identification Code}
\label{sec:lacewing}

LACEwING is a frequentist observation space kinematic moving group identification code. Using the spatial and kinematic information available about a target object ($\alpha$, $\delta$, \replaced{$Dist.$}{$Dist$}, \replaced{$\mu_{RA}$}{$\mu_{\alpha}$}, \replaced{$\mu_{DEC}$}{$\mu_{\delta}$} and $\gamma$), it determines the probability that the object is a member of each of the known NYMGs (Table \ref{tab:grouptable}). As with other moving group identification codes, LACEwING is capable of estimating memberships for stars with incomplete kinematic and spatial information.

LACEwING works in right-handed Cartesian Galactic coordinates, with UVW space velocities and XYZ space positions, where the U/X axis is toward the Galactic center. The matrices of \citet{Johnson1987} transform observational equatorial coordinates into UVW and XYZ, and vice versa. The LACEwING code takes kinematic models of the NYMGs (represented as freely-oriented triaxial ellipsoids in UVW and XYZ space), predicts the observable values for members of each group at the $\alpha$ and $\delta$ of the input star, and compares directly to the measured observable quantities. 

Producing a membership probability from the goodness-of-fit value is a complex task, because most of the NYMGs considered here (particularly those under 100 Myr) are found within a very small range of space velocities, typified by the \citet{Zuckerman2004} ``good box'', with boundaries U=(0,$-$15) km s$^{-1}$, V=($-$10,$-$34) km s$^{-1}$, W=($+$3,$-$20) km s$^{-1}$. Young stars are also significantly less common than older field stars (see Section \ref{sec:field}). Thus, the potential for confusion between the groups and field is potentially large, and different for each NYMG. The LACEwING code accounts for these factors with equations, derived from a large simulation of the Solar Neighborhood, that convert goodness-of-fit parameters to membership probabilities.

A functional implementation of LACEwING requires three components:
\begin{itemize}
\item A code to predict proper motion, radial velocity and distances based on $\alpha$ and $\delta$; compare the predictions to input stellar data; and produce goodness-of-fit values, described in Section \ref{sec:design}.
\item Per-NYMG equations to transform goodness-of-fit values into membership probabilities. The simulation of the Solar Neighborhood and method for producing the equations is given in Section \ref{sec:calibration}.
\item Spatial and velocity distributions that describe each of the moving groups, used for predictions and calibration. The process of creating and vetting these stars is explained in Section \ref{sec:bonafides}, and uses data from the Catalog of Suspected Nearby Young Stars described in Section \ref{sec:catalog}.
\end{itemize}

\subsection{The LACEwING Algorithm}
\label{sec:design}

Within the LACEwING framework, each NYMG is represented by two triaxial ellipsoids described by base values UVW and XYZ, semi-major and semi-minor axes ABC and DEF, and Tait-Bryant angles UV, UW, VW, XY, XZ, and YZ, which are used to rotate the ellipses around the W, V, U and Z, Y, X axes, respectively. This is unlike the the BANYAN II \citep{Gagne2014a} rotation order U,V,W (X,Y,Z)\explain{This sentence moved out of a footnote}.

For each NYMG, we use 100,000 Monte Carlo iterations within the triaxial ellipsoids to generate a realistic spread of UVW velocities. Using inverted matrices from \citet{Johnson1987}, we convert the UVW velocities into $\mu_{\alpha}$, $\mu_{\delta}$, and $\gamma$ for a simulated star at the $\alpha$ and $\delta$ of the target, at a standard distance, which we have chosen as 10 parsecs in analogy to absolute magnitude. The lengths of the proper motion vectors are later used to estimate the kinematic distance, which can be compared to a parallax measurement if it exists.

As an example, we use the bona-fide $\beta$ Pic member \object{AO Men}\explain{Added object tag} (K4\added{~}Ve). At the \replaced{RA}{$\alpha$} and \replaced{DEC}{$\delta$} of AO~Men and a distance of 10 parsecs, an object sharing the mean UVW velocity of $\beta$ Pic should have $\mu_{\alpha}$=-45 mas yr$^{-1}$, $\mu_{\delta}$=289 mas yr$^{-1}$, and $\gamma$=$+$16 km s$^{-1}$.

With the predicted values of $\mu_{\alpha}$, $\mu_{\delta}$, and $\gamma$, and at least some measured kinematic data on the star, the code derives (up to) four of the following metrics:
\begin{enumerate}
\item Proper Motion. The code splits the measured proper motion into components parallel ($\mu_{\parallel}$) and perpendicular ($\mu_{\perp}$) to the predicted proper motion vector; the best match is where the perpendicular component is zero. Our goodness-of-fit metric for proper motions is:
\begin{equation}
\psi_{\mu}^2 = \frac{\mu_{\perp\textrm{star}}^2}{\sigma_{\mu_{\perp\textrm{star}}}^2 + \sigma_{\mu_{\perp\textrm{pred}}}^2}
\end{equation}
If $\mu_{\parallel\textrm{star}}$ has the opposite sign from $\mu_{\parallel{\textrm{pred}}}$ and is more than 1$\sigma$ from zero, we add 1000 to $\psi_{\mu}$ to ensure a poor goodness-of-fit metric.

For our example object AO~Men, the real proper motion ($\mu_{\alpha}$=-8.3 mas yr$^{-1}$ and $\mu_{\delta}$=72.0 mas yr$^{-1}$) is along a vector very similar to the predicted object, leading to a perpendicular component of 2.7 mas yr$^{-1}$and a $\psi_{\mu}$ of 0.05$\sigma$.

\item Distance. Kinematic distance is derived from a ratio of the star's measured proper motion and the predicted proper motion calculated for a member at 10 parsecs: 
\begin{equation}
\frac{Dist_\textrm{pred}}{10} = \frac{\mu_{para,\textrm{pred}}}{\mu_{para,\textrm{star}}} 
\end{equation}
If a trigonometric parallax distance exists, the resulting goodness-of-fit metric is 
\begin{equation}
\psi_{Dist}^2 = \frac{Dist_\textrm{star}^2 - Dist_{\textrm{pred}}^2}{\sigma_{Dist,\textrm{star}}^2 + \sigma_{Dist,\textrm{pred}}^2}
\end{equation}

For our example object AO~Men, the two parallel components of the proper motion are 72.4 mas yr$^{-1}$ (real) and 293.5 mas yr$^{-1}$ (predicted), which yields a ratio of 4.05 and a distance of 40.5 pc. The actual distance (measured by {\it Hipparcos}, \citealt{van-Leeuwen2007}\added{-- we are not using the Gaia DR1 results from \citealt{GaiaCollaboration2016}}) is 38.6 pc.

\item Radial Velocity. The code compares the measured radial velocity to the predicted $\gamma$. In this case, the goodness-of-fit metric is 
\begin{equation}
\psi_{\gamma}^2 = \frac{\gamma_{\textrm{star}}^2 - \gamma_{\textrm{pred}}^2}{\sigma_{\gamma,\textrm{star}}^2 + \sigma_{\gamma,\textrm{pred}}^2}
\end{equation}

For AO~Men, the predicted $\gamma$ of an ideal member of $\beta$ Pic is $+$16.04, while the measured $\gamma$ of AO Men is $+$16.25. The difference between the two is 0.1$\sigma$.

\item Position. With either trigonometric parallax or kinematic distance, the code uses the $\alpha$, $\delta$, and distance to determine how near the star is to the moving group or cluster. As the moving groups are defined with freely-oriented ellipses, this requires a matrix rotation to bring a 100,000-element Monte Carlo approximation of the stellar position uncertainty into the coordinate system of the moving group. The goodness-of-fit metric is:
\begin{equation}
\psi_{Pos}^2 = \frac{X_\textrm{star}^2}{\sigma_{X}^2 + \sigma_{D}^2} + \frac{Y_\textrm{star}^2}{\sigma_{Y}^2 + \sigma_{E}^2} + \frac{Z_\textrm{star}^2}{\sigma_{Z}^2 + \sigma_{F}^2}
\end{equation}

AO~Men and its measured parallax put it 30 pc (2.1$\sigma$) from the center of the $\beta$ Pic moving group.

\end{enumerate}
While the last test for spatial position is not a standard convergence test, it is useful for preventing the code from identifying members of spatially concentrated groups (e.g. Hyades, TW Hya) in physically unreasonable locations.

Each of the four goodness-of-fit metrics has a different characteristic median value when analyzing bona-fide members, and that value varies slightly between groups. The median $\psi_{\mu}$ value is particularly small ($\sim$0.02) when matching bona-fide members to their correct groups, while $\psi_{Dist}$ and $\psi_{\gamma}$ are much larger ($\sim$0.5), and $\psi_{Pos}$ was generally around 2. All goodness-of-fit values tended to be significantly larger when matching stars to the wrong groups, although $\psi_{\mu}$ was again the most sensitive discriminator.

\subsection{LACEwING Calibration}
\label{sec:calibration}
At this stage, the LACEwING code has produced up to four different goodness-of-fit metrics estimating the quality of match between a star and one of the NYMGs. We combine these metrics in quadrature:
\begin{equation}
\psi^2 \times N_{\textrm{metrics}}= \psi_{\mu}^2+\psi_{Dist}^2+\psi_{\gamma}^2+\psi_{\textrm{Pos}}^2
\end{equation}
All tests indicated that the code recovers more members when each metric is given equal weight, as shown here.

We now want to derive functions to transform the goodness-of-fit values into membership probabilities. This is not simple, because the groups overlap with each other to different degrees. Some, like Ursa Major, have distinct UVW velocities that do not overlap with any other group. Others, like the Hyades, have unique and compact XYZ spatial distributions. Most, however, overlap in UVW and XYZ with other moving groups.

The additional complication is that there are seven different possible combinations of data, each with its own goodness-of-fit value that we need to calibrate separately, for each moving group considered:
\begin{enumerate}
\item $\alpha$, $\delta$, $\mu_{\alpha}$ and $\mu_{\delta}$
\item $\alpha$, $\delta$, \replaced{Dist.}{$Dist$}
\item $\alpha$, $\delta$, $\gamma$
\item $\alpha$, $\delta$, $\mu_{\alpha}$ and $\mu_{\delta}$, \replaced{Dist.}{$Dist$}
\item $\alpha$, $\delta$, $\mu_{\alpha}$ and $\mu_{\delta}$, $\gamma$
\item $\alpha$, $\delta$, \replaced{Dist.}{$Dist$}, $\gamma$
\item $\alpha$, $\delta$, $\mu_{\alpha}$ and $\mu_{\delta}$, \replaced{Dist.}{$Dist$}, $\gamma$
\end{enumerate}
To quantify these probabilities, we generate a simulation of the Solar Neighborhood \replaced{which}{that} we run through the LACEwING algorithm. The results of the simulation will be used to determine the relationship between goodness-of-fit and membership probability.

When generating the simulation, a random number generator first assigns the new star to one of the groups, in proportion to the population of each group as derived in Section \ref{sec:bonafides}. \replaced{Thus, more field stars are generated than TW Hya members.}{Thus, more field stars will be generated than TW Hya members because the field is more populous.}

To generate UVW and XYZ values for a simulated star, we use the freely oriented ellipse parameters of the moving groups (from Section \ref{sec:bonafides}). If the assigned group is one of the nearby young moving groups, a UVW velocity and XYZ position is generated under the assumption that the velocity and spatial distributions are both Gaussian. If the assigned group is the field, the UVW velocities are assumed to be Gaussian distributions, but the XY positions are drawn from a uniform distribution truncated at semimajor and semiminor axes D and E, and the Z position above the plane is drawn from an exponential distribution with a scale height of 300 parsecs \citep{Bochanski2010} truncated at semiminor axes F.

Once a star has been generated, the UVW and XYZ values are converted into equatorial coordinates $\alpha$, $\delta$, $\pi$, $\mu_{\alpha}$, $\mu_{\delta}$, and $\gamma$. Measurement errors are generated in the form of Gaussian distributions scaled to position to 0.05\arcsec, $\mu$ to 5 mas yr$^{-1}$, $\pi$ to 0.5 mas, and $\gamma$ to 0.5 km s$^{-1}$, and added to the equatorial values. The simulated stars are then run through the LACEwING algorithm, comparing them to every moving group using all seven combinations of data. The goodness-of-fit values are then recorded.

To produce the actual relations that yield membership percentages, we then take all of the stars compared to a particular moving group {\it X}, using a combination of data {\it Y}. The goodness-of-fit values are binned into 0.1 goodness-of-fit value bins. Within each bin, we calculate the fraction of stars in that bin that were actually generated as members (``true'' members) of moving group {\it X}.

\begin{figure}
\center
\includegraphics[angle=0,width=0.5\textwidth]{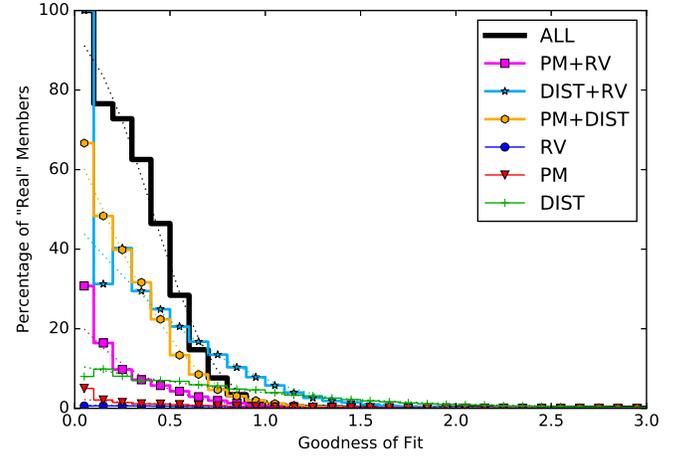}
\caption{Goodness-of-fit histograms for $\beta$ Pic in field star mode, and associated fitted Gaussian CDFs (curved lines). \added{Higher curves mean that the estimated probability of membership in $\beta$ Pic is greater.}}
\label{fig:goodness_of_fit}
\end{figure}

These functions are best described by Gaussian Cumulative Distribution Functions \deleted{(CDFs)} with normalized amplitude. The fraction of stars that were ``true members''\deleted{)} as a function of goodness-of-fit value \added{(for $\beta$ Pic)} is shown in Figure \ref{fig:goodness_of_fit} as histograms, along with the fitted functions. The coefficients of the Gaussian \replaced{CDFs}{Cumulative Distribution Functions} are saved and used to derive membership probabilities given a goodness-of-fit value. 

\begin{figure}
\center
\includegraphics[angle=0,width=0.5\textwidth]{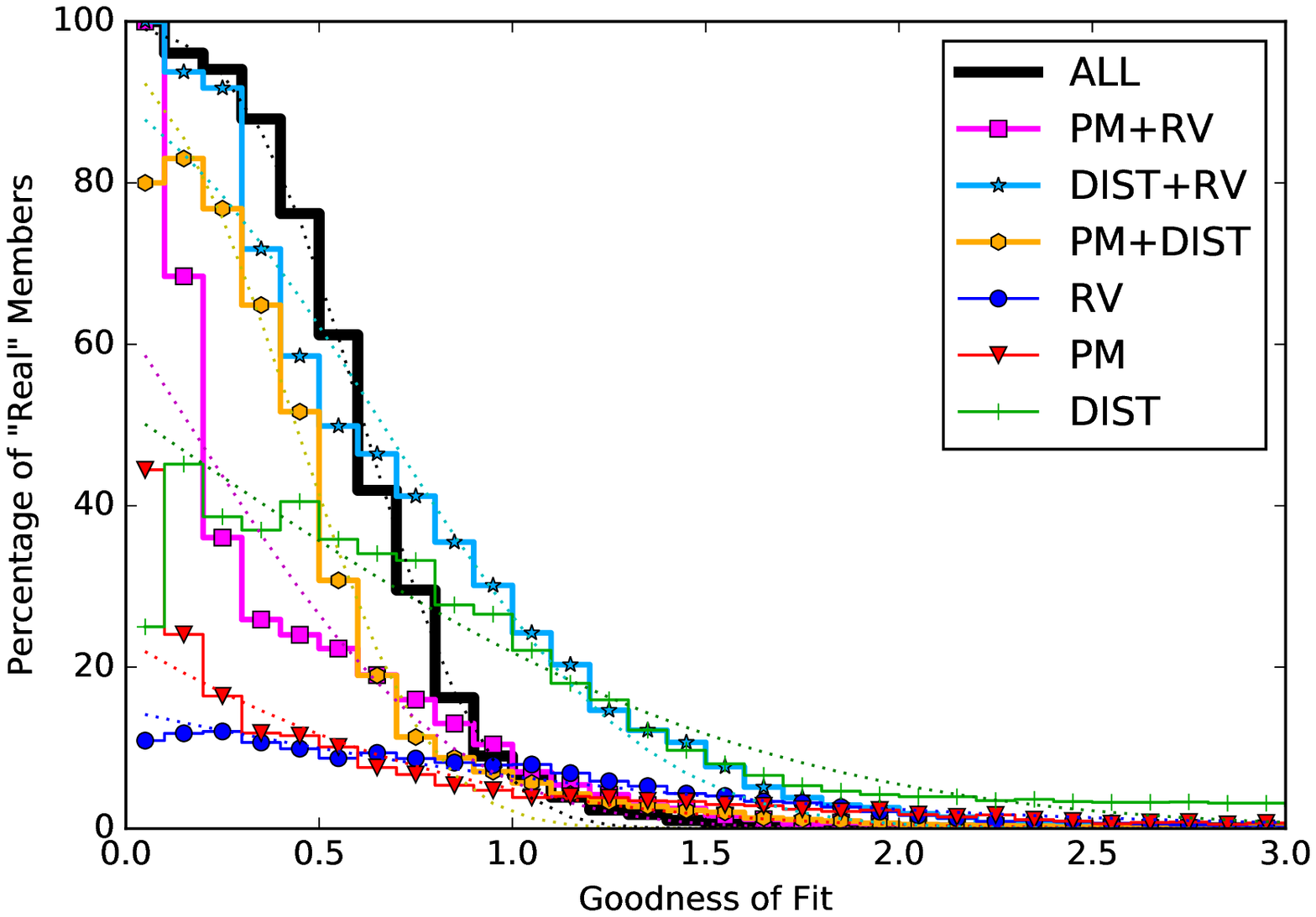}
\caption{Same as Figure \ref{fig:goodness_of_fit}, but for $\beta$ Pic in young star mode. \deleted{Higher curves mean that the estimated probability of membership in $\beta$ Pic is greater.}}
\label{fig:goodness_of_fit_young}
\end{figure}

In order to handle the case where a star is already known to be young (and the probability of membership in an NYMG is higher), there is a second calibration of LACEwING. In this mode, a subset containing all the NYMG members and an equally-sized portion of the field stars (to account for the young field, see Section \ref{sec:conclusions}) are retained, for a 1:1 ratio of NYMG members:young field is selected. This subset is binned into 0.1 goodness-of-fit bins, fractions of stars are calculated in the same way as the above ``field star'' sample, and a different set of Gaussian \replaced{CDF}{Cumulative Distribution Function} coefficients is produced (\added{the ``young star'' curves for $\beta$ Pic are shown in }Figure \ref{fig:goodness_of_fit_young}). 

\begin{figure}
\center
\includegraphics[angle=0,width=0.5\textwidth]{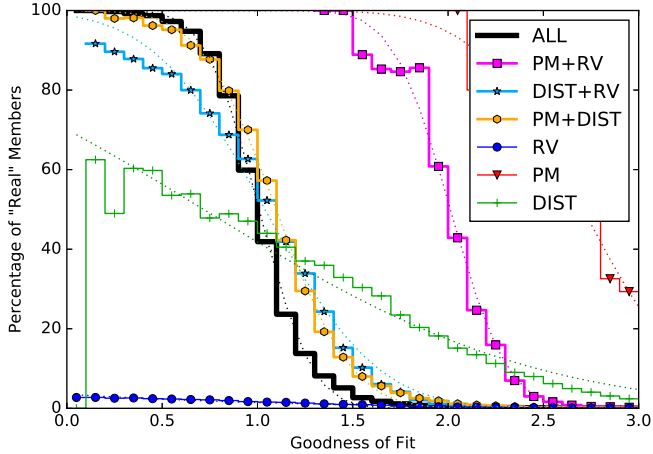}
\caption{Same as Figure \ref{fig:goodness_of_fit}, but for Coma Ber\deleted{enices}. Coma Ber\deleted{encies} has a unique space velocity and spatial position that allows us to identify members more confidently.}
\label{fig:goodness_of_fit_comaber}
\end{figure}

These curves are different for different moving groups; the results of Coma Ber\deleted{erenices} in Field Star mode are shown in Figure \ref{fig:goodness_of_fit_comaber}. With 16 groups, 7 combinations of data, and 2 modes of operation, there are 224 sets of coefficients that make up this implementation of LACEwING. 

For AO Men, our combined goodness-of-fit value was 0.52. In the field star mode histogram (Figure \ref{fig:goodness_of_fit}) and the case of having complete kinematic data, 3632 of the 8 million stars scored between 0.5 and 0.6, and only 1032 of those (28\%) were generated as members of $\beta$ Pic\deleted{)}. If we use the young star mode histograms (Figure \ref{fig:goodness_of_fit_young},) 61\% of the stars in the 0.5-0.6 bin were genuine members of $\beta$ Pic. Using the actual curve fits to compute the probabilities of $\beta$ Pic membership for AO Men yields 26\% (field star mode) and 57\% (young star mode), with an additional 3\% (field star) / 13\% (young star) chance that it is actually a member of Columba according to those curves.

The membership probabilities given by LACEwING are ultimately the complement of the contamination probability: the probability that a star {\it is} a member of a given group and not something else.
This is a subtly different question than, ``what group {\it X} is star {\it Z} a member of?'' The latter question involves comparing different NYMGs directly, and is much more difficult to answer. 
When interpreting LACEwING probabilities, it is important to keep in mind that LACEwING does not force all membership probabilities to add up to 100\% in the way that BANYAN and BANYAN II do; each of the probabilities of membership is an independent assessment of ``Group {\it X} or not Group {\it X}?'' connected only by the fact that all the probability coefficients are derived from the same simulation. Probabilities indicated by LACEwING may add up to more than 100\% if the uncertainties on the input parameters are larger than the typical values in the simulation.
In practice, taking the NYMG that is matched with the highest membership probability is an excellent means of identifying memberships. 

It is important to generate enough simulated stars to properly sample the membership function for even the smallest group - a minimum of roughly 1000 stars are necessary to fit the Gaussian CDF correctly. For the particular implementation of LACEwING presented in this paper, the smallest group was $\eta$ Cha (6 members, out of 40,902), which required at least 6.8 million \added{simulated} stars; 8 million were calculated.

\begin{figure}
\center
\includegraphics[angle=0,width=0.5\textwidth]{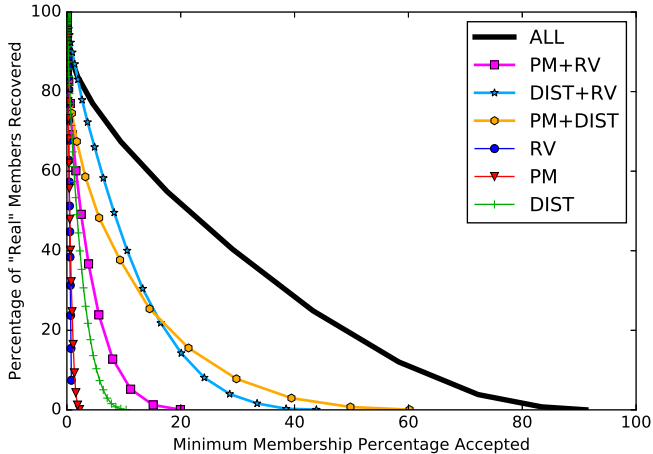}
\caption{Percentage of ``true'' members of $\beta$ Pic (from the simulation) recovered as a function of minimum acceptable membership probability.}
\label{fig:recovery}
\end{figure}

\replaced{Consulting}{From} Figure \ref{fig:goodness_of_fit}, \deleted{we see that }in field star \added{mode and with all kinematic data, }even the best possible match to $\beta$ Pic (goodness-of-fit = 0) has only a 92\% probability of membership in $\beta$ Pic. That is, 8\% of objects that match $\beta$ Pic perfectly are not members. With only proper motion, the maximum possible probability of membership in $\beta$ Pic is only 5\%. The young star mode \replaced{fares better}{yields higher probabilities}; Figure \ref{fig:goodness_of_fit_young} shows a maximum probability of 100\% for $\beta$ Pic members with all information, and a maximum membership probability of 23\% for only proper motion. If we plot the cumulative number of $\beta$ Pic members recovered as a function of minimum probability accepted (Figure \ref{fig:recovery}) we see that we have to set a minimum threshold of 10\% membership probability to recover 90\% of the members in the best possible case where all kinematic data \replaced{is}{are} available.

\begin{figure}
\center
\includegraphics[angle=0,width=0.5\textwidth]{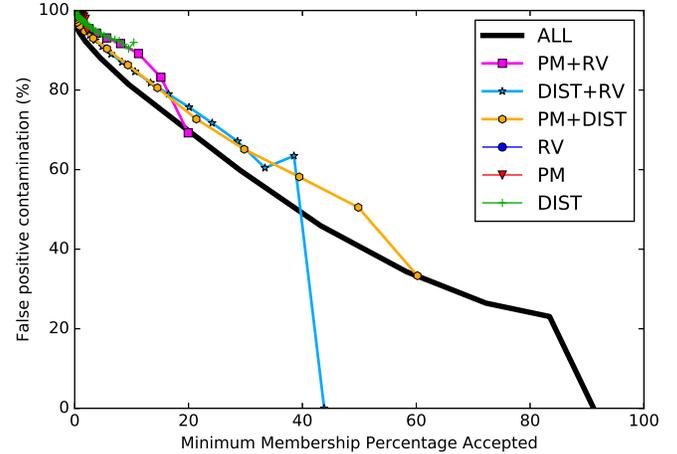}
\caption{Percentage of false positives in a dataset\deleted{,} as a function of the minimum accepted membership percentage in $\beta$ Pic.}
\label{fig:falsepositive}
\end{figure}

\begin{figure}
\center
\includegraphics[angle=0,width=0.5\textwidth]{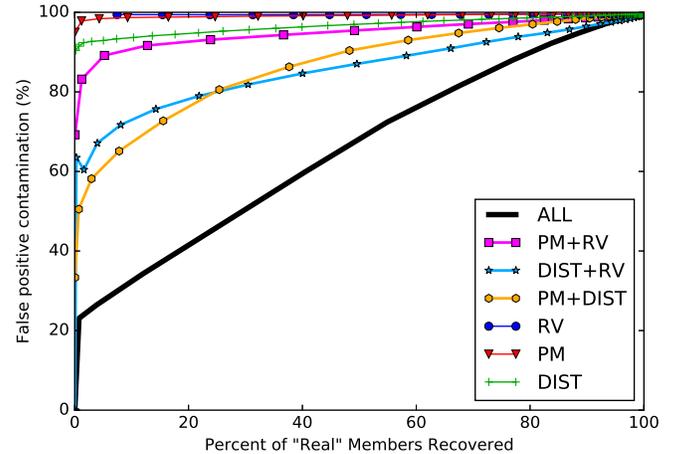}
\caption{The $\beta$ Pic recovery rate of LACEwING as a function of the false positive rate.\deleted{Legend follows Figure \ref{fig:goodness_of_fit}.}}
\label{fig:falserecovery}
\end{figure}

Setting a low membership probability cutoff means a much larger false positive contamination, as shown in Figure \ref{fig:falsepositive}. Selecting an adequate cutoff requires balancing the recovery rate with the contamination rate (Figure \ref{fig:falserecovery}\replaced{,}{;} similar figures describing the BANYAN II Bayesian models can be found in Figures 5 and 6 of \citealt{Gagne2014a}\deleted{, describing the BANYAN II Bayesian models.}) The false positive rates highlight the danger of using kinematics alone to identify young stars: any kinematically selected survey needs to use other spectroscopic and photometric youth indicators to weed out false positives.

Rough guidelines used throughout the rest of this work, and in the stand-alone version of LACEwING, are that probabilities of 66\% and higher are high probabilities of membership, 40-66\% and above are moderate probability, 20-40\% are low probability, and below 20\% is too low to consider meaningful.

Technical details on using LACEwING, recalibrating LACEwING, and incorporating it into other codes are given in the Appendix.

\section{The TRACEwING Epicyclic Traceback Code}
\label{sec:tracewing}

What we wish to accomplish with tracebacks is to identify and reject stars that could not possibly have been near the formation site of a moving group at the time of formation. They may still fall within the UVW velocity (and even XYZ spatial) dispersion of the current distribution, but if we track their position as a function of time they will end up far from the rest of the members.

\subsection{Principles of Traceback}

The TRACEwING code uses an epicyclic approximation of Galactic orbital motion \citep{Makarov2004} to trace the positions of stars back in time using their current measured motion, in increments of 0.1 Myr. It compares the positions of single objects to an NYMG, which is represented by stored freely oriented ellipse parameters fit using the same process used in Section \ref{sec:design}. Based on the equal-volume-radius of the group, the TRACEwING code presents a single-valued representation of how close the target is to the moving group as a function of time.

TRACEwING is essentially two separate steps, carried out by two different programs:
\begin{enumerate}
\item A program that uses the epicyclic kinematic approximation to trace {\it all} bona-fide members of an NYMG back in time, and fits freely-oriented ellipsoids to the ensemble at each time step, saving the parameters for future use.
\item A program that uses the epicyclic kinematic approximation to trace a single star back in time, and compares its positions to saved moving group ellipsoids at each time step.
\end{enumerate}

\subsection{Design of TRACEwING}
With epicyclic traceback, the effects of Galactic orbital motion are approximated by use of sine and cosine functions, controlled by Oort constants and a vertical oscillation parameter. For TRACEwING, we use the equations of position (relative to the Sun, as a function of time in Myr) given in \citet{Makarov2004} and reproduced here: 

\begin{equation}
X = X_0 + U_0\kappa^{-1} \sin{\kappa T} + (V_0 -2AX_0) (1-\cos{\kappa T}) (2B)^{-1}
\end{equation}
\begin{multline}
Y = Y_0 - U_0 (1-\cos{\kappa T})(2B)^{-1} \\ + V_0(A\kappa T - (A - B)\sin{\kappa T}) (\kappa B)^{-1} \\- 2X_0 A (A-B)(\kappa T - \sin{\kappa T}) (\kappa B)^{-1}
\end{multline}
\begin{equation}
Z = Z_0 \cos{\nu T} + W_0\nu^{-1}\sin{\nu T}\\
\end{equation}
In the above equations, A and B are the Oort constants (from \citet{Bobylev2010}, A=$+$0.0178 km s$^{-1}$ pc$^{-1}$, B=$-$0.0132 km s$^{-1}$ pc$^{-1}$), $\kappa$ is the planar oscillation frequency, $\sqrt{-4B\times(A-B)}$, and $\nu$ is the vertical oscillation frequency, $2\pi/85$ Myr$^{-1}$ \added{\citep{Makarov2004}}. 
The approximation deviates noticeably from an unperturbed linear traceback motion after ten million years.

For the first step of TRACEwING, we calculate 4000 Monte Carlo tracebacks for each bona-fide member, in time increments of 0.1 Myr. At each time step, freely-oriented ellipses are fit to the positions of the members of the moving group for each of the 4000 Monte Carlo trials, and averaged to produce a mean position, extent, and orientation of the group at that time. These parameters are saved for comparison to individual stars.

\begin{figure}
\center
\includegraphics[angle=0,width=.5\textwidth]{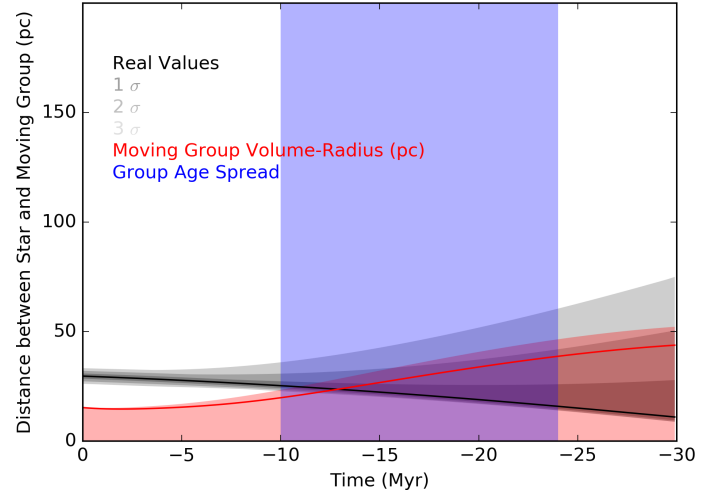}
\caption{Separation between the star AO Men and the center of the $\beta$ Pic moving group (black line), as a function of time. The dark, medium, and light gray regions show the 1, 2, and 3-$\sigma$ uncertainties on the separation between AO Men and $\beta$ Pic. The red line and red region show the volume-equivalent radius of $\beta$ Pic itself (and its 1-$\sigma$ uncertainty). The blue region represents the range of ages of $\beta$ Pic. AO Men's kinematics (black line and dark gray region) place it plausibly within the radius of $\beta$ Pic (red region) at the time of formation (blue region).}
\label{fig:traceback}
\end{figure}

For the second step, we take a target object of interest, which may be any object with full kinematic information - star, brown dwarf, or planet. To determine the potential memberships of the target object, we generate and trace back 20,000 trials within each of the 1$\sigma$, 2$\sigma$, and 3$\sigma$ uncertainties on their observational (equatorial) positions and motions. At each timestep, the distance between the object and the previously calculated position of the moving group (both in Cartesian Galactic XYZ coordinates) is calculated, and an equal-volume radius ($\frac{4}{3}\pi r^{3}=\frac{4}{3}\pi abc$) is used as the effective radius of the group. The targets are then visually classified by whether the 1$\sigma$, 2$\sigma$, or 3$\sigma$ positions potentially place them within the effective radius of the group at the time of formation. The traceback of the bona-fide $\beta$ Pic member AO Men is displayed in Figure \ref{fig:traceback}, and shows a star that was plausibly within the confines of $\beta$ Pic at the time of formation.

\subsection{Traceback Limitations}
\label{sec:traceback_limitations}

Epicyclic approximations do not take into account the gravitational influence of other stars, molecular clouds, Gould's Belt, or the Galactic disk and bar itself. This is most pronounced in open clusters, where the stars themselves are gravitationally bound to each other, but sets limits on the reliability of the technique for moving groups as well. 

To quantify the limits of the technique, we perform two tests. First, we simulate a ``real'' moving group of stars, move it forward in time, ``observe'' it, and trace it back in time to see what a genuine NYMG of various ages should look like traced back to its origin. Second, we select unrelated field stars with a velocity and spatial distribution similar to known NYMGs, and move it back in time as a ``fake'' NYMG. The upper limit on reliability of the traceback technique is the point at which the ``real'' moving group is indistinguishable in volume from the ``fake'' one.

Our ``real'' moving group \replaced{of stars is 50 stars}{is a simulated group of 50 stars} with a Gaussian velocity dispersion of 1.5 km s$^{-1}$ \citep{Preibisch2008} typical of the Sco-Cen star-forming region, and a uniform spatial distribution with a radius of 5 pc. These values are perhaps smaller than most clusters, but provide a best-case scenario. The 50 stars were moved forward in time in steps of 0.1 Myr using the epicyclic approximation and their positions (with the mean position subtracted off, so all stars were near the origin at the final epoch) were saved at 5, 8, 12, 25, 45, 50, 125, 250, 400, and 500 Myr intervals. To observe the stars, we generated new UVW velocities using \replaced{their most recent implied motion: }{the individual stars' change of position over the last 0.1 Myr of the simulated time range, e.g.:}
\begin{equation}
U=0.9778\frac{\textrm{km~Myr}}{\textrm{s}^{-1}~\textrm{pc}}\Delta{X}\frac{\textrm{pc}}{\textrm{0.1 Myr}}
\end{equation}
\added{We then }converted all UVWXYZ values to equatorial coordinates, and applied randomly generated ``observational'' errors: 0.5 mas $\pi$~uncertainties, 10 mas yr$^{-1}$ $\mu$ uncertainties in each axis, and 1 km s$^{-1}$ $\gamma$ uncertainties. These collections of stars \replaced{are run back in time 5, 8, 12, 25, 45, 50, 125, 250, 400, or 500 Myr to determine}{were run back in time same as before to determine} the apparent size at formation. To test the trivial case (perfect information), we \deleted{have} added no observational errors to the generated cluster of stars, and traced them back in time.

To assemble the group of field stars, we searched the Extended {\it Hipparcos} catalog of \citet{Anderson2012} to find stars with $\mu$, $\pi$, and $\gamma$ distributed according to the present-day median \added{velocity dispersion and spatial distribution} parameters of our unbound moving groups, with a velocity dispersion of 1.6 km s$^{-1}$, and spatial distribution within 11.5 pc. Our fake moving group is centered on XYZ=(-5,-5,20) pc and UVW=(-5,-5,-5) km s$^{-1}$, a well-populated region of velocity space not populated by any known NYMG. The 15 selected stars are given in Table \ref{tab:fake_cluster_table}.

\tabletypesize{\tiny}
\begin{deluxetable}{lrrrrrrrr}
\setlength{\tabcolsep}{0.01in}
\tablewidth{0pt}
\tablecaption{Stars in Fake Cluster\label{tab:fake_cluster_table}}
\tablehead{
  \colhead{HIP}      &
  \multicolumn{2}{c}{$\alpha$} &
  \multicolumn{2}{c}{$\delta$} &
  \colhead{$\pi$}     &
  \colhead{$\mu_{\alpha}$} &
  \colhead{$\mu_{\delta}$} &
  \colhead{$\gamma$}  \\
  \colhead{}          &
  \colhead{(deg ICRS)}&
  \colhead{(mas)}     &
  \colhead{(deg ICRS)}&
  \colhead{(mas)}     &
  \colhead{(mas)}     &
  \multicolumn{2}{c}{(mas yr$^{-1}$)} &
  \colhead{(km s$^{-1}$)} }
\startdata
3025  & 009.63263762 & 0.29 & -20.29659788 & 0.21 &  3.97$\pm$0.39 & $+$22.04$\pm$0.36 &  $+$7.69$\pm$0.25 & $+$7.7$\pm$2.9 \\
6206  & 019.88915070 & 0.55 & -39.36272354 & 0.42 & 26.19$\pm$0.75 &  $-$4.57$\pm$0.57 & $+$50.96$\pm$0.47 & $+$2.2$\pm$0.3 \\
8497  & 027.39662951 & 0.19 & -10.68618052 & 0.17 & 43.13$\pm$0.26 &$-$148.10$\pm$0.70 & $-$95.70$\pm$0.70 & $-$1.8$\pm$0.9 \\
42753 & 130.69264237 & 0.49 & +31.86270165 & 0.29 & 19.90$\pm$0.56 & $-$33.68$\pm$0.53 & $-$40.85$\pm$0.38 & $+$5.1$\pm$0.2 \\
60406 & 185.78502901 & 0.97 & +25.85139632 & 0.67 & 11.55$\pm$1.12 & $-$10.54$\pm$0.98 &  $-$7.75$\pm$0.69 & $+$1.0$\pm$0.6 \\
60797 & 186.90988135 & 0.37 & +25.91212788 & 0.28 & 12.58$\pm$0.44 & $-$10.92$\pm$0.55 &  $-$8.79$\pm$0.35 & $-$0.2$\pm$0.7 \\
62805 & 193.04843110 & 0.88 & +25.37351431 & 0.79 & 13.40$\pm$1.17 & $-$11.79$\pm$0.78 &  $-$8.03$\pm$0.71 & $-$4.4$\pm$3.4 \\
66657 & 204.97196970 & 0.28 & -53.46636266 & 0.31 &  7.63$\pm$0.48 & $-$15.30$\pm$0.36 & $-$11.72$\pm$0.36 & $+$3.0$\pm$2.5 \\
68634 & 210.73708216 & 0.37 & +14.97533527 & 0.32 & 37.27$\pm$0.54 & $-$58.12$\pm$0.38 &  $+$3.26$\pm$0.33 & $-$8.9$\pm$0.3 \\
69732 & 214.09656815 & 0.11 & +46.08791912 & 0.12 & 32.94$\pm$0.16 &$-$187.31$\pm$0.14 &$+$159.05$\pm$0.11 & $-$7.9$\pm$1.6 \\
69917 & 214.62984573 & 0.28 & +52.03332847 & 0.33 & 10.27$\pm$0.38 & $-$17.31$\pm$0.34 &  $-$2.96$\pm$0.36 &$-$10.0$\pm$4.3 \\
73941 & 226.64663091 & 0.24 & +36.45596430 & 0.28 & 33.52$\pm$0.36 & $-$64.47$\pm$0.26 & $+$40.55$\pm$0.32 & $-$5.8$\pm$0.1 \\
75363 & 231.01012301 & 0.63 & -27.30511484 & 0.29 & 35.02$\pm$0.65 &  $-$2.44$\pm$0.68 & $-$44.73$\pm$0.52 & $-$6.6$\pm$0.3 \\
79375 & 243.00000643 & 1.27 & -10.06418349 & 0.95 & 20.96$\pm$1.36 &  $-$9.70$\pm$1.00 & $-$14.40$\pm$1.00 & $-$5.1$\pm$1.0 \\
97070 & 295.91455287 & 0.34 & +57.04265613 & 0.39 & 12.86$\pm$0.39 &  $+$3.83$\pm$0.33 & $+$24.12$\pm$0.50 &$-$27.3$\pm$0.2 \\
\hline
\enddata
\tablecomments{All stars in the eXtended {\it Hipparcos} Catalog \citep{Anderson2012} with full kinematic information within 11.5 pc of XYZ=(-5,-5,20) pc and 1.6 km s$^{-1}$ of UVW=(-5,-5,-5) km s$^{-1}$.}
\end{deluxetable}

\begin{figure}
\center
\includegraphics[angle=0,width=.5\textwidth]{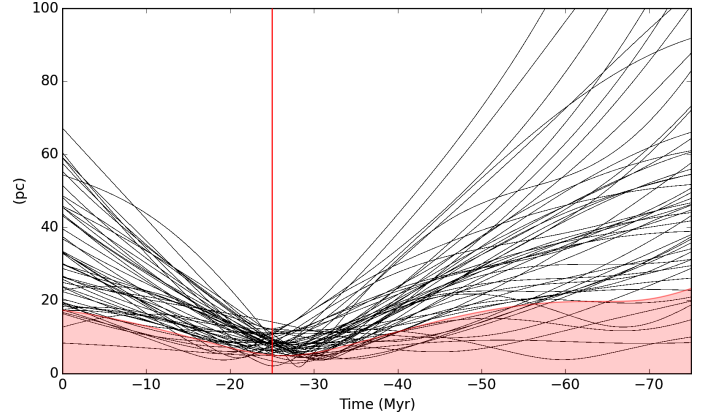}
\caption{Separations between generated ``real'' stars moved forward 25 Myr and the center of their simulated moving group when traced back, as a function of time. The one-sigma volume equivalent radius, representing the volume of the ellipse fit to the calculated positions of the bona-fide member draws, is represented by the red curve; the time of formation is shown by the vertical red line.}
\label{fig:perfect25}
\end{figure}

\begin{figure}
\center
\includegraphics[angle=0,width=.5\textwidth]{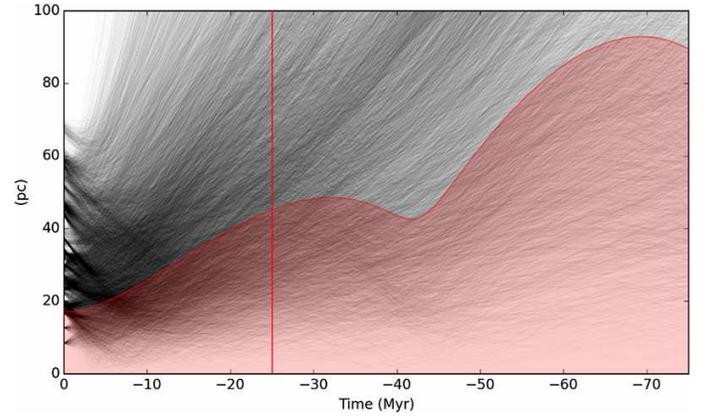}
\caption{Same as Figure \ref{fig:perfect25}, but with realistic measurement errors added to the sample before tracing the stars back to their position 25 Myr in the past. The stars no longer converge to a small effective radius in the past, and do not exhibit any kind of minimum at or near the time of formation 25 Myr ago.}
\label{fig:sample25}
\end{figure}

\begin{figure*}
\center
\includegraphics[angle=0,width=\textwidth]{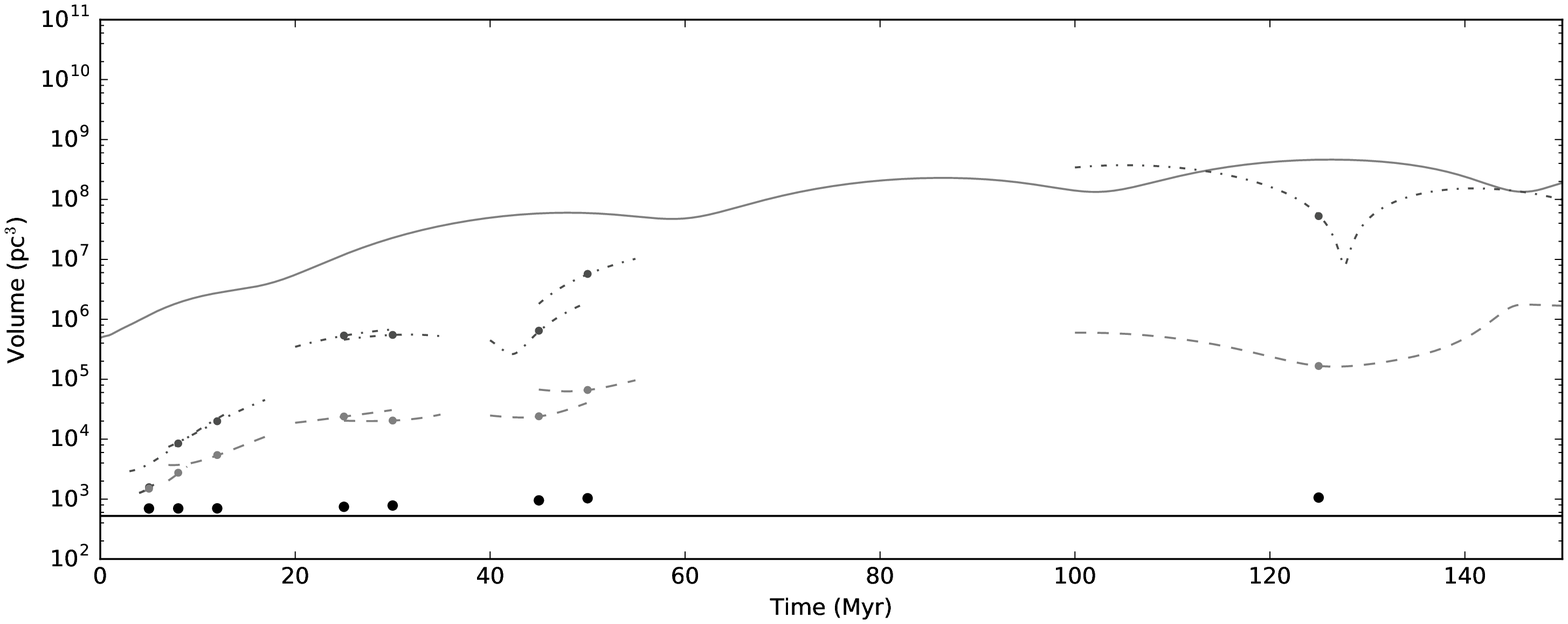}
\includegraphics[angle=0,width=\textwidth]{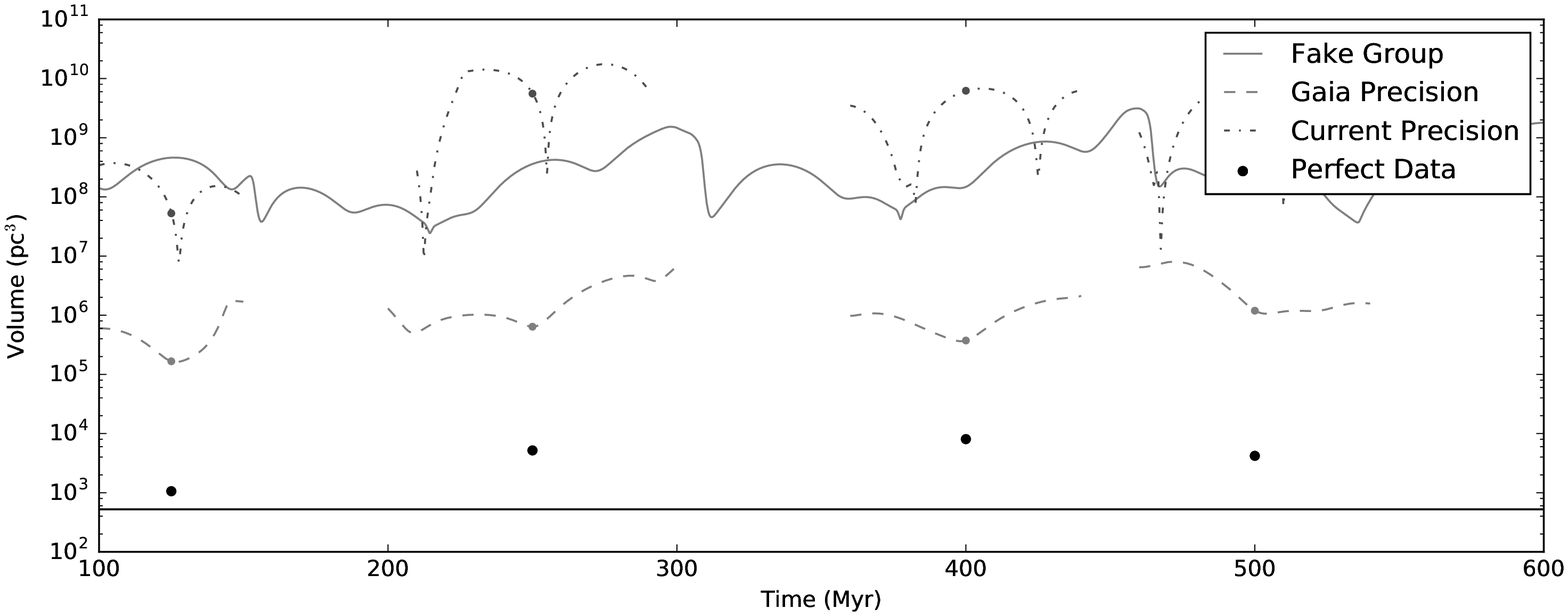}
\caption{Demonstration of the volume-at-formation of our simulated moving group as a function of time-since-formation and observational precision, on top from 0-150 Myr; on bottom, from 100-600 Myr.
The cases with current precision kinematics are shown with dash-dot lines surrounding the formation time of the group. The {\it Gaia}-precision cases are shown with dashed lines. For perfect data (no errors), the volumes at the times of formation are shown as black points. The fake moving group of field stars is shown as a solid curve.}
\label{fig:traceback_explanation}
\end{figure*}

In the trivial case with no observational uncertainties (Figure \ref{fig:perfect25}), the synthetic group traces back to having its smallest volume (although it is larger than the 5 pc initial radius) near the actual time of formation. In the more realistic case (Figure \ref{fig:sample25}), the large measurement errors mean that the minimum size of the simulated moving group appears to be right now, and the apparent effective radius of the simulated moving group at time of formation is 40 pc. In Figure \ref{fig:traceback_explanation} we plot the estimated volume-at-formation of the simulated moving group at various ages, and the fake moving group of field stars moved back in time.

\begin{figure}
\center
\includegraphics[angle=0,width=.5\textwidth]{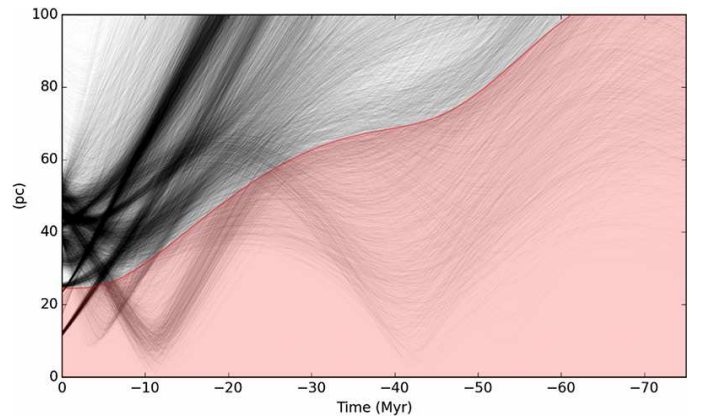}
\caption{Same as Figure \ref{fig:sample25}, but showing the results of our selected ``fake'' sample of field stars when they are traced back in time.}
\label{fig:fakegroup}
\end{figure}

In the case of the fake moving group made up of unrelated field stars (Figure \ref{fig:fakegroup}), the volume is {\it smaller} than the simulated moving group (Figure \ref{fig:traceback_explanation}) after 125 Myr. The epicyclic approximation's Galactic shear causes the simulated group to have a larger present-day velocity dispersion than is currently expected for the known NYMGs, suggesting that we are missing outlying members of the known groups. Crucially, these stars should trace back \replaced{TOWARDS}{towards} the origin of the moving group as they do in our simulated moving group example, unlike the outliers we intend to remove from current membership lists.

\begin{figure}
\center
\includegraphics[angle=0,width=.5\textwidth]{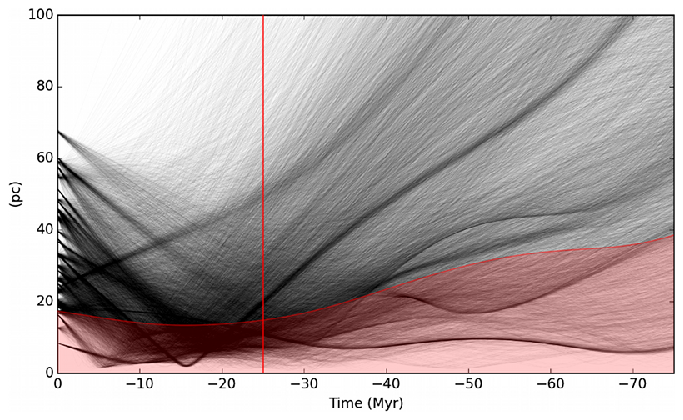}
\caption{Same as Figure \ref{fig:sample25}, but with measurement errors consistent with {\it Gaia} astrometry for $Gaiamag\sim$15 stars.}
\label{fig:gaia25}
\end{figure}

The {\it Gaia} mission will provide astrometry that is several magnitudes more precise than is currently available for stars brighter than \replaced{$Gaiamag \approx$20.7}{$G \approx$20.7}. To determine the effectiveness of {\it Gaia} data, we have repeated our first test using \replaced{typical {\it Gaia} measurement uncertainties for}{measurement uncertainties expected for} \replaced{$Gaiamag\sim15$}{$G\sim15$} sources \citep{GaiaCollaboration2016} of 50 $\mu$as for parallaxes, 40 $\mu$as for positions, and 25 $\mu$as yr$^{-1}$ for each axis of proper motions. The simulated {\it Gaia}-precision data demonstrate significant improvements. The size of the simulated burst of stars shrinks by a factor of 1000 (roughly linearly with the increased precision). As late as 25 Myr (Figure \ref{fig:gaia25}) there is still a minimum (although not the correct radius) in the spatial distribution of the stars near the actual age of the group, and the radius is now relatively constant over time.

We can conclude a few things about the efficacy of tracebacks from these tests, presented here in no particular order:

\begin{enumerate}
\item The effectiveness of tracebacks is almost entirely limited by current measurement precision\added{ of the parameters $\alpha$, $\delta$, $\pi$, $\mu$, and $\gamma$}. After roughly 125 Myr, the cumulative effects of measurement uncertainties make it impossible to distinguish between a group of stars spreading out from a single point of origin, and an unphysical collection of field stars on roughly parallel tracks. We cannot, therefore, comment on the existence of any moving groups older than 125 Myr based on epicyclic tracebacks alone. As measurement uncertainties shrink, we will gradually approach the ``perfect'' case (Figure \ref{fig:perfect25}) where genuinely related stars will trace back to smaller volumes than unrelated stars selected with a similar velocity dispersion.
\item We cannot say anything about the membership of objects that trace back to within the boundary of the moving group at the time of formation - with the currently available measurement precision they could be true members or coincidental field stars. However, stars that do not trace back to possibly be within the confines are a different case. We found that only \replaced{six percent}{6\%} of simulated members in our 45 Myr old sample were not plausibly within the boundary of the moving group (at the 1$\sigma$ positional uncertainty level) 45 Myr in the past. In contrast, far more than 6\% of actual moving group members (Section \ref{sec:bonafides}) did not trace back to within the confines of their purported moving group, suggesting that the objects are not real members of the group. These are easily identifiable nonmembers, and as data precision improves, we expect to find more of these objects.
\item Very few of the stars in the fake moving group are consistent with possibly being in the center of the moving group (or really, anywhere except the edges of the group), while in the simulated moving group, nearly {\it all} the members are consistent with being in the center of the NYMG. This would seem to be one difference between an actual NYMG and an unphysical selection of stars. 
\end{enumerate}

\section{The Catalog of Suspected Young Stars}
\label{sec:catalog}

Calibrating and testing LACEwING requires kinematic information on genuine members. For this implementation of LACEwING, the bona-fide sample (Section \ref{sec:bonafides}) and proof-of-concept lithium sample (Section \ref{sec:lithium}) come from a catalog of all known and suspected young stars maintained by the authors.

\begin{figure}
\center
\includegraphics[angle=0,width=0.5\textwidth]{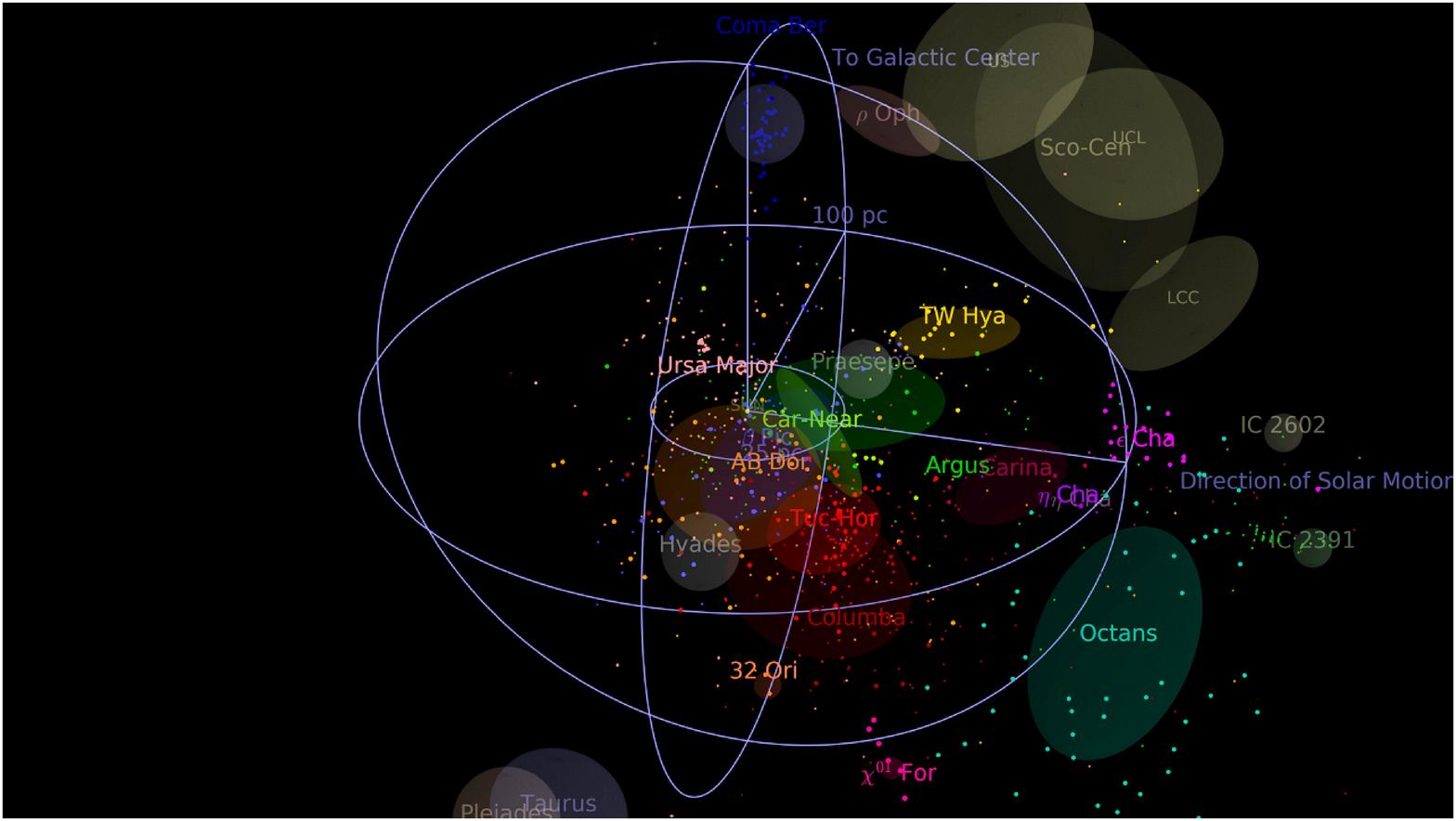}
\caption{\added{A 3D representation of the positions of all NYMG stars in the Catalog of Suspected Nearby Young Stars with either parallaxes or kinematic distances, as seen from a vantage point at $\alpha=200,\delta=+20,Dist=200$ pc. Ellipsoids of the moving groups, and the final bona-fide members shown as larger points, are taken from Section \ref{sec:bonafides}. Rough positions of other open clusters and star-forming regions within $\sim$200 pc are shown for scale and orientation only. (An animation of this figure is available at \url{https://youtu.be/XQ92I5OF3_U}).}}
\label{fig:catalog_movie}
\end{figure}

The catalog \citep{Riedel2016d} is intended to contain basic information on every star, planetary-mass object, and brown dwarf in all nearby (\added{$Dist <$}100 pc) star systems ever reported as young, to provide a single resource for studying the individual and ensemble properties of young stars. It currently contains 5350 objects. Through careful literature searches, this catalog can reasonably be considered complete for membership in the NYMGs published through 2015 January. A similar effort has not been made for the Hyades and Coma Ber open clusters, and they cannot be considered complete, nor does the catalog necessarily contain all information known about the young targets. \added{An animated 3D representation of the catalog is shown in Figure \ref{fig:catalog_movie} plotted in XYZ coordinates.}

As the table is quite large, a list of its headers is given in Table \ref{tab:catalog}, and the full machine-readable table is available  online\footnote{\url{https://github.com/~ariedel/young_catalog}} in CDS format, a comma-separated value file, an OpenDocument Spreadsheet, and an Office Open XML spreadsheet. The catalog was constructed from a wide variety of source papers that reported young stars\added{, identified through a literature search and from other papers collecting datasets \citep{Zuckerman2013,Alonso-Floriano2015}. The full list is}\deleted{ and their properties,} given in Table \ref{tab:sourcetable}.

\startlongtable
\tabletypesize{\tiny}
\begin{deluxetable}{ll} 
\setlength{\tabcolsep}{0.04in}
\tablewidth{0pt} 
\tablecaption{Source Papers\label{tab:sourcetable}}
\tablehead{
 \colhead{Citation} &
 \colhead{Groups Included\tablenotemark{a}} }
 
\startdata
\citet{Eggen1991} & IC 2391 Supercluster \\
\citet{Barrado-y-Navascues1998} & Castor \\
\citet{Makarov2000} & Car-Vela \\
\citet{van-den-Ancker2000} & Capricornus \\
\citet{Montes2001a} & Multiple \\
\citet{Montes2001b} & Multiple \\
\citet{King2003} & Ursa Major \\
\citet{Ribas2003} & Castor \\
\citet{Zuckerman2004} & Multiple \\
\citet{Mamajek2005} & TW Hya \\
\citet{Casewell2006} & Coma Ber \\
\citet{Lopez-Santiago2006} & Multiple \\
\citet{Moor2006} & Multiple \\
\citet{Zuckerman2006} & Car-Near \\
\citet{Guenther2007} & \deleted{$\epsilon$ Cha, }Multiple \\
\citet{Kraus2007} & Coma Ber \\
\citet{Makarov2007} & Multiple \\
\citet{Platais2007} & IC 2391 \added{Cluster}\\
\citet{Stauffer2007} & Pleiades \\
\citet{Kirkpatrick2008} & TW Hya \\
\citet{Mentuch2008} & Multiple \\
\citet{Teixeira2008} & TW Hya \\
\citet{Torres2008} & Multiple \\
\citet{Da-Silva2009} & Multiple \\
\citet{Guillout2009} & Other \\
\citet{Lepine2009} & $\beta$ Pic \\
\citet{Shkolnik2009} & Other \\
\citet{Teixeira2009} & $\beta$ Pic, TW Hya \\
\citet{Caballero2010} & Castor \\
\citet{Lopez-Santiago2010} & Multiple \\
\citet{Lopez-Santiago2010b} & $\eta$ Cha \\
\citet{Maldonado2010} & Multiple \\
\citet{Murphy2010} & $\eta$ Cha \\
\citet{Nakajima2010} & Multiple \\
\citet{Rice2010} & $\beta$ Pic \\
\citet{Schlieder2010} & $\beta$ Pic, AB Dor \\
\citet{Yee2010} & $\beta$ Pic \\
\citet{Kiss2011} & Multiple \\
\citet{Riedel2011} & Argus \\
\citet{Rodriguez2011} & TW Hya, Sco-Cen \\
\citet{Shkolnik2011} & TW Hya \\
\citet{Wahhaj2011} & AB Dor \\
\citet{Zuckerman2011} & Multiple \\
\citet{Kastner2012} & $\epsilon$ Cha \\
\citet{McCarthy2012} & $\beta$ Pic, AB Dor \\
\citet{Mishenina2012} & Other \\
\citet{Murphy2012} & Other \\
\citet{Nakajima2012} & Multiple \\
\citet{Schlieder2012a} & $\beta$ Pic, AB Dor \\
\citet{Schlieder2012c} & $\beta$ Pic, AB Dor \\
\citet{Shkolnik2012} & Multiple \\
\citet{Schneider2012a} & TW Hya \\
\citet{Schneider2012b} & TW Hya \\
\citet{Xing2012} & Other \\
\citet{Barenfeld2013} & AB Dor \\
\citet{Delorme2013} & Tuc-Hor \\
\citet{De-Silva2013} & Argus \\
\citet{Eisenbeiss2013} & Her-Lyr \\
\citet{Hinkley2013} & Other \\
\citet{Liu2013b} & $\beta$ Pic \\
\citet{Malo2013} & Multiple \\
\citet{Mamajek2013} & Other \\
\citet{Moor2013} & Multiple \\
\citet{Murphy2013} & $\eta$ Cha, $\epsilon$ Cha \\
\citet{Rodriguez2013} & Multiple \\
\citet{Schneider2013} & Other \\
\citet{Zuckerman2013} & Oct-Near \\
\citet{Weinberger2013} & TW Hya \\
\citet{Binks2014} & $\beta$ Pic \\
\citet{Casewell2014} & Coma Ber, Hyades \\
\citet{Klutsch2014} & Multiple \\
\citet{Kraus2014a} & Tuc-Hor \\
\citet{Gagne2014a} & Multiple \\
\citet{Gagne2014b} & TW Hya \\
\citet{Gagne2014c} & Argus \\
\citet{Malo2014a} & Multiple \\
\citet{Malo2014b} & \deleted{$\beta$ Pic, }Multiple \\
\citet{Mamajek2014} & $\beta$ Pic \\
\citet{McCarthy2014} & AB Dor \\
\citet{Riedel2014} & Multiple \\
\citet{Rodriguez2014} & $\beta$ Pic \\
\citet{Schneider2014} & Other \\
\citet{Gagne2015} & Multiple \\
\citet{Murphy2015} & Octans \\
E.E. Mamajek (2016, private communication) & Multiple \\
\enddata
\tablecomments{The papers that make up the membership of the catalog of young stars, along with the groups considered.}
\tablenotetext{a}{Papers marked ``Multiple'' consider multiple groups; papers marked ``Other'' consider nearby young stars but do not identify them as members of any groups.}
\end{deluxetable}

It is assumed that the relevant results of earlier papers not on this list (\replaced{e.x.}{e.g.} \citealt{de-la-Reza1989}; \citealt{Kastner1997}; \citealt{Webb1999}; \citealt{Torres2000}; \citealt{Song2002}; \citealt{Torres2003}) have been superceded by or included in the more recent papers on the NYMGs (Table \ref{tab:sourcetable}).

Within the catalog, 1312 of the 5350 total objects have never been reported as members of any group, including groups not believed to be real (Castor), classical pre-{\it Hipparcos} moving groups like the Local Association, or more distant groups like Upper Centaurus Lupus or Cham{\ae}leon I. These nonmembers fall into four categories:
\begin{itemize}
\item Objects with ambiguous membership, that match multiple groups equally well, as reported by \citet{Moor2006}, \citet{Malo2013,Malo2014a,Malo2014b}, and \citet{Gagne2014a}.
\item Young objects that do not match any known group, as reported by \citet{Shkolnik2009,Shkolnik2011}, \citet{Maldonado2010}, \citet{Murphy2012}, \citet{Gagne2014a}, \citet{Schneider2013}, \citet{Riedel2014}, and \citet{Schneider2014}.
\item Young objects reported in papers that did not consider group memberships, as reported by \citet{Guillout2009}, \citet{Mishenina2012}, and \citet{Xing2012}.
\item Field (and variants like ``young disk'') objects considered in papers on young stars and never reported as young, particularly from \citet{Montes2001a}, \citet{Makarov2007}, \citet{Shkolnik2009}, \citet{Lopez-Santiago2010}, \citet{Maldonado2010}, \citet{Rodriguez2011}, \citet{Shkolnik2011}, \citet{McCarthy2014}, \citet{Gagne2014a}, \citet{Riedel2014}, and \citet{Klutsch2014}.
\end{itemize}

All data relevant to population studies of stellar youth and membership have been taken from these source papers. Care has been taken to homogenize the data as much as possible: upper limit flags have been added, H\deleted{-}$\alpha$ equivalent width (EW) has been standardized as negative when in emission, and lithium EW is uniformly recorded in milliangstroms.

\subsection{Survey Sources}
\label{sec:surveys}

The catalog is supplemented by additional data sources drawn from large surveys. This aids in providing useful information about the stars, and strengthens the consistency of the source data.

\subsubsection{Astrometry}

Positions and Proper motions were preferentially sourced from the following ICRS catalogs tied to the {\it Hipparcos} reference frame:
\begin{enumerate}
\item {\it Hipparcos} 2 \citep{van-Leeuwen2007}, 1915 objects, 0.1--3.0 mas position precision, 0.1--5.0 mas yr$^{-1}$ $\mu$ precision.
\item UCAC4 \citep{Zacharias2013}, 1527 objects, 10--100 mas \added{$\alpha\delta$} precision, 1--10 mas yr$^{-1}$ $\mu$ precision.
\item \replaced{TYCHO-2}{Tycho-2} \citep{Hoeg2000}, 906 objects, 10--100 mas \added{$\alpha\delta$} precision, 1--10 mas yr$^{-1}$ $\mu$ precision.
\item PPMXL \citep{Roeser2010}, 712 objects, 50--100 mas \added{$\alpha\delta$} precision, 5--20 mas yr$^{-1}$ $\mu$ precision
\item 2MASS and 2MASS-6X \citep{Skrutskie2006}. 286 objects, 60--120 mas \added{$\alpha\delta$} precision, no $\mu$
\item SDSS DR9 \citep{Ahn2012} 4 objects, 80--200 mas \added{$\alpha\delta$} precision, no $\mu$.
\item Source papers (from Table \ref{tab:sourcetable}). These mostly provide proper motions (5--20 mas yr$^{-1}$ $\mu$ precision) for 261 stars found in 2MASS or 2MASS-6X. The other 23 have proper motions copied from a primary star.
\end{enumerate}
There are still 43 objects (35 of which are in the Pleiades) that have no reported proper motions. Parallaxes were sourced from papers listed in Table \ref{tab:surveytable}. Where available, parallaxes for objects in multiple systems and multiple observations of the same stars were combined into weighted mean system parallaxes, under the assumptions that the published uncertainties are accurate and that the parallax of every component of the system is the same to within measurement errors.

\startlongtable
\tabletypesize{\tiny}
\begin{deluxetable}{ll} 
\setlength{\tabcolsep}{0.04in}
\tablewidth{0pt} 
\tablecaption{Survey Papers\label{tab:surveytable}}
\tablehead{
 \colhead{Citation} &
 \colhead{Data Used}}
 
\startdata
\citet{Houk1975} & Spectral Types \\
\citet{Houk1978} & Spectral Types \\
\citet{Houk1982} & Spectral Types \\
\citet{Andersen1991} & Multiplicity \\
\citet{Gliese1991} & Spectral Types \\
\citet{Hoffleit1991} & \deleted{Bright Star }Catalog Names \\
\citet{Kirkpatrick1991} & Spectral Types \\
\citet{Cannon1993} & \deleted{Henry Draper }Catalog Names \\
\citet{Gatewood1993} & $\pi$ \\
\citet{Gatewood1995b} & $\pi$ \\
\citet{Gatewood1995c} & $\pi$ \\
\citet{Reid1995} & Spectral Types \\
\citet{van-Altena1995} (YPC4) & $\pi$, Opt phot. \\
\citet{Covino1997} & $\gamma$, Li \\
\citet{Benedict1999} & $\pi$ \\
\citet{Soederhjelm1999} & $\pi$\tablenotemark{a} \\
\citet{Voges1999} (RASS-BSC) & X-ray phot. \\
\citet{Weis1999} & $\pi$ \\
\citet{Barbier-Brossat2000} (GCMRV) & $\gamma$ \\
\citet{Benedict2000} & $\pi$ \\
\citet{Ducati2001} & Opt phot. \\
\citet{Hoeg2000} (TYCHO-2) & pos., $\mu$, Opt phot. \\
\citet{Voges2000} (RASS-FSC) & X-ray phot. \\
\citet{Dahn2002} & $\pi$ \\
\citet{Gizis2002} & $\gamma$ \\
\citet{Henry2002} & Spectral Types \\
\citet{Nidever2002} & $\gamma$ \\
\citet{Torres2002} & $\pi$\tablenotemark{a}\\
\citet{Cutri2003} (2MASS) & pos., NIR phot. \\
\citet{Song2003} & $\gamma$ \\
\citet{Thorstensen2003} & $\pi$ \\
\citet{McArthur2004} & $\pi$ \\
\citet{Pourbaix2004} (SB9) & Multiplicity \\
\citet{Vrba2004} & $\pi$ \\
\citet{Costa2005} & $\pi$ \\
\citet{Jao2005} & $\pi$ \\
\citet{Lepine2005a} & $\mu$ \\
\citet{Soderblom2005} & $\pi$ \\
\citet{Valenti2005} & $\gamma$, v$\sin$i \\
\citet{Benedict2006} & $\pi$ \\
\citet{Gontcharov2006} & $\gamma$ \\
\citet{Gray2006} & Spectral Types \\
\citet{Henry2006} & $\pi$, Opt phot. \\
\citet{Torres2006} & $\gamma$, H$\alpha$, Li, v$\sin$i \\
\citet{Biller2007} & $\pi$ \\
\citet{Close2007} & Spectral Types \\
\citet{Daemgen2007} & Multiplicity \\
\citet{Gizis2007} & $\pi$, phot. \\
\citet{Kharchenko2007} & $\gamma$ \\
\citet{Scholz2007} & $\gamma$ \\
\citet{van-Leeuwen2007} ({\it Hipparcos}-2) & $\pi$, pos., $\mu$ \\
\citet{Ducourant2008} & $\pi$ \\
\citet{Fernandez2008} & $\gamma$ \\
\citet{Jameson2008} & $\mu$ \\
\citet{Gatewood2009} & $\pi$ \\
\citet{Subasavage2009} & $\pi$ \\
\citet{Teixeira2009} & $\pi$ \\
\citet{Bergfors2010} & Multiplicity \\
\citet{Blake2010} & $\gamma$ \\
\citet{Raghavan2010} & Multiplicity \\
\citet{Riedel2010} & $\pi$, Opt phot. \\
\citet{Roeser2010} (PPMXL) & position, $\mu$, NIR phot. \\
\citet{Shkolnik2010} & Multiplicity, $\gamma$ \\
\citet{Smart2010} & $\pi$ \\
\citet{Stauffer2010} & \deleted{Gliese }Catalog Names \\
\citet{Bianchi2011} (GALEX DR5) & UV phot. \\
\citet{Girard2011} (SPM4) & Opt phot. \\
\citet{Messina2011} & Spectral Types \\
\citet{Moor2011} & $\gamma$ \\
\citet{von-Braun2011} & $\pi$ \\
\citet{Ahn2012} (SDSS DR9) & pos., SDSS phot. \\
\citet{Allen2012} & $\mu$ \\
\citet{Bailey2012} & $\gamma$ \\
\citet{Bowler2012a} & Multiplicity \\
\citet{Bowler2012b} & Multiplicity \\
\citet{Dupuy2012} & $\pi$ \\
\citet{Faherty2012} & $\pi$ \\
\citet{Janson2012} & Multiplicity \\
\citet{Zacharias2013} (UCAC4) & pos., $\mu$, Opt, SDSS, NIR phot.\\
\citet{Bowler2013} & Multiplicity \\
\citet{Cutri2013} (ALLWISE) & MIR phot. \\
\citet{Kordopatis2013} (RAVE DR4) & $\gamma$ \\
\citet{Liu2013a} & $\pi$ \\
\citet{Marocco2013} & $\pi$ \\
\citet{Dieterich2014} & $\pi$, Opt phot. \\
\citet{Dittmann2014} & $\pi$ \\
\citet{Ducourant2014} & $\mu$, $\pi$ \\
\citet{Lurie2014} & $\pi$, Opt phot. \\
\citet{Naud2014} & Multiplicity \\
\citet{Zapatero-Osorio2014a} & $\pi$ \\
\citet{Elliott2015} & $\gamma$ \\
\citet{Henden2015} (APASS DR9) & Opt, SDSS phot. \\
\citet{Mason2015} (WDS) & Multiplicity \\
\citet{Faherty2016} & $\gamma$ \\
\enddata
\tablecomments{Additional Data Sources used in the catalog.}
\tablenotetext{a}{Parallax replaces {\it Hipparcos} data}
\end{deluxetable}

\subsubsection{Radial Velocity}
\label{sec:radialvelocity}
Radial velocities were assumed to apply to all companions within roughly \replaced{2 arcseconds}{2\arcsec}. Given that it is both possible and likely that different members of a multiple star system have different radial velocities, when different components had independently measured $\gamma$, they were combined using a weighted standard deviation to produce a systemic velocity.

Attempts have been made to reduce the double-counting of $\gamma$s, particularly in cases where a later paper cited a $\gamma$ taken from one of the catalogs included here. This has particularly been a problem for \citet{Barbier-Brossat2000} and \citet{Kharchenko2007}, which both contain the General Catalog of Radial Velocities (\citealt{Wilson1953}) where uncertainties were reported as letter codes. \citet{Barbier-Brossat2000} and \citet{Kharchenko2007} recommend a different translation of letter code into km s$^{-1}$ uncertainty than the VizieR version of \citet{Wilson1953} (and papers that cite it directly, such as \citealt{Malo2013}) itself does, leading to nearly-identical $\gamma$s appearing in different sources.

\tabletypesize{\tiny}
\begin{deluxetable}{lll} 
\setlength{\tabcolsep}{0.04in}
\tablewidth{0pt} 
\tablecaption{Radial Velocity Uncertainty Defaults\label{tab:rvprecision}}
\tablehead{
 \colhead{Citation} &
 \colhead{Uncertainty} &
 \colhead{Rationale} }
\startdata
\citet{Eggen1991} & 5 & Comparison to other extant $\gamma$s \\
\citet{Barbier-Brossat2000} & 3.7 & Letter code C unless a code was given \\
\citet{Montes2001a} & 1 & Typical uncertainty in paper \\
\citet{Montes2001b} & 1 & Typical uncertainty in paper \\
\citet{Gontcharov2006} & 1 & Typical uncertainty in catalog \\
\citet{Torres2006} & 1 & Cited agreement with \citet{Nordstroem2004} \\
\citet{Kharchenko2007} & 3.7 & Letter code C unless a code was given \\
\citet{Guillout2009} & 1 & Typical uncertainty in paper \\
\citet{Maldonado2010} & 1 & Typical uncertainty in paper \\
\citet{Murphy2010} & 2 & Typical uncertainty in paper \\
\citet{Schlieder2010} & 2 & Typical uncertainty in paper \\
\citet{Schneider2012a} & 2 & Typical uncertainty in paper \\
\citet{De-Silva2013} & 1 & Subsequent to \citet{Torres2006} \\
\citet{Malo2013} & 1 & Typical uncertainty in paper \\
\citet{Malo2014a} & 1 & Typical uncertainty in paper \\
\citet{Malo2014b} & 1 & Typical uncertainty in paper \\
\citet{Elliott2015} & 1 & Subsequent to \citet{Torres2006} \\
\enddata
\end{deluxetable}

In many cases, $\gamma$s have been published without uncertainties. Because our weighted standard deviations require an uncertainty, we have invented them where necessary, and flagged them with 'e' in our source tables. Radial velocities originating from \citet{Wilson1953} with quality codes had uncertainties assigned according to the quality notes as suggested in the table notes; where no quality code was available, we have set the errors to 3.7 km s$^{-1}$, equivalent to letter code ``C''. Most other papers were given 1 km s$^{-1}$ uncertainties, as per Table \ref{tab:rvprecision}.

\subsubsection{Photometry}
\label{sec:photometry}
Photometry came from numerous sources, and were applied in a set order, presented here in decreasing order of preference:
For optical data:
\begin{enumerate}
\item Photometry from source papers (except \citealt{van-Altena1995})
\item Southern Proper Motion (SPM4) catalog, CCD second epoch measurements only \citep{Girard2011} - $B$,$V$ only.
\item The American Association of Variable Star Observers Photometric All Sky Survey (APASS DR9, \citealt{Henden2016}), where all stars have been proper-motion corrected to epoch 1 Jan 2011, roughly the midpoint of the survey. $B$,$V$ only.
\item APASS DR6 data, as incorporated into the Fourth United States Naval Observatory Compiled Astrometric Catalog (UCAC4, \citealt{Zacharias2013}) - $B$,$V$ only.
\end{enumerate}
The APASS DR9 data did not completely replace UCAC4's APASS DR6 data due to a better position solution and cross-matching done when UCAC4 incorporated APASS DR6. 

SDSS $u'g'r'i'z'$ photometry was sourced from
\begin{enumerate}
\item The Sloan Digital Sky Survey (SDSS9, \citealt{Ahn2012})
\item APASS DR9 \citep{Henden2016}, corrected for proper motion to 1 Jan 2011. ($g'r'i'$ only)
\item UCAC4 \citep{Zacharias2013} ($g'r'i'$ only)
\end{enumerate}
Near-infrared data was sourced from our source papers, if they deblended photometry (only \citealt{Riedel2014}) or from the Two Micron All-Sky Survey \citep{Skrutskie2006}. Mid-infrared data in WISE $W1,W2,W3,W4$ photometric bands was sourced from the ALLWISE catalog \citep{Cutri2013}, which supersedes WISE All-Sky data \citep{Cutri2012}.

X-ray data \replaced{was}{were} extracted from the {\it ROSAT} All-Sky Survey's bright star catalog \citep{Voges1999} and faint star catalog \citep{Voges2000} using an aperture of 25\arcsec~around all targets, after they were corrected by proper motion to their 1 Jan 1991 positions (the rough median date of the survey). \replaced{UV}{Ultraviolet} data from GALEX was extracted from the All-Sky Imaging Survey and Medium-Depth Imaging Survey \added{\citep{Bianchi2011}} after correcting all stars to their 1 Jan 2007 positions. 

Deblending magnitudes is possible \citep{Riedel2014} but cannot be done systematically for all stars. The source of the majority of our multiplicity information, the Washington Double Star Catalog (WDS; \citealt{Mason2015}), does not report filters with its delta magnitudes on the most readily available public versions (VizieR, USNO text tables\footnote{\url{http://ad.usno.navy.mil/wds/}, checked 2016 October 5}).

\subsubsection{Multiplicity}
\label{sec:multiplicity}
For the purposes of this paper, the multiplicity information in the catalog is not complete. The fundamental unit of the catalog is intended to be the single object, with one object per entry even if no information is known other than that the object exists.

The question of multiplicity for members of NYMGs is occasionally difficult, as some independent members of moving groups may be picked up by surveys as extremely wide common proper motion binaries \citep{Caballero2009,Caballero2010,Shaya2011}. Apart from well-known binaries, we have set an informal limit of 500\arcsec~for binaries.

\replaced{The Washington Double Star (WDS) catalog \citep{Mason2015}}{WDS} contains information on multiples observed through direct imaging, adaptive optics\added{, micrometery, and speckle interferometry}, and is our primary source for multiplicity information. While there are catalogs of spectroscopic orbits (SB9, \citealt{Pourbaix2004}), there is no comparable central source for general spectroscopic binaries. Thus, all information on other, closer multiples has come from individual survey papers and system notes in WDS.

Companions listed in WDS have been accepted if and only if they have been observed more than once and are still consistent with being common proper motion pairs. Discovery papers generally contain the most reliable information about spectroscopic and visual binaries when discovered, but many papers included here are compilations themselves, or deal with systems known elsewhere.

\section{Input Membership Data}

\subsection{A New Bona-fide Sample}
\label{sec:bonafides}

LACEwING requires kinematic models of the NYMGs. These are UVW and XYZ ellipsoids fit to genuine members of the groups. To create a list of bona-fide members, we have pulled previously-identified bona-fide members from the Catalog of Suspected Nearby Young Stars (Section \ref{sec:catalog}). We then filtered out probable interlopers from the samples using the TRACEwING code (Section \ref{sec:tracewing}). The resulting filtered samples of bona-fide moving group members were then fit with ellipsoids.

\subsubsection{Initial Member Data}

The starting point for our membership list are 546 stars from published lists of high confidence members, most notably the BANYAN series of papers (\citealt{Malo2013}, \citealt{Gagne2014a}, and subsequent).
\begin{itemize}
\item BANYAN papers \citep{Malo2013}, with additions and subtractions from \citet{Gagne2014a,Gagne2015} and \citet{Malo2014a,Malo2014b}. These papers list bona-fide members of TW Hya, $\beta$ Pic, Tuc-Hor, Columba, Carina, Argus, and AB Dor. Bona-fide members are ``all stars with a good measurement of trigonometric distance, proper motion, Galactic space velocity and other youth indicators such as H\deleted{-}$\alpha$ emission, X-ray emission, appropriate location in the Hertzprung-Russel diagram, and lithium absorption'' \citep[page 2]{Malo2013}; in practice the youth indicator for most targets is X-ray emission.
\item \citet{King2003} performed a thorough kinematic, activity, and isochronal analysis of the Ursa Major moving group, which concluded with a list of 60 nearly assured members, which were broken into a nucleus of 14 systems, and 46 other members. We adopt the 14 nucleus members as the bona-fide members of Ursa Major.
\item \citet{Eisenbeiss2013} conducted an analysis of Her-Lyr using gyrochronology, isochrone fits, lithium abundances, and chromospheric activity, concluding with the identification of seven ``canonical'' members, which we adopt as bona-fide members.
\item \citet{Murphy2010} analyzed the $\eta$ Cha open cluster and reconsidered membership for the \replaced{group}{cluster} using proper motions, surface gravity measurements, activity, and lithium. We adopt their list of $\eta$ Cha members as bona-fide members. Not all of them have trigonometric parallaxes.
\item \citet{Murphy2013} analyzed the $\epsilon$ Cha moving group using techniques similar to those used for $\eta$ Cha. We adopt their list of $\epsilon$ Cha members as bona-fide members. Not all of these stars have measured trigonometric parallaxes either.
\item \citet{Murphy2015} studied the Octans moving group using spectroscopy, photometry, and fast rotation. We adopt their list of Octans members as bona-fide members. None of them have measured trigonometric parallaxes.
\item \citet{Casewell2006} and \citet{Kraus2007} examined the Coma Ber open cluster for low mass members, using proper motions and photometry for isochrone fits. We adopt their ``known'' members as bona-fide members of Coma Ber.
\item \citet{Zuckerman2006} proposed a new moving group Car-Near. We take all reported members as bona-fide members.
\item \citet{Zuckerman2013} proposed a new moving group, Oct-Near. We have taken all probable members as bona-fide members.
\item E. E. Mamajek (private communication) supplied a list of members of the 32 Ori and $\chi^{01}$ For moving groups \citep{Mamajek2015}. We have taken all members rated as likely or definitive as bona-fide members.
\end{itemize}

We have made several alterations to this list. \object{PX Vir} is listed by \citet{Eisenbeiss2013} as a canonical member of Her-Lyr and by both \citet{Malo2013} and \citet{Gagne2014a} as a bona-fide member of AB Dor. It has been made a member of AB Dor (in the final analysis, it is a bona-fide member of AB Dor, and a bad match to Her-Lyr). \citet{Gagne2014a} has erroneous entry for \object[GJ 2079]{GJ~2079AB} (HIP~50156) as a bona-fide member of both $\beta$ Pic and Carina, when it is a modest probability member of Carina (J. Gagn{\'e} 2016, private communication) and was removed from the bona-fide sample.

Several papers have removed targets from these lists. \citet{Hinkley2013}, \citet{Barenfeld2013}, and \citet{McCarthy2014} have run more detailed analyses that have ruled out, or at least cast doubt on, members of Columba and AB Dor. The BANYAN papers \citep{Gagne2014a,Malo2014a} have removed those and other targets from their bona-fide lists, but we have retained all targets that were solely rejected for large uncertainties or discrepancies with their {\it XYZUVW} model.

Another major difference with previous bona-fide lists is that in the production of the catalog, we have reconsidered whether stars are parts of bound systems (\object{AU Mic}+\object[AT Mic]{AT Mic AB}; \object[bet01 Tuc]{$\beta^{01}$ Tuc AB}+\object[bet02 Tuc]{$\beta^{02}$ Tuc AB}+\object[bet03 Tuc]{$\beta^{03}$ Tuc AB}; \citealt{Mason2015}). We only include the system primary in our kinematic analysis, and therefore have fewer systems in our initial and cleaned bona-fide samples.

\subsubsection{Bona-Fide candidate filtering}

\begin{figure}
\center
\includegraphics[angle=0,width=.5\textwidth]{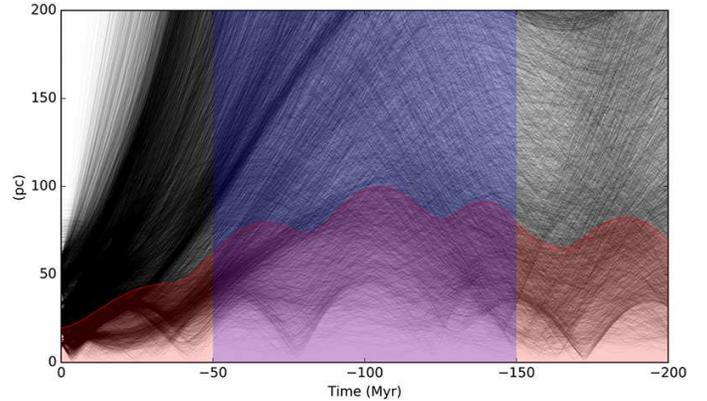}
\caption{An epicyclic traceback of the entire AB Doradus moving group, showing the separations between each bona-fide member (represented by 5000 random draws) and the center of the group, as a function of time.}
\label{fig:AB_Dor_traceback}
\end{figure}

\begin{figure}
\center
\includegraphics[angle=0,width=.5\textwidth]{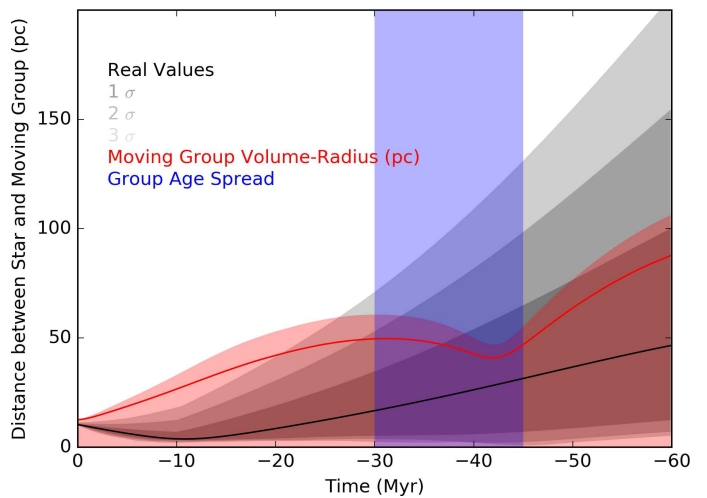}
\caption{Same as Figure \ref{fig:traceback} except showing CPD-64~17 traced back relative to Tuc-Hor.}
\label{fig:traceback_CPD-64-00017}
\end{figure}

\begin{figure}
\center
\includegraphics[angle=0,width=.5\textwidth]{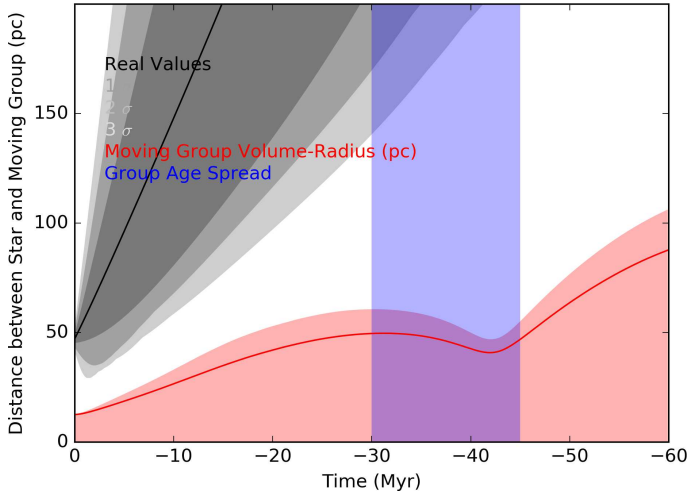}
\caption{Same as Figure \ref{fig:traceback} except showing HIP 104308 traced back relative to Tuc-Hor.}
\label{fig:traceback_HIP104308}
\end{figure}

The bona-fide list of members was filtered using TRACEwING (Section \ref{sec:tracewing}) to identify and remove outliers: stars that could not possibly have been in the same location as the rest of the group at the time of formation. Moving groups were generated from the bona-fide list (Figure \ref{fig:AB_Dor_traceback}), and then every member of the moving group was traced back to the rest of the group individually. Outliers were defined as being more than 2-$\sigma$ from the location of the NYMG at all times between the minimum and maximum reasonable age for the group, as collected in Table \ref{tab:grouptable}\deleted{)}. As an example, Figure \ref{fig:traceback_CPD-64-00017} shows the Tuc-Hor bona-fide member \object[CPD-64 17]{CPD-64~17} within the confines of Tuc-Hor over the entire range of quoted ages (30-45 Myr), while Figure \ref{fig:traceback_HIP104308} shows the Tuc-Hor non-member \object[HIP 104308]{HIP~104308} nowhere near Tuc-Hor at any time. After the end of a filtering step, the moving group was recalculated with the refined member list. The process of filtering outliers was repeated until the moving group was self-consistent, which took three or fewer iterations. This reduced the bona-fide list to 297 systems.

\begin{figure*}
\center
\includegraphics[angle=0,width=\textwidth]{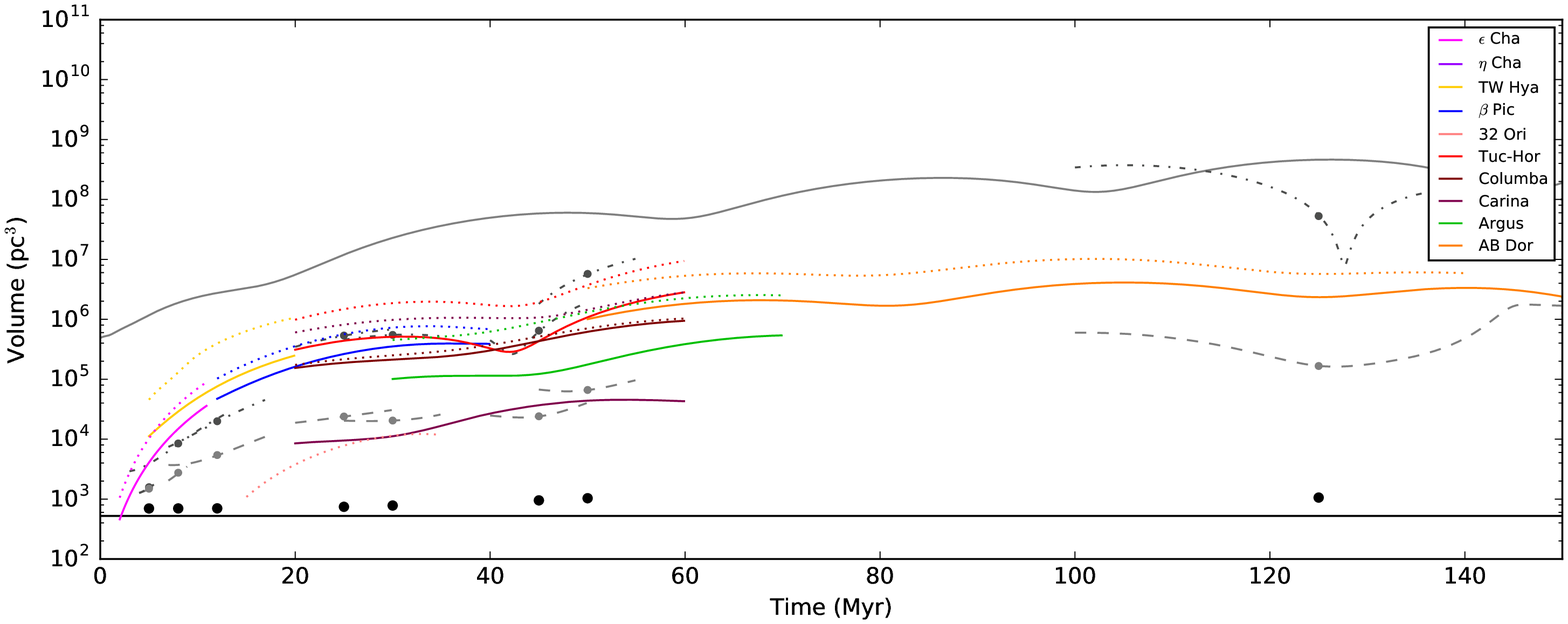}
\includegraphics[angle=0,width=\textwidth]{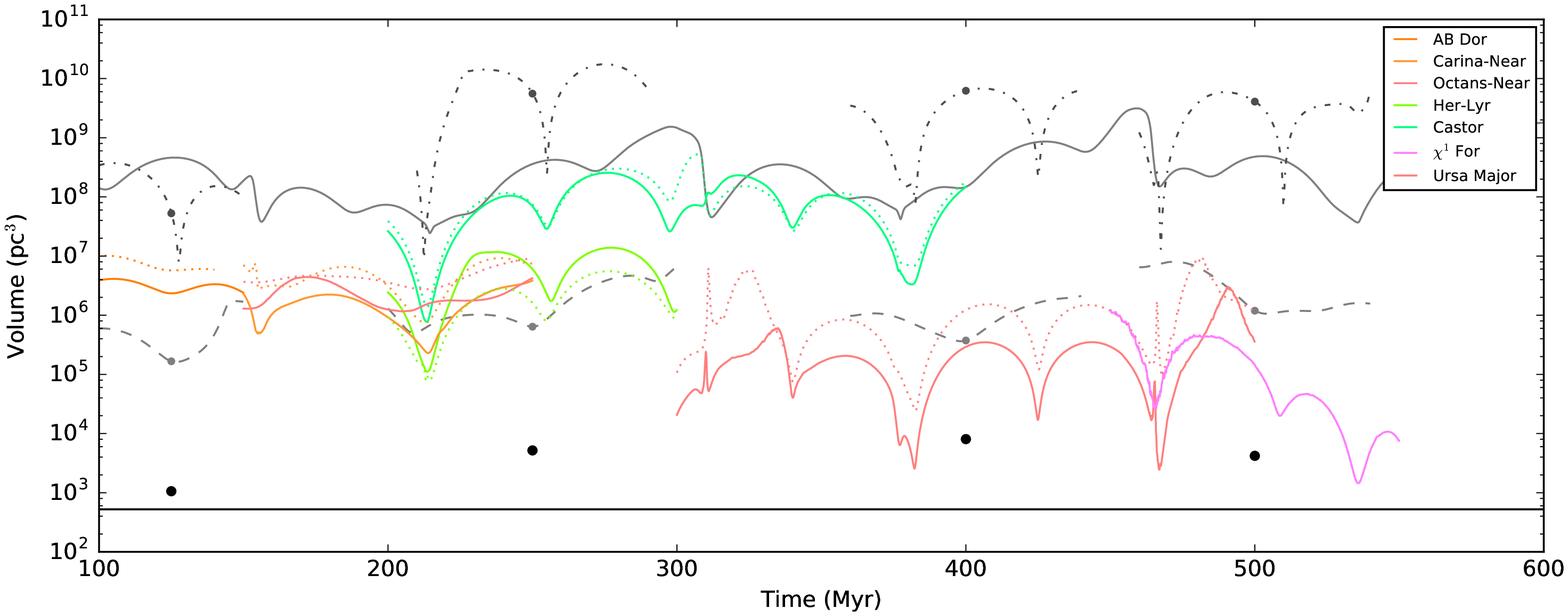}
\caption{Volumes of the NYMGs around their supposed times of formation, before (dashed) and after (solid) outliers have been removed, plotted on Figure \ref{fig:traceback_explanation}. All NYMGs are smaller than the simulated ``real'' moving group with current-precision kinematics (dash-dot curves); some are smaller than the ``real'' moving group with {\it Gaia}-precision kinematics (dashed curves); none are the size of the ``real'' moving group with no uncertainties (points). All, except Castor, are smaller than the ``fake'' moving group constructed from field stars (thick gray line).}
\label{fig:traceback_groups}
\end{figure*}

The volumes of the NYMGs near their reported times of formation are shown in Figure \ref{fig:traceback_groups}, overplotted on Figure \ref{fig:traceback_explanation}. With currently available data, the final volume of $\beta$ Pic is actually smaller than was calculated for our synthetic 25 Myr old cluster (Section \ref{sec:traceback_limitations}), which suggests that it is consistent with being a genuine product of a single burst of star formation. The same is true of AB Dor (125 Myr), suggesting that it too is consistent with being a real moving group, although AB Dor is also close to indistinguishable from the fake cluster of field stars (Section \ref{sec:traceback_limitations}). The traceback results for AB Doradus in \citet{McCarthy2014} are similar in implied volume to the TRACEwING traceback of AB Doradus in Figure \ref{fig:AB_Dor_traceback}, despite using different epicyclic parameters. This again suggests that the limiting factor in both cases is data precision.

The complete table of bona-fide members (including discarded non-members) is given in the Catalog of Suspected Nearby Young Stars (Section \ref{sec:catalog}), where they are flagged in column ``Bonafide'' with `B' (for bona-fide system primaries), `R' (for rejected system primaries), or `X' (for bona-fide systems without sufficient information that could not be filtered with tracebacks or used to construct kinematic models); lowercase `b', `r', and `x' flags indicate companions to the respective systems. 

\subsection{Moving Group Properties}

\begin{figure*}
\center
\includegraphics[angle=0,width=\textwidth]{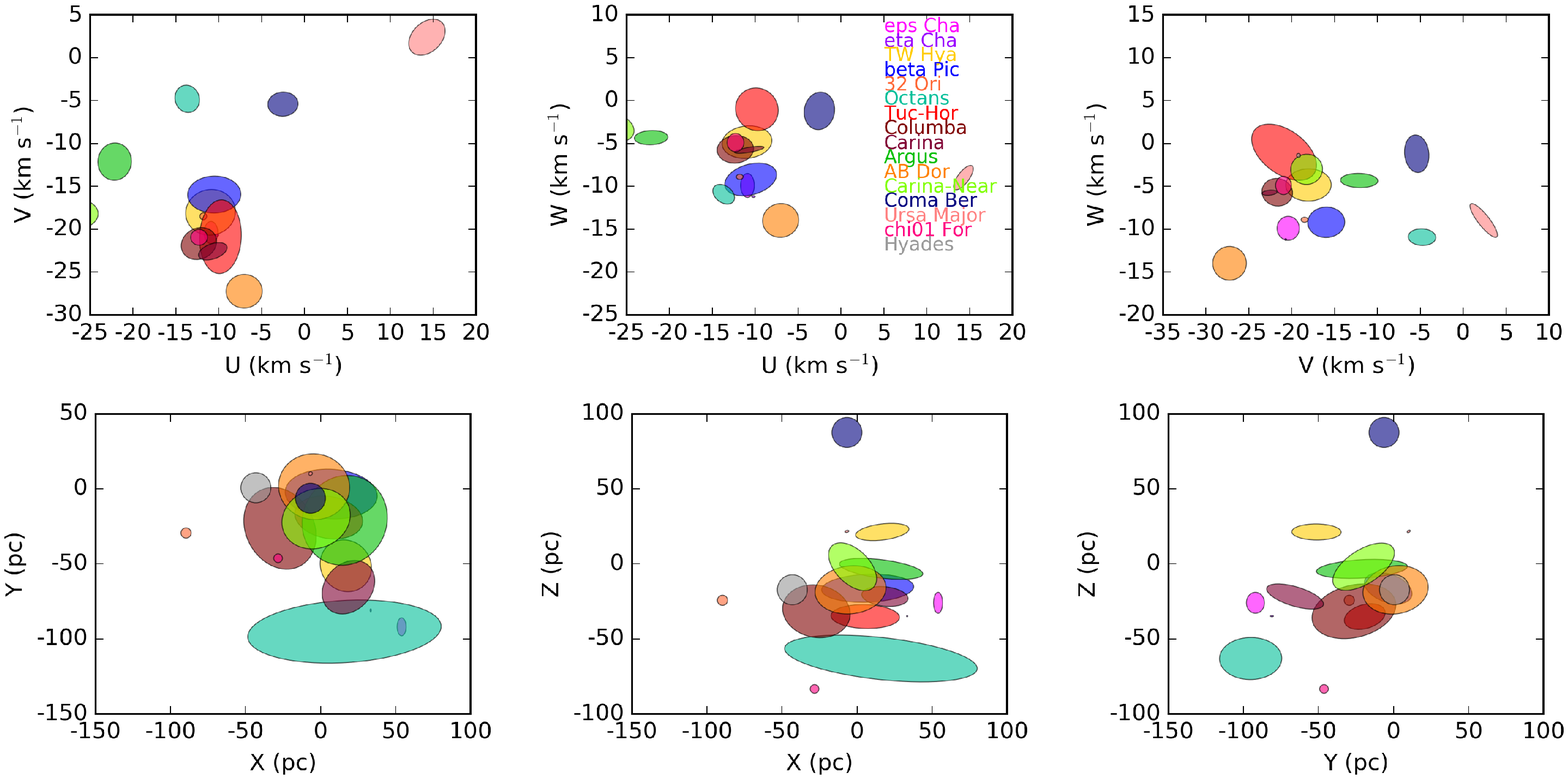}
\caption{UVW and XYZ 2-D plots showing the ellipses of the moving groups and open clusters used in LACEwING. The Hyades are not within the UVW plots. UMa is barely visible on the XYZ plot due to its compact dimensions (considering only the core of Ursa Major). $\eta$ Cha is barely visible on both plots due to its small spatial and kinematic dispersion. Note that the ellipses shown here are 2-D projections of the freely oriented 3-D ellipses used by LACEwING.}
\label{fig:ellipses}
\end{figure*}

As outlined in Section \ref{sec:lacewing}, the LACEwING code relies upon triaxial ellipsoid representations of the NYMGs, and an assessment of the population size. Most of the moving group ellipses were created by fitting the filtered selection of bona-fide stars. The fitting routine assumes that the groups are triaxial ellipsoids with orthogonal axes. The routine finds the UV plane angle with a linear fit to the projected UV data, de-rotates the data to align that axis with the Cartesian plane, and then repeats the process for the UW plane angle and the VW plane angle. Standard deviations are fit to the de-rotated data and are taken as the axis dimensions of the ellipse. This process is repeated for 10,000 Monte Carlo iterations. The final ellipse parameters are the average of this process, and are shown in Table \ref{tab:moving_group_properties}. Two-dimensional projections of the ellipsoids are shown in Figure \ref{fig:ellipses}.

\subsubsection{Groups Whose properties did not Come from Ellipse Fitting}

Seven of the sixteen groups were not fit with our normal process. The ellipse fitting process requires a minimum of four stars with full kinematic information. Five groups lacked sufficient numbers of stars with complete information, and their properties had to be taken instead from other sources: $\eta$ Cha \citep{Murphy2010}, $\epsilon$ Cha \citep{Murphy2013}, 32 Ori (E.E. Mamajek 2016, private communication), $\chi^{01}$ For (E.E. Mamajek 2016, private communication), and the Hyades \citep{Roeser2011}. They are thus represented by axis-locked ellipses, and as shown in Table \ref{tab:moving_group_properties}, all rotation angles are set to 0.

The properties of Coma Ber are a mix of ellipse fit to our bona-fide sample for the UVW space motions, and conversion of values from \citet{van-Leeuwen2009} for the XYZ space position and tidal radius.

None of the stars in Octans have trigonometric parallaxes, but \citet{Murphy2015} published estimated UVW values and distances for each member. We fit ellipsoids to the UVW velocities and computed XYZ position ellipses using the kinematic distances, $\alpha$, and $\delta$. Results for both Coma Ber and Octans are thus a hybrid of our work and others.

\subsubsection{The Field population}
\label{sec:field}
\begin{figure*}
\center
\includegraphics[angle=0,width=\textwidth]{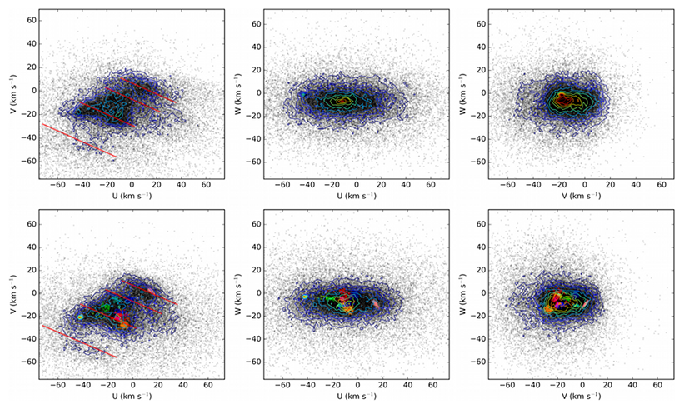}
\caption{UVW projection plots of (top) 39870 stars from the XHIP catalog and (bottom) 40000 stars drawn from our kinematic distributions. Moving Groups are colored following Figure \ref{fig:ellipses}\added{, and their points have been enlarged and darkened to make them visible. Red lines in the leftmost plots represent the streams as defined by \citet{Skuljan1999}.}}
\label{fig:UVWsimulation}
\end{figure*}

As shown in the top panels of Figure \ref{fig:UVWsimulation}, the kinematics of the solar neighborhood as plotted from the \replaced{XIP}{XHIP} catalog \citep{Anderson2012} have a complex structure. Of the NYMGs and open clusters, only the Hyades is readily visible; the remainder of the structures are thought to be the result of Galactic resonances. 

We have replicated the structures with a by-eye fit of seven ellipsoid components, which correspond to (in terms of the \citealt{Skuljan1999} groups), Sirius, Coma Berenices (leading), Coma Berenices (trailing), Pleiades (leading), Hyades (trailing), $\zeta$ Hercules, and a broad generic field population. Following \citet{Skuljan1999}, all groups are inclined by 25 degrees to the coordinate axes, with the exception of the Pleiades branch at -25 degrees, and the unrotated field population.

\begin{figure*}
\center
\includegraphics[angle=0,width=\textwidth]{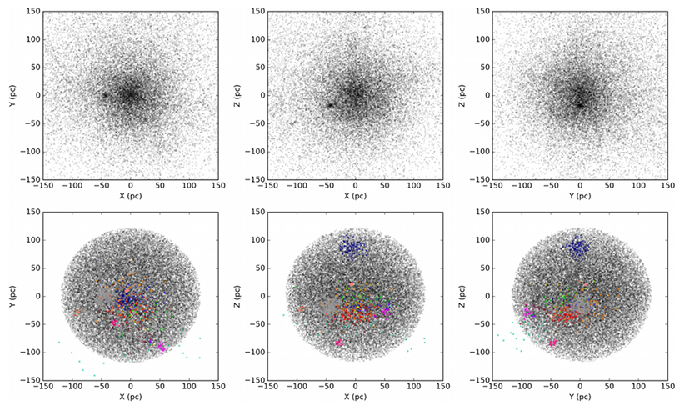}
\caption{Same as Figure \ref{fig:UVWsimulation} except with XYZ projection plots.}
\label{fig:XYZsimulation}
\end{figure*}

The bottom panels of Figure \ref{fig:UVWsimulation} demonstrate our synthetic field population, where the top panels plot stars with $<50$\% parallax uncertainties and $\gamma$ measurements with uncertainties from the XHIP catalog. Figure \ref{fig:XYZsimulation} shows the same plot for the XYZ population, which we have modeled in X and Y as a uniform distribution truncated at a distance of 120 pcs (to accommodate groups like Octans and $\epsilon$ Cha that extend beyond 100 pc), and in Z as an exponential with a scale height of 300 pcs, again truncated at a distance of 120 pcs.

\subsection{Relative populations of groups}
In order to provide an appropriate simulation of members, we must also consider the relative populations of the groups. For these purposes we returned to the Catalog of Suspected Nearby Young Stars \added{(Section \ref{sec:catalog}) }and considered all members of the groups, beyond just the groups that survived our bona-fide vetting process above. 

There are two major sources of incompleteness that must be considered here: First, the more recently-discovered groups and older groups (where stars are less obviously youthful) have not been searched as completely for low\replaced{-}{ }mass stars as younger and longer-known groups. For example, none of the 14 known members of the rarely-studied $\chi^{01}$ For group ($\sim$500 Myr) have M dwarf primaries, but 31 of the 38 known members of TW Hya ($\sim$10 Myr) have M dwarf primaries. Second, the continued reliance on the highly magnitude limited {\it Hipparcos} and Tycho-2 catalogs makes it hard to identify members of more distant groups.

To account for this bias, we count only the B, A, F, G, and K primaries (and evolved versions of the same) when tallying membership in these groups. This relies upon three further assumptions: First, that all the groups are similarly complete down to spectral type K7; this will cause more rarely studied groups like 32 Ori and $\chi^{01}$ For to be underpredicted. Second, that the initial mass function of all of these groups has the same slope; this will overpredict membership in groups with known top-heavy mass functions like the Hyades \citep{Goldman2013}, and \replaced{underpredict in any groups with a bottom-heavy mass function.}{in theory underpredict membership in any group with a bottom-heavy mass function.} Third, that a K7 spectral type refers to the same mass at the age of every group. This is harder to quantify given the differences in evolutionary tracks between different stellar evolution codes, but should cause younger groups (whose members will be cooler\deleted{ and redder}) like TW Hya and $\beta$ Pic to be slightly underpredicted relative to older groups.

For the field star population, we analyze the Catalog of Suspected Nearby Young Stars for lithium-rich (see Section \ref{sec:lithium}) and bona-fide members, and find that 271 systems that are bona-fide or lithium rich are predicted to be within 25 \replaced{parsecs}{pc} of the Sun. Given the density of star systems within 5 \replaced{parsecs}{pc} (52 systems) there should be 6500 systems within 25 pc of the Sun (a sample which is highly incomplete, with only 2184 systems currently known according to \added{T. J. }Henry et al. in prep); a ratio of just under 24:1. We therefore take the ratio of young star systems to field stars to be 25:1. Given a further result (see Section \ref{sec:young_field}) that less than half of the young stars are in moving groups, we take the ratio of young field stars to moving group and open cluster members as 1:1. The ratio of all field stars (young and old) to moving group members is therefore 50:1. The population of field stars (split across the seven subgroups, Section \ref{sec:field}) is thus set at 50 times the combined number of all moving group and open cluster members.

\begin{deluxetable*}{l r ccc ccc ccc}
\setlength{\tabcolsep}{0.02in}
\tablewidth{0pt}
\tabletypesize{\tiny}
\tablecaption{Moving Group Kinematic Properties\label{tab:moving_group_properties}}
\tablehead{
	\colhead{} &
        \colhead{} &
	\multicolumn{3}{c}{Values} &
	\multicolumn{3}{c}{Axes} &
	\multicolumn{3}{c}{Angles\tablenotemark{a}} }
\startdata
\hline
Group & Number & U & V & W & A & B & C & UV & UW & VW \\
Name  & \deleted{O}BAFGK & (km s$^{-1}$) & (km s$^{-1}$) & (km s$^{-1}$) & (km s$^{-1}$) & (km s$^{-1}$) & (km s$^{-1}$) & (rad) & (rad) & (rad) \\
\hline
$\epsilon$ Cha  &  17 & -10.9 & -20.4 & -9.9 & 0.8 & 1.3 & 1.4 & 0 & 0 & 0 \\
$\eta$ Cha      &   6 & -10.2 & -20.7 & -11.2 & 0.2 & 0.1 & 0.1 & 0 & 0 & 0 \\
TW Hya          &   7 & -10.954 & -18.036 & -4.846 & 3.043 & 2.332 & 1.703 & 0.227 & 0.022 & 0.098 \\
$\beta$ Pic     &  34 & -10.522 & -15.964 & -9.2 & 3.167 & 2.039 & 1.609 & 0.020 & 0.045 & 0.238 \\
32 Ori          &  12 & -11.8 & -18.5 & -8.9 & 0.4 & 0.4 & 0.3 & 0 & 0 & 0 \\
Octans          &  22 & -13.673 & -4.8 & -10.927 & 1.749 & 1.678 & 1.029 & 0.32 & -0.52 & 0.241 \\
Tuc-Hor         &  63 &  -9.802 & -20.883 & -1.023 & 4.01 & 2.883 & 1.458 & -0.042 & -0.588 & 0.568 \\
Columba         &  52 & -12.311 & -21.681 & -5.694 & 2.321 & 1.43 & 1.322 & 0.470 & -0.142 & 0.329 \\
Carina          &  22 & -10.691 & -22.582 & -5.746 & 1.763 & 0.532 & 0.178 & 0.341 & 0.092 & 0.044 \\
Argus           &  38 & -22.133 & -12.122 & -4.324 & 1.992 & 1.755 & 0.774 & -0.088 & -0.026 & 0.002 \\
AB Dor          &  86 &  -7.031 & -27.241 & -13.983 & 2.136 & 1.929 & 1.859 & 0.041 & 0.050 & 0.182 \\
Car-Near        &  10 & -27.020 & -18.255 & -3.021 & 3.044 & 1.819 & 1.147 & 0.023 & 0.149 & -0.286 \\
Coma Ber        & 104 &  -2.512 & -5.417 & -1.204 & 1.868 & 1.364 & 1.876 & 0.057 & 0.106 & -0.202 \\
Ursa Major      &  55 & 14.278 & 2.392 & -8.987 & 2.64 & 0.594 & 0.407 & -0.799 & -0.766 & 0.500 \\
$\chi^{01}$ For  &  14 & -12.29 & -20.95 & -4.9 & 0.98 & 0.92 & 1.07 & 0 & 0 & 0 \\
Hyades          & 260 & -41.1 & -19.2 & -1.4 & 0.23 & 0.23 & 0.23 & 0 & 0 & 0 \\
Field (Sirius)  &5800 &    8 & 2 & -7.25 & 12 & 6 & 9 & -0.436 & 0 & 0 \\
Field (Coma1)   &4400 & -10 & -8 & -7.25 & 9 & 6 & 7 & -0.436 & 0 & 0 \\
Field (Coma2)   &2700 & 15 & -18 & -7.25 & 14 & 7 & 7 & -0.436 & 0 & 0 \\
Field (Hyades)  &5800 & -32 & -17 & -7.25 & 12 & 6 & 9 & 0.5 & 0 & 0 \\
Field (Pleiades)&5800 & -12 & -24 & -7.25 & 10 & 6 & 9 & -0.436 & 0 & 0 \\
Field ($\zeta$ Her)& 800 & -35 & -48 & -7.25 & 14 & 6 & 8 & -0.436 & 0 & 0 \\
Field         & 14800 & -11.1 & -25 & -7.25 & 50 & 25 & 25 & 0 & 0 & 0 \\
\hline
Group &  & X & Y & Z & D & E & F & XY & XZ & YZ \\
Name & & (pc) & (pc) & (pc) & (pc) & (pc) & (pc) & (rad) & (rad) & (rad) \\
\hline
$\epsilon$ Cha  &  &  54 & -92 & -26 & 3 & 6 & 7 & 0 & 0 & 0 \\
$\eta$ Cha      &  &  33.4 & -81 & -34.9 & 0.4 & 1 & 0.4 & 0 & 0 & 0 \\
TW Hya          &  &  16.816 & -51.33 & 21.194 & 17.9 & 7.681 & 5.197 & -0.847 & 0.034 & 0.183 \\
$\beta$ Pic     &  &   7.075 & -3.509 & -16.277 & 30.736 & 16.323 & 7.186 & -0.06 & 0.043 & -0.337 \\
32 Ori          &  & -89.634 & -29.47 & -24.34 & 3.4 & 3.4 & 3.4 & 0 & 0 & 0 \\
Octans          &  &  15.913 & -95.179 & -63.138 & 64.92 & 20.831 & 13.888 & 0.059 & -0.107 & 0.08 \\
Tuc-Hor         &  &   5.477 & -19.146 & -35.177 & 22.83 & 13.179 & 6.713 & -0.175 & -0.043 & 0.287 \\
Columba         &  & -27.056 & -26.369 & -31.674 & 22.794 & 24.357 & 15.479 & 0.497 & 0.074 & 0.317 \\
Carina          &  &  18.582 & -65.598 & -21.795 & 16.127 & 12.938 & 3.63 & -0.747 & 0.27 & -0.24 \\
Argus           &  &  16.075 & -21.027 & -3.39 & 28.119 & 25.709 & 6.084 & -0.434 & -0.104 & -0.018 \\
AB Dor          &  &  -4.323 & 1.391 & -17.372 & 23.857 & 21.531 & 15.014 & -0.16 & 0.085 & 0.221 \\
Car-Near        &  &  -2.961 & -19.919 & -1.955 & 23.641 & 10.619 & 5.037 & -1.176 & -0.532 & 1.006 \\
Coma Ber        &  &  -6.706 & -6.308 & 87.522 & 10 & 10 & 10 & 0 & 0 & 0 \\
Ursa Major      &  &  -6.704 & 10.134 & 21.622 & 1.388 & 0.763 & 0.251 & 0.559 & 0.312 & 0.682 \\
$\chi^{01}$ For  &  & -28.3 & -46.3 & -83.4 & 2.9 & 2.9 & 2.9 & 0 & 0 & 0 \\
Hyades          &  & -43.1 & 0.7 & -17.3 & 10 & 10 & 10 & 0 & 0 & 0 \\
Field (Sirius)  &  & -0.18 & 2.1 & 3.27 & 120 & 120 & 120 & 0 & 0 & 0 \\
Field (Coma1)   &  & -0.18 & 2.1 & 3.27 & 120 & 120 & 120 & 0 & 0 & 0 \\ 
Field (Coma2)   &  & -0.18 & 2.1 & 3.27 & 120 & 120 & 120 & 0 & 0 & 0 \\
Field (Hyades)  &  & -0.18 & 2.1 & 3.27 & 120 & 120 & 120 & 0 & 0 & 0 \\
Field (Pleiades)&  & -0.18 & 2.1 & 3.27 & 120 & 120 & 120 & 0 & 0 & 0 \\
Field ($\zeta$ Her)&  & -0.18 & 2.1 & 3.27 & 120 & 120 & 120 & 0 & 0 & 0 \\
Field           &  & -0.18 & 2.1 & 3.27 & 120 & 120 & 120 & 0 & 0 & 0 \\
\hline
\enddata
\end{deluxetable*}

\subsection{Lithium Sample}
\label{sec:lithium}

As a proof of concept, we use LACEwING on a sample of stars with detectable lithium. Given that kinematics and activity alone are not sufficient to say \added{that} an object is young, our process of examining only spectroscopically young objects reduces the possibility that we are attempting to reproduce erroneous membership assignments.

\begin{figure}
\center
\includegraphics[angle=0,width=0.5\textwidth]{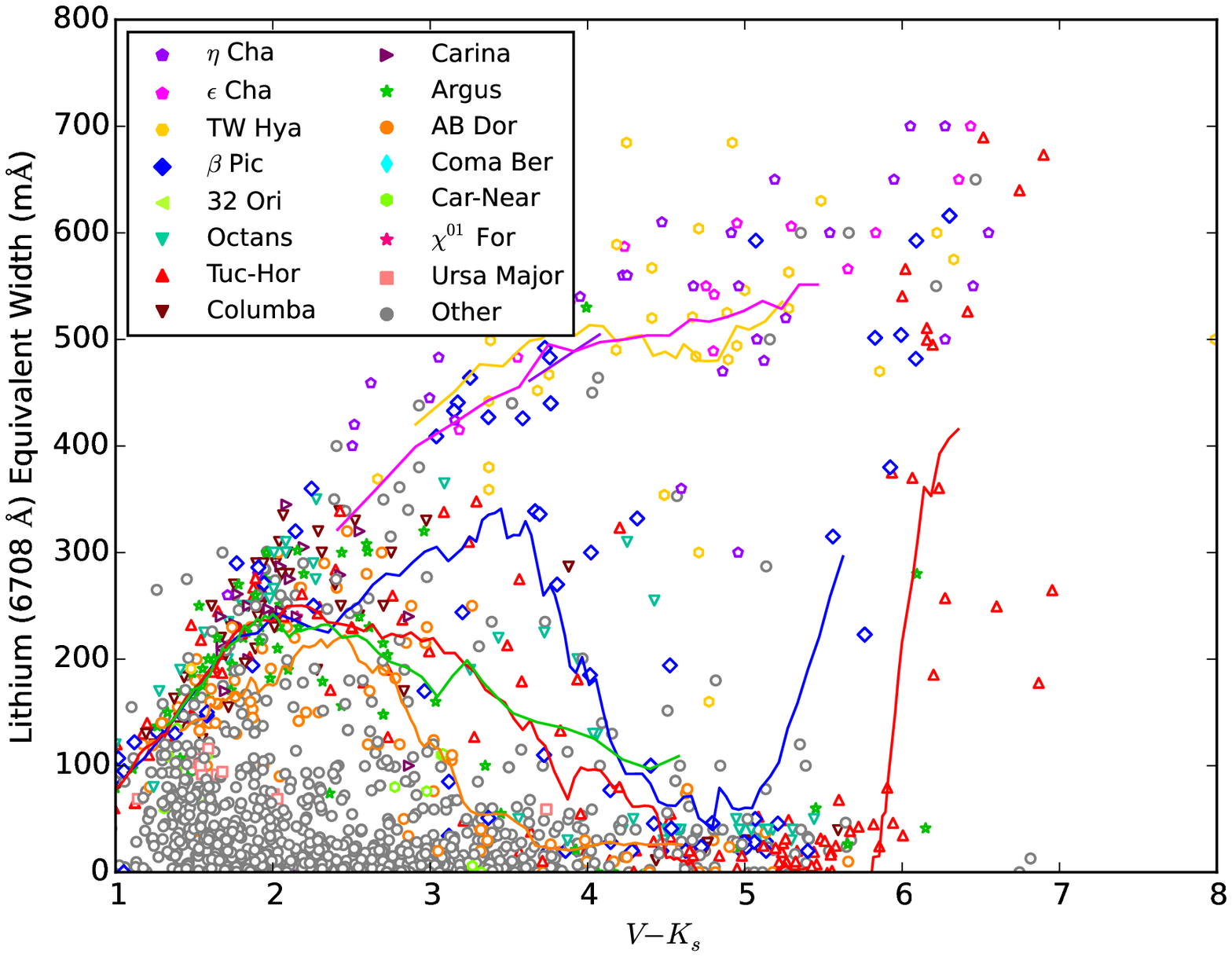}
\caption{EW(Li) from the Catalog of Suspected Nearby Young Stars. Lines are 15-element moving averages.}
\label{fig:lithium}
\end{figure}

The presence of lithium (specifically the red-optical 6708\AA~doublet) is one of the most reliable spectroscopic methods of identifying young stars \citep{Soderblom2010}. Lithium is fused at lower temperatures than hydrogen, and nearly all of it is primordial. Any amount of lithium present in a stellar spectrum (unless the object is an asymptotic giant branch star) is a sign that the object has not fused it yet, either because the object is very young, or because it is a $<60 M_{\textrm{Jup}}$ brown dwarf that never reaches sufficiently high temperatures \citep{Rebolo1992}. The exact age at which lithium is depleted varies based on the mass of the object \citep{Yee2010}, and spans ages typical of the nearby young moving groups. In Figure \ref{fig:lithium}, we plot all lithium measurements from the Catalog of Suspected Nearby Young Stars, with 15-element moving averages tracing out the lithium depletion as a function of age\replaced{- a}{. A}t the age of $\epsilon$ Cha, barely any lithium is gone\replaced{;}{,} while by the age of AB Dor, there is a very clear lithium depletion. 

Although our survey of papers reporting lithium is not complete, 1877 of the 5350 stars in our catalog of suspected young stars (Section \ref{sec:catalog}) have at least one attempt to measure their Li {\sc I} \added{$\lambda$}6707.8\AA~EW \added{doublet} (EW(Li)), as shown in Figure \ref{fig:lithium}. EW(Li) is frequently reported without uncertainties on the measurements. We have created a lithium sample by selecting objects with EW(Li) $>$ max(10 m\AA, 2 $\sigma_{\textrm{Li}}$) from the Catalog of Suspected Nearby Young Stars (Section \ref{sec:catalog}). Where no uncertainty is quoted, the uncertainty is set at 50 m\AA. This matches the limits of several major papers reporting lithium, including \citet{Eisenbeiss2013}, \citet{Kraus2014b}, and \citet{Rodriguez2013}; the largest upper limits reported by \citet{Malo2013} and \citet{Moor2013} are 46 m\AA.

Objects that are companions (according to the WDS and other resources in Section \ref{sec:catalog}) to stars in the lithium sample are also included, yielding a total of 1179 stars. From there, we considered only the primaries of the systems that have at least one lithium-detected member (including 15 primaries that do not themselves have measured lithium) so as not to count star systems more than once, resulting in a list of 1037 star systems.

Objects that were members of groups not within 100 parsecs were removed. This includes the members of IC 2391 identified by \citet{Torres2008} as members of Argus, and a wide variety of objects from \citet{Guenther2007} that belong to more distant regions (e.g. Cham{\ae}leon, $\rho$ Oph\added{iuchi}, \replaced{Sco-Cen}{Scorpius-Centaurus}). This reduced the size of the lithium sample to 930 systems.

This sample substantially overlaps with the bona-fide sample (Section \ref{sec:bonafides}): 152 systems are in both lists. Some genuinely young stars at the fully convective boundary around spectral type M3~V are excluded from this sample because lithium is fully depleted at their ages, while some more massive stars still maintain lithium even at old ages (e.g. $\alpha$ Cen AB, \citealt{Mishenina2012}). The sample can be found in the ``LiSample'' column of the Catalog of Suspected Nearby Young Stars, where system primaries that qualify for the lithium sample are flagged with `L', system primaries that qualify for the lithium sample but have membership in a more distant group are flagged with `F', and primaries that do not have lithium but have a companion that qualifies are flagged with `A'. As with the bona-fide sample, lowercase letters follow the same rules but denote companions.

\section{LACEwING performance comparison} 
\label{sec:results}

In this section we present the results of running LACEwING on the bona-fide and lithium samples (Section \ref{sec:lithium}) and compare its performance (and algorithm) to other available moving group identification codes.

\subsection{Code and Methodology Comparison}
\label{sec:comparison}

\subsubsection{Public Codes}

LACEwING does computations in observation space, directly comparing the proper motion and radial velocity vectors. Like BANYAN II, LACEwING treats groups as six-dimensional freely oriented ellipsoids and has a separate set of calibrations for use when stars are already known to be young. Unlike BANYAN II, groups have different population sizes relative to each other, and the field population to young star ratio is 25:1 rather than 4.1:1 as stated by \citet{Gagne2014a}. 

Each publicly available code for finding moving groups uses a different algorithm and is based on different bona-fide member lists.

\paragraph{BANYAN\footnote{\url{http://www.astro.umontreal.ca/~malo/banyan.php}, checked 2016 October 7}} \citep{Malo2013}\added{. It} uses Bayesian statistical techniques to provide an assessment of the moving group membership of a target object. It works in observation-space, directly testing against predictions of $\alpha$, $\delta$, $\mu_{\alpha}$, $\mu_{\delta}$, $\pi$, and $\gamma$ (a version also exists that considers $I_{Cousins}-J$ photometry, but this is not publicly available) to identify objects based on both position and motion, with an additional component comparing XYZ positions. It has models that test for TW Hya, $\beta$ Pic, Tuc-Hor, Columba, Carina, Argus, AB Dor, and a Field (Old) population. The output probability is relative to all other groups, such that all percentages sum to 100\%.

\paragraph{BANYAN II\footnote{\url{http://www.astro.umontreal.ca/~gagne/banyanII.php} checked 2016 October 7}} \citet{Gagne2014a}\added{. It} is a modification of BANYAN with new kinematic models and methods of treating them. It works in observational space, and uses $\alpha$, $\delta$, $\mu_{\alpha}$, $\mu_{\delta}$, $\pi$, and $\gamma$ to determine matches based on UVW space velocities and XYZ space positions. Like BANYAN, a version of BANYAN II exists that considers $J-K_S$ photometry, but is not publicly available. It has models that evaluate membership in TW Hya, $\beta$ Pic, Tuc-Hor, Columba, Carina, Argus, AB Dor, a young field, and an old field, where the field is based on a Besan\c{c}on model \citep{Robin2003}. Unlike BANYAN, the kinematic models of the moving groups are represented as freely oriented \replaced{3D}{three-dimensional} ellipsoids rather than axis-locked moving groups. Also unlike BANYAN, the field population is expected to outnumber field objects at a rate of 4.1:1, though all moving groups are expected to be the same relative size as each other. BANYAN II has separate probability measurements if the object is known to be young.

\paragraph{Rodriguez convergence code\footnote{\url{http://dr-rodriguez.github.io/CPCalc.html} checked 2016 October 7}} in \citet{Rodriguez2013}\replaced{. It} uses the convergent points of six groups (TW Hya, $\beta$ Pic, Chameleon-Near, Tuc-Hor, Columba, and AB Dor) and their scalar velocities to predict memberships based on $\alpha$, $\delta$, $\mu_{\alpha}$, and $\mu_{\delta}$. Its probabilities are based on a mathematical comparison of the expected proper motion vector to a computed one. While the code predicts \replaced{Dist.}{$Dist$} and $\gamma$, it does not use either value in computations.

\subsubsection{Other Codes}

The convergence method used in \citet{Montes2001a} is an implementation of techniques used by \citet{Eggen1995b}. It uses the velocity contributions V$_{\textrm{tangential}}$ = 4.74$\mu\pi^{-1}$ and V$_{\textrm{radial}}$ to determine the total velocity V$_{\textrm{Total}}$. Using the angular separation between the star and the convergent point ($\lambda$), it splits V$_{\textrm{Total}}$ into velocity components parallel (V$_{\textrm{T}}$) and perpendicular to (PV) the convergent vector. It also splits the parallel velocity vector V$_{\textrm{T}}$ into an RV component $\rho_{\textrm{C}}$. The two metrics of membership reported are whether PV $<$ 0.1V$_{\textrm{T}}$ and whether $\rho_{\textrm{C}}$ and V$_{\textrm{radial}}$ differ by less than 4-8 km s$^{-1}$- no percentages are calculated. As used in \citet{Montes2001a}, the code considers membership in the Local Association, Hyades Supercluster, IC 2391 Supercluster, Ursa Major moving group, and Castor moving group. 

The convergent point method used by \citet{Mamajek2005} is much more similar to the Rodriguez convergence code.

The codes used by \citet{Lepine2009,Schlieder2010,Schlieder2012a,Schlieder2012c} and \citet{Kraus2014a} are quite similar to LACEwING. They take the UVW velocity of a moving group and determine the expected proper motion and radial velocity of a member at \replaced{that RA and DEC}{those equatorial coordinates} (at an implied distance of 1 parsec); the proper motion vector is then used to determine a kinematic distance and kinematic RV. These kinematic distances are compared to spectroscopic, photometric or trigonometric distances \citep{Lepine2009} or with spectrophotometric distances and a distance modulus cutoff to enforce the stars falling on the approximately correct isochrone \citep{Kraus2014a}. They lack the XYZ position metric used by LACEwING, and do not convert the proper motion matches or distance matches to membership probabilities. \citet{Lepine2009} only reports members of $\beta$ Pic; \citet{Schlieder2010,Schlieder2012a,Schlieder2012c} report $\beta$ Pic and AB Dor; \citet{Kraus2014a} only considers Tuc-Hor. It is not clear if the codes consider more than one or two groups in their studies.

The moving group identification technique used by \citet{Shkolnik2012}, \citet{Riedel2014} (see also \citealt{Baines2012}) and \citet{Binks2015} are quite similar, and compute proximity to the UVW distributions of stars using the formula 
\begin{multline}
\chi^2 = \frac{1}{3}(\frac{(U_{\textrm{star}} - U_{\textrm{NYMG}})^2}{\sigma_{U\textrm{,star}}^2+\sigma_{U\textrm{,NYMG}}^2} \\
+ \frac{(V_{\textrm{star}} - V_{\textrm{NYMG}})^2}{\sigma_{V\textrm{,star}}^2+\sigma_{V\textrm{,NYMG}}^2} \\
+ \frac{(W_{\textrm{star}} - W_{\textrm{NYMG}})^2}{\sigma_{W\textrm{,star}}^2+\sigma_{W\textrm{,NYMG}}^2}),
\end{multline}
which computes the difference between the star's UVW and the moving group's UVW in units roughly analogous to standard deviations. This technique does not deal with missing information as easily as a convergence-style routine; it is impossible to calculate the UVW without full kinematic information, requiring the use of estimates or repeated calculations over a range.
Differences are mostly in implementation. \citet{Shkolnik2012} studied 14 NYMGs and assumed a 2 km s$^{-1}$ velocity dispersion for all groups; they required a maximum $\chi^2$ value of six and a maximum velocity difference (to avoid identifying stars with large uncertainties) of 5 km s$^{-1}$. They calculated photometric distances where parallaxes were not available; all targets had $\gamma$s. \citet{Binks2015} studied 10 groups with individualized dispersions and set a $\chi^2$ cutoff of 3.78, and had a maximum velocity difference of 5 km s$^{-1}$. \citet{Riedel2014} studied 13 groups with individualized dispersions, and accepted a maximum $\chi^2$ (there, called $\gamma$) value of 4.0. There was no limit on actual velocity difference. Where radial velocity did not exist, they computed UVW values for a range of radial velocities from $-$100 to $+$100 km s$^{-1}$ and took the minimum $\chi^2$ (and corresponding RV) from a polynomial fit as the correct value; all targets had parallaxes. 

The kinematic moving group identification technique in the SACY papers (e.g. \citealt{Torres2008}) is similar to, but simpler than the \citet{Shkolnik2012}, \citet{Riedel2014}, and \citet{Binks2015} method, with $F=(p\times(M_{\textrm{v}}-M_{\textrm{v,iso}})^2 + (U_{\textrm{star}}-U_{\textrm{NYMG}})^2 + (V_{\textrm{star}}-V_{\textrm{NYMG}})^2 + (W_{\textrm{star}}-W_{\textrm{NYMG}})^2)^{\frac{1}{2}}$, where $M_{\textrm{v}}-M_{\textrm{v,iso}}$ is the difference between the absolute V magnitude of the star and the absolute V magnitude of a star of a particular $V-I_C$ color, on the age-appropriate isochrone\added{, and $F$ is the velocity-space separation between a star (utilizing a best-fit $UVW$ for the star if the distance is unknown) and a moving group}. In practice, they claim, the scaling constant $p$ was only non-zero for the Octans moving group, and therefore the equation is generally a simple velocity-separation independent of uncertainties. The cutoff of good values starts at $F$=3.5, but the process used by SACY is to iteratively minimize UVW for a cluster of young stars until all outliers are removed. The team has considered nine groups: $\beta$ Pic, Tuc-Hor, Columba, Carina, TW Hya, $\epsilon$ Cha, Octans, Argus, and AB Dor.

The code used by \citet{Klutsch2014} considers membership in two ways: First, a method using Gaussian representations of the UVW velocities (and their errors, formed by error propagation); second, member-to-member analysis, wherein the stars are compared to both the moving group center and the other known members; this allows for more variation and is potentially more robust at dealing with non-Gaussian distributions of stars.

\subsection{Performance Comparison}

We present the results of comparing LACEwING to the BANYAN, BANYAN II, and Convergence codes in Table \ref{tab:groupstable_bonafide}. For each group, we record the number of members in our \replaced{Bona-Fide}{bona fide} sample (Column 2), the number of objects identified as members of that group by the code (Column 3), and the number of objects identified as members that were genuine members (Column 4); with the corresponding number of false positives (Column 5).

For instance, LACEwING (in young star mode) identifies 24 $\epsilon$ Cha members in the bona-fide sample, and all 24 of its recovered members are known members of $\epsilon$ Cha: a 0\% false positive rate, but not a 100\% recovery rate given that there were 27 bona-fide members of $\epsilon$ Cha to find. \replaced{To}{In} contrast, LACEwING (in both modes) did recover all 11 bona-fide members of $\chi^{01}$ For, and did not identify other stars as $\chi^{01}$ For members.

Looking at false positive rates is not as useful for the \replaced{Lithium}{lithium} sample (Table \ref{tab:groupstable_lithium}), because 518 out of the 930 lithium-rich stars are not previously known members of moving groups, and many are likely to actually be members \replaced{that simply haven't been seen before.}{that have not been investigated kinematically before.}

\startlongtable
\begin{deluxetable}{lccccl}
\setlength{\tabcolsep}{0.02in}
\tablewidth{0pt}
\tabletypesize{\scriptsize}
\tablecaption{Comparison of Completion in the Bona Fide Data Set\label{tab:groupstable_bonafide}}
\tablehead{
  \colhead{Group}      &
  \colhead{Real}       &
  \colhead{Identified}  &
  \colhead{Recovered} &
  \colhead{False Positive} & 
  \colhead{Code}       \\
  \colhead{(1)}             &
  \colhead{(2)}             &
  \colhead{(3)}             &
  \colhead{(4)}             &
  \colhead{(5)}             &
  \colhead{(6)}             }
\startdata
\hline
$\epsilon$ Cha & 27 & 24 & 24 & 0 & L-Y \\
               &    & 22 & 22 & 0 & L-F \\
\hline
$\eta$ Cha     &  2 &  2 & 2 & 0 & L-Y \\
               &    &  2 & 2 & 0 & L-F \\
\hline
TW Hya         & 17 & 15 & 15 & 0 & L-Y \\
               &    & 10 & 10 & 0 & L-F \\
               &    & 16 & 16 & 0 & B1 \\
               &    & 16 & 16 & 0 & B2-Y \\
               &    & 16 & 16 & 0 & B2-F \\
               &    & 19 & 6 & 13 & C \\
\hline
$\beta$ Pic    & 28 & 22 & 22 & 0 & L-Y \\
               &    & 14 & 14 & 0 & L-F \\
               &    & 43 & 27 & 16 & B1 \\
               &    & 25 & 25 & 0 & B2-Y \\
               &    & 25 & 25 & 0 & B2-F \\
               &    & 38 & 14 & 24 & C \\
\hline
32 Ori         & 10 &  9 & 9 & 0 & L-Y \\
               &    &  8 & 8 & 0 & L-F \\
\hline
Octans         & 45 & 44 & 44 & 0 & L-Y \\
               &    & 17 & 17 & 0 & L-F \\
\hline
Tuc-Hor        & 32 & 30 & 29 & 1 & L-Y \\
               &    & 31 & 30 & 1 & L-F \\
               &    & 30 & 29 & 1 & B1 \\
               &    & 31 & 30 & 1 & B2-Y \\
               &    & 31 & 30 & 1 & B2-F \\
               &    & 70 & 28 & 42 & C \\
\hline
Columba        & 16 & 14 & 13 & 1 & L-Y \\
               &    & 10 & 10 & 0 & L-F \\
               &    & 29 & 15 & 14 & B1 \\
               &    & 15 & 14 & 1 & B2-Y \\
               &    & 15 & 14 & 1 & B2-F \\
               &    & 34 & 8 & 26 & C \\
\hline
Carina         &  4 &  4 & 4 & 0 & L-Y \\
               &    &  3 & 3 & 0 & L-F \\
               &    &  4 & 4 & 0 & B1 \\
               &    &  7 & 4 & 3 & B2-Y \\
               &    &  7 & 4 & 3 & B2-F \\
\hline
Argus          &  6 &  8 & 5 & 3 & L-Y \\
               &    &  5 & 5 & 0 & L-F \\
               &    &  8 & 6 & 2 & B1 \\
               &    &  7 & 6 & 1 & B2-Y \\
               &    &  7 & 6 & 1 & B2-F \\
\hline
AB Dor         & 35 & 33 & 30 & 3 & L-Y \\
               &    & 26 & 26 & 0 & L-F \\
               &    & 35 & 35 & 0 & B1 \\
               &    & 30 & 30 & 0 & B2-Y \\
               &    & 30 & 30 & 0 & B2-F \\
               &    & 74 & 28 & 46 & C \\
\hline
Car-Near       &  9 &  3 & 3 & 0 & L-Y \\
               &    &  2 & 2 & 0 & L-F \\
\hline
Coma Ber       & 45 & 33 & 33 & 0 & L-Y \\
               &    & 31 & 31 & 0 & L-F \\
\hline
Ursa Major     &  5 &  5 & 5 & 0 & L-Y \\
               &    &  5 & 5 & 0 & L-F \\
\hline
$\chi^{01}$ For & 11 & 11 & 11 & 0 & L-Y \\
               &    & 10 & 10 & 0 & L-F \\
\hline
Hyades         & 0  &  0 & 0 & 0 & L-Y \\
               &    &  0 & 0 & 0 & L-F \\
\hline
\enddata
\tablecomments{Comparison of the recovery rates of LACEwING in Young star (L-Y) and Field star (L-F) calibration to BANYAN (B1), BANYAN II in Young star (B2-Y) and Field star (B2-F) calibration, and the Convergence (C) code. In this context, ``membership'' means the highest probability was for the group in question.}
\end{deluxetable}

\startlongtable
\begin{deluxetable}{lccccl}
\setlength{\tabcolsep}{0.02in}
\tablewidth{0pt}
\tabletypesize{\scriptsize}
\tablecaption{Comparison of completion in the Lithium Data Set\label{tab:groupstable_lithium}}
\tablehead{
  \colhead{Group}      &
  \colhead{Real}       &
  \colhead{Identified}  &
  \colhead{Recovered} &
  \colhead{False Positive} & 
  \colhead{Code}       \\
  \colhead{(1)}             &
  \colhead{(2)}             &
  \colhead{(3)}             &
  \colhead{(4)}             &
  \colhead{(5)}             &
  \colhead{(6)}             }
\startdata
\hline
$\epsilon$ Cha & 30 & 30 & 25 & 5 & L-Y \\
               &    & 27 & 24 & 3 & L-F \\
\hline
$\eta$ Cha     & 15 & 17 & 14 & 3 & L-Y \\
               &    & 15 & 14 & 1 & L-F \\
\hline
TW Hya         & 31 & 28 & 25 & 3 & L-Y \\
               &    & 11 & 11 & 0 & L-F \\
               &    & 34 & 27 & 7 & B1 \\
               &    & 28 & 27 & 1 & B2-Y \\
               &    & 28 & 27 & 1 & B2-F \\
               &    & 65 & 12 & 53 & C \\
\hline
$\beta$ Pic    & 44 & 24 & 19 &  5 & L-Y \\
               &    & 13 & 12 &  1 & L-F \\
               &    & 77 & 36 & 41 & B1 \\
               &    & 38 & 26 & 12 & B2-Y \\
               &    & 37 & 26 & 11 & B2-F \\
               &    & 78 & 20 & 58 & C \\
\hline
32 Ori         &  0 &  1 &  0 &  1 & L-Y \\
               &    &  0 &  0 &  0 & L-F \\
\hline
Octans         & 22 & 31 & 21 & 10 & L-Y \\
               &    &  9 &  9 &  0 & L-F \\
\hline
Tuc-Hor        & 68 & 61 & 55 &  6 & L-Y \\
               &    & 46 & 43 &  3 & L-F \\
               &    & 54 & 52 &  2 & B1 \\
               &    & 53 & 50 &  3 & B2-Y \\
               &    & 53 & 50 &  3 & B2-F \\
               &    & 94 & 49 & 45 & C \\
\hline
Columba        & 41 & 35 & 15 & 20 & L-Y \\
               &    & 12 &  8 &  4 & L-F \\
               &    & 70 & 32 & 38 & B1 \\
               &    & 33 & 19 & 14 & B2-Y \\
               &    & 33 & 19 & 14 & B2-F \\
               &    & 79 & 14 & 65 & C \\
\hline
Carina         & 21 & 19 & 8 & 11 & L-Y \\
               &    &  8 & 4 &  4 & L-F \\
               &    & 17 & 8 &  9 & B1 \\
               &    & 19 & 7 & 12 & B2-Y \\
               &    & 19 & 7 & 12 & B2-F \\
\hline
Argus          & 28 & 23 &  7 & 16 & L-Y \\
               &    &  8 &  4 &  4 & L-F \\
               &    & 24 & 11 & 13 & B1 \\
               &    & 12 &  6 &  6 & B2-Y \\
               &    & 11 &  6 &  5 & B2-F \\
\hline
AB Dor         & 65 & 51 & 36 & 15 & L-Y \\
               &    & 25 & 23 &  2 & L-F \\
               &    & 62 & 50 & 12 & B1 \\
               &    & 27 & 24 &  3 & B2-Y \\
               &    & 25 & 24 &  1 & B2-F \\
               &    &139 & 48 & 91 & C \\
\hline
Car-Near       &  4 &  3 & 1 & 2 & L-Y \\
               &    &  0 & 0 & 0 & L-F \\
\hline
Coma Ber       &  0 &  1 &  0 &  1 & L-Y \\
               &    &  0 &  0 &  0 & L-F \\
\hline
Ursa Major     & 11 &  3 &  1 &  2 & L-Y \\
               &    &  1 &  1 &  0 & L-F \\
\hline
$\chi^{01}$ For & 1 &  3 &  1 &  2 & L-Y \\
               &    &  1 &  1 &  0 & L-F \\
\hline
Hyades         & 0  & 11 &  0 & 11 & L-Y \\
               &    &  0 &  0 &  0 & L-F \\
\hline
\enddata
\tablecomments{Same as Table \ref{tab:groupstable_bonafide}, \added{but }for the lithium sample.}
\end{deluxetable}

\added{The comparisons in Tables \ref{tab:groupstable_bonafide} and \ref{tab:groupstable_lithium} are not `recoveries'} in the traditional sense, as there are no widely accepted correct answers (apart from, perhaps, the bona-fide member list). LACEwING, BANYAN, and BANYAN II are roughly tied in terms of accuracy, except with the AB Dor moving group\added{,} where BANYAN clearly outperforms all other codes.

\replaced{Tests of LACEwING while under development demonstrated that the recovery rate decreased as the number of moving groups being tested for increased although the false positive rate dropped as well.}{Tests of LACEwING while under development show that as more groups are added to the kinematic model, the recovery rate drops, but the false positive rate drops as well. While developing LACEwING, attempts were made to strike a compromise between recovery rates and false positive rates.}

\section{The NYMGs}
\label{sec:NYMGs}
In the course of this study, we have collected notes about the moving groups themselves, both generally and in terms of what our TRACEwING and LACEwING analyses say about the existence and properties of the moving groups themselves.

\subsection{$\epsilon$ Cham\ae leontis}
This group was discovered by \citet{Mamajek2000} during examination of the $\eta$ Cha open cluster. For a time, it was assumed (e.g. \citealt{Torres2008}) to be co-eval with the open cluster and likely part of the same star forming event; \citet{Murphy2013} recently used evolutionary models that suggest it is younger than $\eta$ Cha. The hottest (and likely most massive) member of $\epsilon$ Cha is \object[eps Cha]{$\epsilon$ Cha} itself, a B9V star.

\subsection{$\eta$ Cham\ae leontis Open Cluster}
First discovered by \citet{Mamajek1999}, $\eta$ Cha is the smallest and third-closest open cluster to the Sun. The hottest member of the open cluster is \object[eta Cha]{$\eta$ Cha} itself, a B9V star whose companions form most of the members of the open cluster. Only two members of $\eta$ Cha ($\eta$ Cha and \object{RS Cha}) have parallaxes, so our moving group parameters are taken from \citet{Murphy2010}.

\subsection{TW Hydrae}
Discovered by \citet{de-la-Reza1989}, TW Hya was the second known nearby young moving group discovered (after Ursa Major), and TW Hya itself is the closest classical T Tauri star to the Sun. Its UVW velocity and projected sky position are similar to the Lower Centaurus Crux \deleted{(LCC) }region of the Sco-Cen star forming complex, and the group itself extends from roughly 30 pc from the Sun to the near edge of \replaced{LCC}{Lower Centaurus Crux}. Many \replaced{suggested}{proposed} TW Hya members are now thought to actually be part of that background group. The hottest member of TW Hya is not actually \object{TW Hya} itself (K6Ve), but \object{TWA 11} (HR 4796), an A0V star.

\subsection{32 Orionis}
This group was first noticed by \citet{Mamajek2007} as a small knot of stars around the B5V+B7V binary star \object{32 Ori}; very little has been studied about this group of stars. Where most of the NYMGs have been linked to origins in the Sco-Cen star forming region, 32 Ori may have a different origin: \citet{Bouy2015} proposed a different arrangement of gas near the Sun where instead of Gould's belt, there are three parallel streams, one of which would stretch from 32 Ori to the Orion Nebula Complex.

\subsection{$\beta$ Pictoris}
Discovered by \citet{Barrado-y-Navascues1999}, $\beta$ Pic is one of the moving groups that effectively surrounds the Sun. The most massive member of the moving group is \object{HR 6070} (A0V), though our tracebacks actually rejected it as a member. The hottest member of $\beta$ Pic not rejected by our tracebacks is \object[bet Pic]{$\beta$ Pic} itself, an A5V star with a planet and prominent debris disk.

\subsection{Capricornus}
Discovered by \citet{van-den-Ancker2000}, Capricornus had only two proposed members, one of which \added{(\object[BD-17 6127]{BD-17 6127 AB})} is now identified as a $\beta$ Pic member\added{; the other, \object{HD 356823}, matches no known group}. No attempt was made to consider this moving group for inclusion in LACEwING.

\subsection{Cham{\ae}leon-Near}
First published by \citet{Zuckerman2004}, most of this moving group's members are now thought to be part of Argus and $\epsilon$ Cha. This group was never considered for inclusion in LACEwING.

\subsection{Octans}
\label{sec:octans}
First proposed by \citet{Torres2008}, Octans is one of the most distant moving groups. Based on our ellipse fits, it is also the largest (and therefore least dense) moving group, and LACEwING has difficulty differentiating its members from field stars. Very few papers have studied Octans, apart from \citet{Murphy2015}. The hottest star in Octans is \object{HD 36968}, an F2 star. Every other group except Carina and Car-Near has a hotter member in the B5-A5 range, suggesting that there is either something different about Octans, or \replaced{ a hotter member remains}{hotter members remain} to be identified.

\subsection{Tucana-Horologium}
\label{sec:tuc-hor}
The Horologium moving group was discovered by \citet{Torres2000}, followed by the Tucana moving group \citep{Zuckerman2001a}, and then the realization that they were two parts of the same group \citep{Zuckerman2001b}. Thanks primarily to the work of \citet{Kraus2014a}, Tuc-Hor has the most known members of any NYMG, though most are M dwarfs. Considering only the \replaced{OBAFGK}{BAFGK} members, Tuc-Hor is still likely smaller (63 \replaced{OBAFGK}{BAFGK} members) than AB Dor (86 \replaced{OBAFGK}{BAFGK} members). The hottest star in Tuc-Hor is \object[alf Pav]{$\alpha$ Pav}, a B2IV star, although the \object[bet01 Tuc]{$\beta$ Tuc} sextuple system (B9V+A0V+A2V+A7V+unknowns) may be the most massive.

For most NYMGs, their members have generally non-existent probabilities of membership in other NYMGs. This is not true for Tuc-Hor: a large fraction of Tuc-Hor members also have low but significant probabilities of membership in Columba, and somewhat less significant probability of membership in $\beta$ Pic.

\subsection{Columba}
\label{sec:columba}
The Columba moving group was first announced by \citet{Torres2008} along with the Carina moving group, as subdivisions of a larger ``GAYA'' complex that also contained Tuc-Hor. While Columba does have a differing UVW velocity from Tuc-Hor and Carina, its spatial location at one end of the Tuc-Hor \replaced{(Carina is on the other end)}{(see Figure \ref{fig:catalog_movie})} and persistently similar age to the other two groups continue to suggest that this may not be distinct from Tuc-Hor. LACEwING finds that many Columba bona-fide members have low probabilities of membership in Tuc-Hor. The most massive member of Columba is \object{HR 1621} (B9V). The known planet host \citep{Marois2008} and putative Columba member \citep{Baines2012} \object{HR 8799} survived both TRACEwING filtering and LACEwING analysis as a bona-fide member of Columba, despite the fact that its high mass (as an A5V star) and position far from other known members is at odds with the idea of mass segregation.

\subsection{Carina}
\label{sec:carina}
As mentioned in Section \ref{sec:columba}, Carina was first discovered by \citet{Torres2008} as part of the ``GAYA'' complex, a larger group that included Tuc-Hor and Columba; \replaced{Carina sits on the opposite end of Tuc-Hor from Columba.}{like Columba, Carina sits physically adjacent to Tuc-Hor.} A preliminary version of LACEwING used in \citet{Riedel2016a} and \citet{Faherty2016} actually excluded Carina because, with an earlier version of the Catalog, tracebacks removed all but two stars from Carina (see Section \ref{sec:herlyr}). Carina has a spatial volume only slightly larger than $\eta$ Cha, $\epsilon$ Cha, and Ursa Major. The most massive member of Carina is \object{HD 83096}, an F0V star (which is lower-mass than most moving groups' largest members; see also Section \ref{sec:octans}).

Most members of Carina have low\added{, but non-zero,} probabilities of membership in Columba or Tuc-Hor. It is also a fairly poorly-recovered group by all moving group codes studied; the best recovery in the lithium sample was 8 of 21 members, by both BANYAN and LACEwING in \replaced{Field}{Young} Star mode.

\subsection{Carina-Vela}
\label{sec:car-vel}
Discovered by \citet{Makarov2000}, Car-Vel was suspected of being related to the IC 2391 open cluster in much the same way as \added{the} \deleted{Eggen's} IC 2391 Supercluster \added{\citep{Eggen1991}}, although they share no members. Most of the stars thought to be in Car-Vel are now assigned as field stars, members of Carina, or members of Argus. This group was never considered for inclusion in LACEwING.

\subsection{Argus}
\label{sec:argus}
Argus was first identified by \citet{Torres2008}\deleted{,} as an improvement on the Car-Vel moving group, with which it shares some members. Argus shares a UVW velocity with the IC 2391 open cluster and is thought to be the product of the same star-forming event \citep{De-Silva2013}. This association is problematic: with the cluster at a distance of \replaced{$\sim$140}{$\sim$150} pc\added{ \citep{Caballero2008}}, it is not clear how the stars could have reached the vicintity of the Sun yet have space velocities parallel to IC 2391 at the present day. Argus is also problematic in that \citet{Bell2015} found the members to not be co-eval, and failed to compute an age for the NYMG. The hottest member of Argus is \object[eps Pav]{$\epsilon$ Pav} (A0V).

The problematic nature of Argus may also be exhibited in the lithium sample study, where none of the codes managed to find even half of the 28 lithium-detected members of Argus.

\subsection{AB Dor}
\label{sec:abdor}
AB~Dor was first identified by \citet{Torres2003} as ``AnA'' and by \citet{Zuckerman2004a} as AB~Dor (and possibly pre-discovered by \citet{Asiain1999} as subgroup B4.) AB~Dor has a space velocity distribution encompassing that of the Pleiades, and has been thought to be a product of the same star-forming event \citep{Luhman2005,Ortega2007}. Though initially thought to be a younger population at 50 Myr \citep{Close2005}, it is now believed to be as old as 150 Myr \citep{Bell2015}, which would make it older than the Pleiades. AB Dor is the largest of the NYMGs based on number of systems with \replaced{OBAFGK}{BAFGK} primaries. However, AB Dor is old enough that lithium and surface gravity are less effective indicators for \replaced{low-mass}{low mass} objects, and chemical tagging studies \citep{Barenfeld2013,McCarthy2014} have determined that perhaps as many as half of AB Dor members may be interlopers. AB Dor is effectively an all-sky moving group. The hottest member of AB Dor is \object{Alnair} (B6V).

LACEwING recovers all but \replaced{8}{eight} bona-fide members of AB Dor in Field star mode, and all but \replaced{5}{five} members in young star mode; in all cases the rejected stars match no other moving group. In both cases, it failed to recover Alnair and \object[del Scl]{$\delta$ Scl} (the hottest member \added{that }it does recover is \object{HR 1014}, A3V).

\subsection{Carina-Near}
\label{sec:carina-near}
Car-Near was identified by \citet{Zuckerman2006} as a nearby older population of stars. The hottest member of Car-Near is \object{HR 3070} (F1), which is cooler than most group's hottest members (see Sections \ref{sec:octans} and \ref{sec:carina}). 

Car-Near is not well recovered by LACEwING due to its large volume and small membership. One member (\object{LP 356-14}) is identified in both young and field star modes as a member of Argus, the group whose UVW velocity Car-Near most closely resembles. HR 3070 is an Argus member in Young Star mode. The rest either match Car-Near or no group at all. \added{Despite this poor recovery, Car-Near still appears to be a real group. It does not have an unusually large present-day volume like Oct-Near (Section \ref{sec:octans-near}) and corresponding complete failure of recovery, its members are largely only members of the group itself, and it does produce self-consistent tracebacks, unlike Her-Lyr (Section \ref{sec:herlyr}). It seems most likely that other members of Car-Near will be found, increasing the spatial density and therefore the recovery rate.}

\subsection{Octans-Near}
\label{sec:octans-near}
Oct-Near was identified by \citet{Zuckerman2013} as a potential very-nearby association of stars (including \object[EQ Peg]{EQ~Peg AB} at 6.2 pc) with similar velocities to Octans and Castor. Though considered for inclusion in LACEwING, Oct-Near posed two problems for inclusion. First, as presented in \citet{Zuckerman2013}, the group had multiple apparent ages between 30 and 200 Myr. Second, the present-day moving group ellipses were more than three times the size of the other moving groups and the resulting groups so sparse that LACEwING could not recover any of the supposed bona-fide members. Accordingly, it is not included in this implementation of LACEwING. The hottest member of Octans-Near was \object{34 Psc} (B9~V).

\subsection{Hercules-Lyra}
\label{sec:herlyr}
Her-Lyr was first identified by \citet{Gaidos1998} and \citet{Fuhrmann2004}, and was comprised almost entirely of nearby stars; it would have been another all-sky moving group. The existence of Her-Lyr has been disputed for some time; as \citet{Mamajek2015} notes, multiple papers (most recently, \citealt{Eisenbeiss2013} have identified members of Her-Lyr, but none have consistent lists of members. Nevertheless, in recognition of the fact that the supposed members of Her-Lyr were a population of lithium-rich stars older than AB Dor, we attempted to retain Her-Lyr. Her-Lyr did appear in the preliminary LACEwING calibration used in \citet{Riedel2016a} and \citet{Faherty2016}, but with data from our updated Catalog of Suspected Nearby Young Stars (Section \ref{sec:catalog}), only one star -- \object{V0439 And} -- remained within the boundaries of Her-Lyr and survived traceback filtering. Her-Lyr is not included in this implementation of LACEwING. The hottest member of Her-Lyr was \object[alf Cir]{$\alpha$ Cir} (A7V), which was not listed as a bona-fide member in \citet{Eisenbeiss2013}.

\subsection{Castor}
Castor was first identified by \citet{Anosova1991} in a relatively wide (6 km s$^{-1}$) search for potential clusters around prominent triple star systems. Though refined in multiple papers over the ensuing 25 years, the existence of Castor has been questioned and debunked by \citet{Mamajek2013} on the grounds that the most prominent members (\object{Vega}, \object{Fomalhaut}, \object{LP 944-020}, and \object{Castor}) were nowhere near each other even 10 Myr ago and could not have formed in the same molecular cloud. This determination was made with a different dataset and a simple linear traceback. The existence of Castor was also disputed by \citet{Zuckerman2013} on the basis of differing ages and large velocity spreads within the group.

We have attempted to trace back Castor ourselves, starting with all 84 members ever identified as Castor members (regardless of current identification) from the Catalog of Suspected Nearby Young Stars with parallaxes and radial velocities. Only 33 stars survived three rounds of TRACEwING filtering, at which point the group included neither Castor itself nor LP~944-020, and was still the largest moving group in terms of volume at formation (Figure \ref{fig:traceback_groups}). Castor's volume at formation is indistinguishable from the size of the fake moving group of field stars. Castor, a sextuple system with four A-type stars, is likely both the hottest and most massive member of the moving group. 

We cannot conclusively rule out the existence of Castor because it is still smaller than our simulated moving group between the ages of 200-400 Myr (covering both the age given by \citealt{Barrado-y-Navascues1998} and the ages of the prominent members collected in \citealt{Mamajek2013}), but we \deleted{strongly }suspect this is further evidence that Castor is not a real moving group, and we have not included it in LACEwING.

\subsection{Ursa Major}
The Ursa Major moving group, often referred to as the Sirius Supercluster in older literature and occasionally referred to as a cluster \citep{Mamajek2015}, was the first moving group discovered, by R. A. Proctor in the late 19th century. \citet{Castro1999} published a chemical tagging analysis that revealed that the group members are identifiably rich in barium. \citet{King2003} published the most recent large-scale study of the group and determined that Sirius is not a likely member. Ursa Major is represented here by only the core members from \citet{King2003}, and is thus one of the smallest moving groups in LACEwING. The hottest member of Ursa Major is \object[eps UMa]{$\epsilon$ UMa} (A0pCr), although \object[Mizar]{Mizar-Alcor}, a sextuplet with five A-type stars, is likely to be more massive.

Ursa Major is another troubling group; the final traceback set of Ursa Major members did not include Mizar-Alcor, and the resulting LACEwING calibration missed 8 of the 11 lithium-rich members in young mode (although LACEwING does recover Mizar-Alcor as a member).

\subsection{Coma Berenices Open Cluster}
Coma Ber (Melotte 111, Collinder 256) is an open cluster \deleted{roughly} 86 pc distant consisting of roughly 195 stars from our limited literature search, of which 104 are \replaced{OBAFGK}{BAFGK} members. Thus, Coma Ber\deleted{enices} is second only to the Hyades in terms of size. Coma Ber was a difficult moving group to add to our simulation of the Solar Neighborhood because it has a very low UVW velocity, and the simulation produced proper motions indistinguishable from zero for many of the simulated members. Fortunately, LACEwING's field population prevents LACEwING from identifying large portions of the sky as Coma Ber members. The hottest member of Coma Ber is \object{AI Com} (A0p).

\subsection{$\chi^{01}$ Fornax}
$\chi^{01}$ For (also known as Alessi 13) was first published in a catalog by \citet{Dias2002}. It has remained obscure for the past decade, but appears to be real (E.E. Mamajek 2106, private communication). The only available age estimates are from  \citet{Pohnl2010} and \citet{Kharchenko2013}, which hover around 500 Myr, though Mamajek believes the group may be younger \citep{Mamajek2016}. For the purposes of traceback we intended to use 450-550 Myr, but only three members of the group have parallaxes\replaced{, and w}{. W}e used the UVW properties from the private communication with E.E. Mamajek instead, and calculated the XYZ positions from the \replaced{RA}{$\alpha$}, \replaced{DEC}{$\delta$}, \replaced{distance}{$Dist.$}, and tidal radius. The hottest member of $\chi^{01}$ For is \object[chi01 For]{$\chi^{01}$ For} (A1V) itself.

\subsection{Hyades Open Cluster}
The Hyades open cluster (Melotte 25, Collinder 50) has been known since antiquity. Based on \replaced{an analysis}{analyses} of \citet{Roeser2011} and \citet{Goldman2013}, there are 724 known members of the Hyades, of which 260 are \replaced{OBAFGK}{BAFGK} members (including giants and white dwarfs). The hottest stellar member (excluding white dwarfs) is \object[tet02 Tau]{$\theta^{02}$ Tau}, an A7III giant. The age is believed to be between 600 \citet{Zuckerman2004} and 800 Myr \citet{Brandt2015}, and is the upper limit of what we consider to be a young group.

The Hyades are represented here by a conversion of the properties in \citet{van-Leeuwen2009} to UVWXYZ.

\subsection{Young Non-members}
\label{sec:young_field}
A large fraction of our lithium sample - 582 of the 930 star systems - do not trace back to any of the NYMGs in young star mode. \replaced{Twenty-Four}{Twenty-four} of these just miss ($>10$\% probability of membership in at least one group) the usual 20\% probability cut, which could be due to only having \deleted{a }proper motion available for membership assignment, but that still leaves 534 of 930 systems (57\%) as non-members of the groups. 
This behavior is not unique to LACEwING. Running the same sample through BANYAN, which had the highest recovery rate of any of the codes, found 589 stars fell into the ``Old'' category. BANYAN II in field star mode identified 369 stars as young field objects and 355 stars as old field objects; in young star mode it identified 720 members of the lithium sample as ``Old''. This is not a perfect comparison, as the BANYAN codes test for fewer groups than LACEwING, and some of the lithium-detected stars are members of groups older than BANYAN's oldest.

If we break down the entire lithium sample, only 487 of the star systems have ever been considered as potential members of groups LACEwING tests for. An additional \replaced{19}{79} have only been considered as members of moving groups LACEwING does not test for (Her-Lyr, Oct-Near, Castor, Car-Vel\added{, Local Association, Hyades Supercluster, IC 2391 Supercluster})\deleted{, and an additional 60 have been identified as members of the pre-{\it Hipparcos} streams (Local Association/Pleiades Moving Group, Hyades Supercluster/Stream, IC 2391 Supercluster)}. The remaining 364 systems have never been considered as members of any specific young group.

This general behavior has been noted before, particularly in surveys that did not filter by kinematic match before following up stars for further observations \citep{Shkolnik2009,Shkolnik2012,Riedel2014,Riedel2016b,Binks2015}. These results have suggested that there is an unidentifiable population of young stars.

If we accept that these stars are young, there are five possibilities for their origins.
\begin{enumerate}
\item The stars did form as part of the known NYMGs, and flaws in our data or models are responsible for the stars that are not associated with a group.
\item The stars did form as part of the known NYMGs, but dynamical interactions (most likely an ejection from a higher-order multiple star system early in its evolution) gave them high relative velocities such that they do not kinematically match the group that they formed with. Chemical tagging would be useful for identifying such systems, provided it is possible to uniquely identify a moving group in that way. Kinematic tracebacks that place a star near another member of a moving group of the appropriate age would be suggestive, but would require advancements in both data precision and technique from what is presented here.
\item The stars did form as part of groups that are known but not nearby. LACEwING only tests for groups currently known to extend within 100 pc of the Sun. The Scorpius-Centaurus and Taurus-Auriga star-forming regions are under 200 pc away. With typical velocity dispersion of 1.5 km s$^{-1}$ \citep{Preibisch2008}; a star could move from 118 pc (the canonical distance to the $\sim$16 Myr old Lower Centaurus Crux region of Scorpius-Centaurus, \citealt{Preibisch2008}) to within 100 pc of the Sun in just over 10 Myr.
\item The stars did form as part of unknown NYMGs \replaced{which}{that} we have not yet discovered. 32 Ori and $\chi^{01}$ For are relatively unexplored groups; the All Sky Young Association \citep{Torres2016} has been announced but no particulars have yet been given. Other, smaller groups that have no members more massive than M dwarfs may yet be hiding in the Solar Neighborhood.
\item The stars did not form as part of any groups at all, and were instead part of a one-off star formation event. The NYMGs already range greatly in size from the $\eta$ Cha open cluster (21 known systems as of 2015 January) to the Tuc-Hor moving group (209 known systems as of 2015 January); it is not clear what the smallest star formation event can be.
\end{enumerate}

\section{Conclusions}
\label{sec:conclusions}

We have introduced the LACEwING moving group identification code, which uses kinematics to determine the probability of membership in 13 NYMGs and \replaced{3}{three} open clusters within 100 pc. We have introduced the TRACEwING epicyclic traceback code, which uses an epicyclic approximation to Galactic orbital motion to trace stars back to their origins. We have also introduced the catalog of suspected young stars, which contains a wide variety of kinematic, spectroscopic, photometric, and membership information on 5350 nearby stars that have been identified in the literature as potentially young.

We have demonstrated that LACEwING produces reliable results consistent with current expectations, for the first time across all known moving groups within 100 parsecs. Despite handling a large number of moving groups, LACEwING's recovery rates are in line with other previously established moving group codes like BANYAN and BANYAN II. By including more moving groups in the kinematic identification, we make it substantially easier to identify members and obtain ages for nearby stars. Uniform and repeatable determinations are now possible for groups with a wide range of ages, covering a wide range of youthful states.

The TRACEwING epicyclic traceback code allows us to identify objects based on their spatial origins, which should be a more fundamental constraint than present-day kinematics, particularly once higher precision data is available. On the issue of populations, it should provide a useful means for testing the spatial formation scenarios of the nearby young moving groups.

The Catalog of Suspected Nearby Young Stars constitutes a valuable resource for studying the large-scale properties of nearby young stars in both an individual and populational basis. It will be maintained as part of a larger database of young stars (\citealt{Hillenbrand2015}, currently under construction), and made available to other researchers for use as a source of data for a wide variety of studies.

The {\it Gaia} mission, due to its precision, accuracy, and extraordinary magnitude range, will make it possible to perform analyses like this on stars beyond 100 pc from the Sun. {\it Gaia} data will allow us to conclusively answer the question of the existence of groups like Castor. {\it Gaia} should make it possible to test the various theories to explain the origins of the young field, and potentially break up the currently known groups into smaller, more physically meaningful groups just as the {\it Hipparcos} mission did for the previously known stellar streams.

Based on the analysis done here, the most certain groups are $\epsilon$ Cha, $\eta$ Cha, TW Hya, $\beta$ Pic, 32 Ori, Tuc-Hor, AB Dor, Coma Ber, and $\chi^{01}$ For; the existence of the Hyades is also not in doubt. More work needs to be done on the rest of these groups. In particular, that includes the Columba and Carina moving groups which may be part of Tuc-Hor; the 32 Ori (and $\chi^{01}$ For) groups that have not been well studied to date; and groups like Octans, Argus, Car-Near, and Oct-Near that appear to be problematic in size or recovery performance.

\section{Acknowledgements}
\acknowledgements

A.R.R. acknowledges support from NSF grant AST-131278, NASA ADAP grant NNX12AD97G, and generous support from the Office of the Provost at the College of Staten Island, City University of New York. The authors wish to thank J. Gagn{\'e}, L. Malo, D.\added{~R.} Rodriguez, and E.\added{~E.} Mamajek for helpful suggestions and commentary, G. Schwarz for help with catalog preparation, J.\added{~L.} McDonald for editorial assistance, and our two referees, J.\added{~A.} Caballero and an anonymous referee, whose suggestions greatly improved the quality and readability of the paper, the code, and the data products.

This publication makes use of data products from the Two Micron All Sky Survey, which is a joint project of the University of Massachusetts and the Infrared Processing and Analysis Center/California Institute of Technology, funded by the National Aeronautics and Space Administration and the National Science Foundation.

This publication makes use of data products from the {\it Wide-field Infrared Survey Explorer}, which is a joint project of the University of California, Los Angeles, and the Jet Propulsion Laboratory/California Institute of Technology, funded by the National Aeronautics and Space Administration.

This research has made use of the Washington Double Star Catalog maintained at the U.S. Naval Observatory. This research was also made possible through the use of the AAVSO Photometric All-Sky Survey (APASS), funded by the Robert Martin Ayers Sciences Fund.
This research has made extensive use of the SIMBAD database and VizieR catalogue access tool operated at CDS, Strasbourg, France. The original description of the VizieR service was published in A\&AS 143, 23.

\software{Astropy \citep{Astropy2013}, Numpy \citep{Walt2011}, Scipy \citep{Jones2001}, Matplotlib \citep{Hunter2007}, LACEwING \citep{Riedel2016c}}

\bibliographystyle{aasjournal} \bibliography{riedel_a}

\appendix
\twocolumngrid
\section{LACEwING manual}

LACEwING is available from the Astronomy Source Code Library: \url{http://ascl.net/1601.011} and also from Github: \url{https://github.com/ariedel} (as of 2016 Dec 5)

What is required to run LACEwING to obtain membership probabilities is as follows:
\begin{itemize}
\item A Python 2.7 interpreter with numpy, astropy, and matplotlib.
\item lacewing.py - the main routines necessary for LACEwING, specifically
\item kinematics.py - Routines to convert between equatorial ($\alpha$, $\delta$, $\pi$) and Galactic (XYZ) coordinates (and also kinematic tracebacks)
\item ellipse.py - ellipse fitting and rotation routines
\item astrometry.py - Proper Motion conversion routines and other miscellaneous important routines.
\item Moving\_Groups\_all.csv - A comma separated value format file with the precalculated moving group parameters
\item The files on Github contain an additional file, Moving\_Groups\_all\_prelim.csv, which contains the preliminary moving groups used by \citet{Faherty2016}. Rename to Moving\_Groups\_all.csv to use.
\end{itemize}

The gal\_uvwxyz function in kinematics.py is a modified version of gal\_uvw originally written by Wayne Landsman for the IDL Astronomy User's Library (under a 2-clause BSD license) and converted to Python 2 by Sergey Koposov.

\subsection{General Usage}
\label{sec:lacewing_usage}

The LACEwING algorithm is available as a function that can be called from other programs (see Section \ref{sec:lacewing_function}), but if it is run directly from the command line it defaults to reading from a comma separated value file.

\begin{spverbatim}
python lacewing.py inputfile [calibration] [output filename] [verbose output] [g.o.f]
\end{spverbatim}

LACEwING (using with the astropy.io.ascii module) looks for a one-line header with any or all of the columns selected below (broadly, they are either common names for the quantities or AAS Journal standards). These can be given in any order; any other headers present in the file are ignored. Most of the errors are optional and if they (and their header) are not present, LACEwING will substitute default values.
\begin{itemize}
\item Name - Ascii, any length; should not contain commas itself.
\item RA,DEC (or RAdeg, DEdeg) - Right ascension and declination in decimal degrees, J2000/ICRS equinox.
\item RAh, RAm, RAs, DE-, DEd, DEm, DEs - Sexagesimal coordinates, J2000/ICRS equinox, split into 7 columns with a separate declination +/- flag. These will only be read if RA and DEC are invalid, empty, or do not exist.
\item eRA,eDEC (or e\_RAdeg, e\_DEdeg) - $RA\cos{DEC}$ and $DEC$ uncertainties in milliarcseconds. These, and their header keyword, are optional and will default to 1000 milliarcseconds.
\item pmRA,pmDEC (or pmRA, pmDE; or pmra and pmdec) - $\mu_{RA\cos{DEC}}$ and $\mu_{DEC}$ in mas yr$^{-1}$, J2000/ICRS equinox.
\item epmRA, epmDEC (or e\_pmRA,e\_pmDE; or epmra, epmdec) - $e\mu_{RA\cos{DEC}}$ and $e\mu_{DEC}$ uncertainties in mas yr$^{-1}$, J2000/ICRS equinox. These, and their header, are optional and will default to 10 mas yr$^{-1}$.
\item pi (or plx) - trigonometric parallax in mas. It is highly recommended that you NOT use photometric parallax estimates.
\item epi (or e\_plx) - trigonometric parallax uncertainty in mas. This must be present in order to use the parallax.
\item rv (or HRV) - radial velocity in km s$^{-1}$.
\item erv (or e\_HRV) - radial velocity uncertainty in km kilometers s$^{-1}$. This must be present in order to use the radial velocity.
\item Note - Any text you wish to have duplicated in the file output, such as a previously known membership assignment.
\end{itemize}

The additional command-line parameters are:
\begin{itemize}
\item calibration - Type ``young'' to switch to the alternative LACEwING calibration where stars are assumed to be young. Defaults to ``field''
\item output filename - Enter an output filename; otherwise LACEwING defaults to lacewing\_output.csv
\item verbose - If this is ``verbose'', LACEwING will output a more detailed report as described below. It defaults to a much more compact output format.
\item g.o.f - If this is set to anything other than ``percentage'', LACEwING outputs the combined goodness-of-fit statistic rather than determining the moving group membership probability.
\end{itemize}

The LACEwING output format, in the regular version, is a comma separated value file containing, in order:
\begin{itemize}
\item Star Name
\item Note (directly from the input)
\item Best-matching Moving Group
\item Membership Probability for the best-matching moving group (unless the 'g.o.f' flag is set on the command line)
\item Kinematic Distance (in parsecs) for the best-matching group
\item Kinematic Distance uncertainty (in parsecs) for the best-matching group
\item Kinematic Radial Velocity (in km s$^{-1}$) for the best-matching group
\item Kinematic Radial Velocity uncertainty (in km s$^{-1}$) for the best-matching group
\item Membership probability for $\epsilon$ Cha
\item Membership probability for $\eta$ Cha
\item Membership probability for TW Hya
\item Membership probability for $\beta$ Pic
\item Membership probability for 32 Ori
\item Membership probability for Octans
\item Membership probability for Tuc-Hor
\item Membership probability for Columba
\item Membership probability for Carina
\item Membership probability for Argus
\item Membership probability for AB Dor
\item Membership probability for Carina-Near
\item Membership probability for Coma Ber
\item Membership probability for Ursa Major
\item Membership probability for $\chi^{01}$ For
\item Membership probability for Hyades
\end{itemize}

The Verbose Output formula is drastically different, consisting of a block of 16 lines detailing the match between the star and every group, one at a time. It is most useful for tracking down the reason why a particular star did not match a particular group.
\begin{itemize}
\item Name
\item Right Ascension (degrees, from the input)
\item Declination (degrees, from the input)
\item Moving Group 
\item Text string: ``PROB='' (or ``SIG='', if the g.o.f. flag has been set on the command line)
\item Membership Probability in the moving group (or combined goodness-of-fit if the g.o.f. flag has been set)
\item Text string: ``PM=''
\item Proper Motion goodness-of-fit metric
\item Kinematic (predicted) $\mu_{RA\cos{DEC}}$ (mas yr$^{-1}$)
\item Kinematic $\mu_{RA\cos{DEC}}$ uncertainty (mas yr$^{-1}$)
\item Kinematic $\mu_{DEC}$ (mas yr$^{-1}$)
\item Kinematic $\mu_{DEC}$ uncertainty (mas yr$^{-1}$)
\item Measured $\mu_{RA\cos{DEC}}$ (mas yr$^{-1}$)
\item Measured $\mu_{RA\cos{DEC}}$ uncertainty (mas yr$^{-1}$)
\item Measured $\mu_{DEC}$ (mas yr$^{-1}$)
\item Measured $\mu_{DEC}$ uncertainty (mas yr$^{-1}$)
\item Text string: ``DIST=''
\item Distance goodness-of-fit metric
\item Kinematic distance (pc)
\item Kinematic distance uncertainty (pc)
\item Measured distance ($\frac{1}{\pi}$, pc)
\item Measured distance uncertainty ($\frac{\sigma_{\pi}}{\pi^2}$, pc)
\item Text string: ``RV=''
\item Radial velocity goodness-of-fit metric (km s$^{-1}$)
\item Kinematic radial velocity (km s$^{-1}$)
\item Kinematic radial velocity uncertainty (km s$^{-1}$)
\item Measured radial velocity (km s$^{-1}$)
\item Measured radial velocity uncertainty (km s$^{-1}$)
\item Text string: ``POS='' (or ``KPOS='', if the spatial position is based on the kinematic distance)
\item Spatial Position goodness-of-fit metric
\item 3D separation between star and center of moving group (pc)
\item 3D separation uncertainty (pc)
\item Note (directly from the input)
\end{itemize}

The lacewing\_summary.py function will summarize verbose output into the regular format.
\begin{verbatim}
python lacewing_summary.py inputfile
\end{verbatim}
The output file will be named $inputfile$.summary.csv

\subsection{LACEwING as a function}
\label{sec:lacewing_function}

LACEwING as a function requires two calls: 
One, lacewing.moving\_group\_loader() returns a list of moving group classes loaded from the Moving\_Group\_all.csv file. This must be done before lacewing.lacewing(), so the resulting list can be passed to lacewing.lacewing(). This is done to avoid the redundancy of reloading the parameters every time.

lacewing.lacewing() accepts data on a single object at a time, and cannot be fed lists, arrays, or tuples. The returned list of moving group classes must be passed to lacewing.lacewing() every time. The first arguments is required.
\begin{itemize}
\item moving\_group - list, moving group class list (as created by lacewing.moving\_group\_loader())
\end{itemize}
The remainder of the items are optional; if not present (or explicitly set to NoneType None) lacewing() will treat them as unknowns.
\begin{itemize}
\item young= - string, if this is ``young'' then lacewing() will use the young star calibration. Otherwise (or if None, or not specified), lacewing() will use the field star calibration.
\item ra=  - float, decimal degrees, J2000/ICRS coordinates
\item era= - float, RA$\cos{DEC}$ uncertainty in decimal degrees 
\item dec= - float, decimal degrees, J2000/ICRS coordinates
\item edec= - float, DEC uncertainty in decimal degree
\item pmra= - float, $\mu_{RA\cos{DEC}}$ in arcseconds yr$^{-1}$
\item epmra= - float, $\mu_{RA\cos{DEC}}$ uncertainty in arcseconds yr$^{-1}$
\item pmdec= - float, $\mu_{DEC}$ in arcseconds yr$^{-1}$
\item epmdec= - float, $\mu_{DEC}$ uncertainty in arcseconds yr$^{-1}$
\item plx= - float, $\pi$ in arcseconds
\item eplx= - float, $\pi$ uncertainty in arcseconds
\item rv= - float, RV in km s$^{-1}$
\item erv= - float, RV uncertainty in km s$^{-1}$
\end{itemize}
Note that lacewing() does not check to see if the uncertainties exist (or are not None), so (for example) supplying rv without erv will cause lacewing() to crash with a TypeError.

The output from lacewing() is a list of dicts, one dict per moving group in the order they appear in Moving\_Group\_all.csv. Each dict contains the following keys (which will be None unless data exists)
\begin{itemize}
\item 'group' - string, group name 
\item 'gof' - float, goodness-of-fit parameter
\item 'probability' float, membership probability (percent)
\item 'pmsig' - float, proper motion match metric
\item 'kin\_pmra' - float, estimated $\mu_{RA\cos{DEC}}$ (arcseconds yr$^{-1}$)
\item 'kin\_epmra' - float, estimated $\mu_{RA\cos{DEC}}$ uncertainty (arcseconds yr$^{-1}$)
\item 'kin\_pmdec' - float, estimated $\mu_{DEC}$  (arcseconds yr$^{-1}$)
\item 'kin\_epmdec' - float, estimated $\mu_{DEC}$ uncertainty (arcseconds yr$^{-1}$)
\item 'distsig' - float, distance match metric
\item 'kin\_dist' - float, expected distance (pc)
\item 'kin\_edist' - float, expected distance uncertainty (pc)
\item 'rvsig' - float, radial velocity match metric
\item 'kin\_rv' - float, expected radial velocity (km s$^{-1}$)
\item 'kin\_erv' - float, expected radial velocity uncertainty (km s$^{-1}$)
\item 'possig' - float, position match metric (using measured distance)
\item 'pos\_esig' - float, position match metric uncertainty
\item 'pos\_sep' - float, distance from moving group center (pc)
\item 'posksig' - float, position match metric (using kinematic distance)
\item 'pos\_eksig' - float, kinematic position match metric uncertainty
\item 'pos\_ksep' - float, kinematic distance from moving group center (pc)
\end{itemize}

There are two example implementations of lacewing.lacewing() in the repository: The default reader within the lacewing.py file, and the sample star generator in lacewing\_montecarlo.py

\subsection{Adding a moving group to LACEwING}

\subsubsection{Step 1: Add the group to Moving\_Groups\_all.csv.}

\paragraph{Generate new moving group parameters}
Create a comma separated value file in the format described above for the default csv loader, containing data on every member of the group.

Run this file through the moving group maker:
\begin{verbatim}
python lacewing_mgpmaker.py inputfile
\end{verbatim}
Also, run this through the 2d version to get parameters for lacewing\_uvwxyz.py
\begin{spverbatim}
python lacewing_mgpmaker2d.py inputfile
\end{spverbatim}

The output files are .csv files named ``Moving\_Group\_{\it Group Name}.dat'' and  ``Moving\_Group\_{\it Group Name}\_2d.dat'' with U, V, W, A, B, C, UV, UW, VW, X, Y, Z, D, E, F, XY, XZ, and YZ values, with uncertainties (which LACEwING ignores) for each value on the second line. The first line is suitable for entering into Moving\_Groups\_all.csv. The 2D versions of those values should be stored in the A2, B2, C2, UV2 (etc) columns. Note that UVW and XYZ values are the same for 2D and 3D projections, so there are no special 2D entries for them.

Alternatively: If you have the more customary UVW and XYZ values (no rotations) from a different source, you can enter those into the file, specifying 0 for all the rotation angles. The same values will apply for 3D and 2D.

Column ``Name'' should contain the name of the group, preferably less than 20 characters, with underscores instead of spaces.

Column ``Number'' is a best-estimate of the current known members of the group.

Column ``Weightednumber'' is a cumulative fraction of stars in each group, 
\begin{equation}
\frac{\sum_{0}^{i}{Number_{group(i)}}}{\sum_{0}^{N}{Number_{group(i)}}}
\end{equation}
 such that all are between 0 and 1, and should be generated from the number of members. 

Column ``uniform'' specifies whether the group should be simulated as (0) a uniform spatial and Gaussian velocity distribution (unused), (1) a Gaussian spatial and Gaussian velocity distribution, like an open cluster, or (2) a spatial distribution with a scale height of 300 parsecs and a Gaussian velocity distribution, like the field population.

You can also add an Age (unused), References (unused), and the fractional RGB color components (for plots, such as lacewing\_uvwxyz.py). Also, fill the membership coefficients section with zeros so that the program will function initially.

If you are removing a group, just delete its line and re-compute the ``Weightednumber'' column.

\paragraph{Step 2: Generate a new simulation}

Generate a new Monte Carlo simulation.

\begin{spverbatim}
python lacewing_montecarlo.py iterations number 
\end{spverbatim}

Where ``iterations'' is the number of stars to draw out of the distributions now listed in Moving\_Groups\_all.csv.  ``number'' is a value to append to the names of the output files so that they will be kept separate, and is a way of more efficiently using computer resources.

Good values for ``number'' are the number of cores you are willing to devote to this process. Good values for ``iterations'' depends on the number of members in the least-populated group. This process needs roughly 1,000 simulated member in total to populate the eventual membership histograms well enough to fit them. So, if you have a quad-core machine and the least-populated group accounts for fractionally 0.0001 of the stars, you need 10 million points and should generate 5 million with four running instances. An Intel Core i7 4700MQ running 8 instances can generate about two million stars per day, and 8,000,000 entries takes up roughly 610 MiB of space per moving group. 

The program will use random number generators to follow the procedure in Section \ref{sec:calibration} and generate simulated stars to run through lacewing.lacewing().

The output will be {\it number} files for each moving group, each containing {\it iterations} entries, with the goodness-of-fit values matching each star to the moving group using all 7 possible combinations of input data, and a record of what group the star was generated as a member of.

Concatenate the files for each group into one file per group. Preferably move them to another sub-folder.

\paragraph{Step 3: Generate and fit probability histograms}

Run lacewing\_percentages.py on the Monte Carlo output.

\begin{spverbatim}
python lacewing_percentages.py montecarlofile [young]
\end{spverbatim}

This program generates the histogram of `percentage of objects at each goodness-of-fit value that are actual members'. Run it on one of the files from the Monte Carlo output to generate a wide variety of plots (as appearing in Section \ref{sec:calibration}) and a file called {\it group}.percentages. This file contains the cumulative Gaussian fit parameters that need to be saved in Moving\_Groups\_all.csv. All of these files are stored in a folder called montecarlo, that is created if it does not exist.

If you specify ``young'', the program will generate the young star calibration; as it combs through the input file it will ignore all but an equal number of field stars, to provide the case where the stars are assumed to be young. These calibrations (in {\it group}.young.percentages) should also be saved in Moving\_Groups\_all.csv, replacing the zeroes put there earlier.

You can now use LACEwING to predict members of your new moving group. 

\subsection{lacewing\_uvwxyz.py}
\label{sec:lacewing_uvwxyz.py}

Requirements:
\begin{itemize}
\item A Python 2.7 interpreter with numpy, astropy, and matplotlib.
\item lacewing.py - the main routines necessary for LACEwING, specifically the csv loader.
\item lacewing\_uvwxyz.py - the UVWXYZ fitting and plotting tool.
\item kinematics.py - Routines to convert between equatorial (RA, DEC, $\pi$) and Galactic (XYZ) coordinates (and also kinematic tracebacks)
\item ellipse.py - ellipse fitting and rotation routines
\item astrometry.py - Proper Motion conversion routines and other miscellaneous important routines.
\item Moving\_Groups\_all.csv - A comma separated value format file with the precalculated moving group parameters
\end{itemize}

\subsubsection{General Usage}
lacewing\_uvwxyz.py calculates XYZ (where parallaxes exist) and UVWXYZ (where all six elements of kinematics exist) values, which are output to an output filename. It also generates 2D projection plots of stars plotted on the UVW and XYZ spatial axes, as shown in Figure \ref{fig:ellipses}.

\begin{spverbatim}
python lacewing_uvwxyz.py inputfile [output filename] [XYZ]
\end{spverbatim}

Input file format is the same as lacewing; it uses the same .csv loader).

The next two arguments are optional:
\begin{itemize}
\item output filename - Enter an output filename; otherwise LACEwING defaults to lacewing\_output.csv
\item XYZ - If this string is ``XYZ'', the output .png image will have a second row of projected XYZ positions.
\end{itemize}

There are two outputs: 
\begin{itemize}
\item One image per star (named {\it starname}\_{\it decimal coordinates}.png) with panels showing the U vs V, U vs W, and V vs W projections of 3D velocity (and, if ``XYZ'' was specified, X vs Y, X vs Z, and Y vs Z projections of 3D space) similar to Figure \ref{fig:ellipses}. 
\item One output comma separated value file with rows of Name, U, eU, V, eV, W, eW, X, eX, Y, eY, Z, eZ values for each star.
\end{itemize}

\subsection{TRACEwING}
\begin{itemize}
\item A Python 2.7 interpreter with numpy, astropy, and matplotlib.
\item tracewing.py - the kinematic traceback tool.
\item lacewing.py - for the csv loader.
\item kinematics.py - Routines to trace the star back in time with epicyclic tracebacks.
\item ellipse.py - ellipse fitting and rotation routines
\item astrometry.py - Proper Motion conversion routines and other miscellaneous important routines.
\item Moving\_Group\_{\it Group Name}\_epicyclic.dat - Stored parameter files for the moving groups. The files on Github have been calculated back to -800 Myr.
\end{itemize}

tracewing.py computes the traceback of stars to a given moving group, which must be present in a saved parameter file. The outputs are .png figures of the star traced back to the NYMG (Figure \ref{fig:traceback}), covering the time between 0 and the end of the plotting time period.

\begin{spverbatim}
python tracewing.py inputfile Moving_Group_Name "epicyclic" minage maxage end_of_plot_range iterations
\end{spverbatim}
Inputfile should be the same file format as LACEwING. The Moving Group Name should be ``beta\_Pic'' if the name is ``Moving\_Group\_beta\_Pic\_epicyclic.dat''. ``epicyclic'' is the only method tested and with moving group data to trace back to. The ages and end of plot range must all be negative, but greater than -800 Myr. The number of iterations should determine the quality of the plot; 10000 is likely more than enough.

\subsection{Generating New Moving Groups}

To create a new NYMG from a group of stars, assemble all the stellar properties of the NYMG in a file {\it inputfile} (same format as lacewing.py and tracewing.py). Then run the following:
\begin{spverbatim}
python tracewing_mgp.py inputfile Moving_Group_Name "epicyclic" minage maxage end_of_plot_range
\end{spverbatim}
tracewing\_mgp.py has all the same requirements as tracewing.py
This will trace back all the stars in the file to -800 Myr, fit ellipses at every 0.1 Myr step, and save the output to a file Moving\_Group\_{\it Group Name}\_epicyclic.dat suitable for use in tracewing.py. It will also attempt to generate a .png figure of all the stellar positions from 0 to the end of the plot range, with a blue bar drawn in over the min and max age of the group, as shown in Figure \ref{fig:AB_Dor_traceback}.

Be aware that this will require several gigabytes of memory, as it must hold N$_\textrm{members}\times1000\times8000$ positions in memory.
\section{Contents of the Catalog of Suspected Nearby Young Stars}

The Catalog of Suspected Young Stars is also be available on Github at \url{https://github.com/ariedel/young_catalog} in its present form, and will be incorporated into the \citet{Hillenbrand2015} database.

\startlongtable
\begin{deluxetable}{llll}
\setlength{\tabcolsep}{0.02in}
\tablewidth{0pt}
\tabletypesize{\tiny}
\tablecaption{Headers of the Catalog of Suspected Nearby Young Stars\label{tab:catalog}}
\tablehead{
  \colhead{Number}    &
  \colhead{Label}     &
  \colhead{Units}     &
  \colhead{Description}}
\startdata
1 & RAh & h & Right Ascension Hours (J2000 E2000) (calculated) \\
2 & RAm & arcmin & Right Ascension Minutes (J2000 E2000) (calculated) \\
3 & RAs & arcsec & Right Ascension Seconds (J2000 E2000) (calculated) \\
4 & DE- & -- & Declination sign (J2000 E2000) (calculated) \\
5 & DEd & deg & Declination Degrees (J2000 E2000) (calculated) \\
6 & DEm & arcmin & Declination Minutes (J2000 E2000) (calculated) \\
7 & DEs & arcsec & Declination Seconds (J2000 E2000) (calculated) \\
8 & Seq & -- & Sequence Number \\
9 & LiSample & -- & [ LlFfAa] Lithium Sample Flag. (1) \\
10 & Bonafide & -- & [ BbRrXx] Bona-fide Sample Flag. (2) \\
11 & MultStars & -- & ? Number of objects in system (0 if object is a secondary). Blank if unknown \\
12 & sepkey & -- & Key for system separation considered here \\
13 & MultType & -- & Type of multiplicity (3) \\
14 & Sep & arcsec & separation in arcseconds \\
15 & SepPA & deg & ? Last known position angle of separation. \\
16 & SepDate & yr & ? Date separation was recorded \\
17 & orbper & -- & Orbital Period \\
18 & orbperunit & -- & Orbital Period Unit (d/yr) \\
19 & r\_Sep & -- & reference for Sep \\
20 & dV & mag & ? Delta magnitude \\
21 & dVFilter & -- & Filter of delta magnitude \\
22 & r\_dV & -- & reference for dV \\
23 & Name & -- & Common Name \\
24 & TYCHO-2 & -- & TYCHO-2 Identifier (I/259) \\
25 & GJ & -- & Gliese-Jahreiss Catalog of Nearby Stars ID (J/PASP/122/885) \\
26 & HD & -- & ? Henry Draper catalog ID (III/135A) \\
27 & HR & -- & ? Bright Star Catalog ID (V/50) \\
28 & DM & -- & Durchmustrung ID (I/122; I/119; I/114; I/108) \\
29 & 1RXS & -- & ROSAT All-Sky Survey ID (IX/10A; IX/29) \\
30 & UCAC4 & -- & Fourth USNO CCD Astrographic Catalog ID (I/322A) \\
31 & PPMXL & -- & PPMXL ID (I/317) \\
32 & 2MASS & -- & Two Micron All Sky Survey ID (II/246; II/281) \\
33 & SDSS & -- & Sloan Digital Sky Survey Photometric Catalog ID (V/139) \\
34 & ALLWISE & -- & AllWISE ID (II/328) \\
35 & YPC & -- & ? General Catalog of Trigonometric Parallaxes ID (I/238A) \\
36 & HIP & -- & ? {\it Hipparcos} ID (I/311) \\
37 & RAdegraw & deg & Raw Right Ascension (J2000) \\
38 & e\_RAdegraw & mas & uncertainty on RAdegraw \\
39 & DEdegraw & deg & Raw Declination (J2000) \\
40 & e\_DEdegraw & mas & uncertainty on DEdegraw \\
41 & JD & d & Epoch of position measurement \\
42 & refPOS & -- & Reference for position \\
43 & RAdeg & deg & Right Ascension (J2000 E2000) (calculated) \\
44 & e\_RAdeg & mas & uncertainty on RAdeg (calculated) \\
45 & DEdeg & deg & Declination (J2000 E2000) (calculated) \\
46 & e\_DEdeg & mas & uncertainty on DEdeg (calculated) \\
47 & plx & mas & ? Weighted Mean Parallax (calculated) \\
48 & e\_plx & mas & ? uncertainty on plx (calculated) \\
49 & HRV & km/s & ? Weighted Mean Heliocentric Radial Velocity (calculated) \\
50 & e\_HRV & km/s & ? uncertainty on HRV (calculated) \\
51 & PredDist & pc & ? Predicted Distance \\
52 & e\_PredDist & pc & ? uncertainty on PredDist \\
53 & PredDist-method & -- & Method for PredDist \\
54 & r\_PredDist & -- & Reference for predDist \\
55 & PredHRV & km/s & ? Predicted Heliocentric Radial Velocity \\
56 & e\_PredHRV & km/s & ? uncertainty on PredHRV \\
57 & PredHRV-method & -- & Method for PredHRV \\
58 & r\_PredHRV & -- & Reference for predicted Radial Velocity \\
59 & pmRA & mas/yr & ? pmRA*cos(DEC) \\
60 & e\_pmRA & mas/yr & ? uncertainty on pmRA \\
61 & pmDE & mas/yr & ? pmDEC \\
62 & e\_pmDE & mas/yr & ? uncertainty on PMDE \\
63 & refPM & -- & Reference for proper motion \\
64 & pm & mas/yr & Total proper motion (calculated) \\
65 & e\_pm & mas/yr & uncertainty on pm (calculated) \\
66 & PA & deg & Proper motion position angle (calculated) \\
67 & e\_PA & deg & uncertainty on PA (calculated) \\
68 & Uvel & km/s & ? Galactic motion U direction (toward Galactic center) \\
69 & e\_Uvel & km/s & ? uncertainty on Uvel \\
70 & Vvel & km/s & ? Galactic motion V direction (toward solar motion) \\
71 & e\_Vvel & km/s & ? uncertainty on Vvel \\
72 & Wvel & km/s & ? Galactic motion W direction (toward Galactic north pole) \\
73 & e\_Wvel & km/s & ? uncertainty on Wvel \\
74 & Xpos & pc & ? Galactic position X (toward Galactic center) \\
75 & e\_Xpos & pc & ? uncertainty on Xpos \\
76 & Ypos & pc & ? Galactic position Y direction (toward solar motion) \\
77 & e\_Ypos & pc & ? uncertainty on Ypos \\
78 & Zpos & pc & ? Galactic position Z direction (toward Galactic north pole) \\
79 & e\_Zpos & pc & ? uncertainty on Zpos \\
80 & refUVWXYZ & -- & Reference for Galactic velocity and position \\
81 & FUV & mag & ? GALEX FUV magnitude \\
82 & e\_FUV & mag & ? uncertainty on FUV \\
83 & NUV & mag & ? GALEX NUV magnitude \\
84 & e\_NUV & mag & ? uncertainty on NUV \\
85 & refUV & -- & Reference for GALEX magnitude \\
86 & u'mag & mag & ? uncertainty on u'mag \\
87 & e\_u'mag & mag & quality of u'mag \\
88 & q\_u'mag & -- & reference for u'mag \\
89 & r\_u'mag & -- & ? SDSS g' magnitude \\
90 & g'mag & mag & ? uncertainty on g'mag \\
91 & e\_g'mag & mag & quality of g'mag \\
92 & q\_g'mag & -- & reference for g'mag \\
93 & r\_g'mag & -- & ? SDSS r' magnitude \\
94 & r'mag & mag & ? uncertainty on r'mag \\
95 & e\_r'mag & mag & quality of r'mag \\
96 & q\_r'mag & -- & reference for r'mag \\
97 & r\_r'mag & -- & ? SDSS i' magnitude \\
98 & i'mag & mag & ? uncertainty on i'mag \\
99 & e\_i'mag & mag & quality of i'mag \\
100 & q\_i'mag & -- & reference for i'mag \\
101 & r\_i'mag & -- & ? SDSS z' magnitude \\
102 & z'mag & mag & ? uncertainty on z'mag \\
103 & e\_z'mag & mag & quality of z'mag \\
104 & q\_z'mag & -- & reference for z'mag \\
105 & r\_z'mag & -- & [ JD] SDSS joint photometry flag (4) \\
106 & jointSDSS & -- & ? Tycho-2 Bt magnitude \\
107 & Btmag & mag & ? uncertainty on Btmag \\
108 & e\_Btmag & mag & ? Tycho-2 Vt magnitude \\
109 & Vtmag & mag & ? uncertainty on Vtmag \\
110 & e\_Vtmag & mag & ? Johnson B magnitude \\
111 & Bmag & mag & ? uncertainty on Bmag \\
112 & e\_Bmag & mag & quality of Bmag \\
113 & q\_Bmag & -- & reference for Bmag \\
114 & r\_Bmag & -- & ? Johnson V magnitude \\
115 & Vmag & mag & ? uncertainty on Vmag \\
116 & e\_Vmag & mag & quality of Vmag \\
117 & q\_Vmag & -- & reference for Vmag \\
118 & r\_Vmag & -- & ? Cousins R magnitude \\
119 & Rmag & mag & ? uncertainty on Rmag \\
120 & e\_Rmag & mag & reference for Rmag \\
121 & r\_Rmag & -- & ? Cousins I magnitude \\
122 & Imag & mag & ? uncertainty on Imag \\
123 & e\_Imag & mag & quality of Imag \\
124 & r\_Imag & -- & reference for Imag \\
125 & jointOpt & -- & [ JD] Optical joint photometry flag (4) \\
126 & Jmag & mag & ? J magnitude \\
127 & e\_Jmag & mag & ? uncertainty on Jmag \\
128 & q\_Jmag & -- & quality of Jmag \\
129 & Hmag & mag & ? H magnitude \\
130 & e\_Hmag & mag & ? uncertainty on Hmag \\
131 & q\_Hmag & -- & quality of Hmag \\
132 & Kmag & mag & ? Ks magnitude \\
133 & e\_Kmag & mag & ? uncertainty on Kmag \\
134 & q\_Kmag & -- & quality of Kmag \\
135 & refJHK & -- & Reference for Infrared photometry \\
136 & jointJHK & -- & [ JD] Infrared joint photometry flag (4) \\
137 & W1mag & mag & ? W1 magnitude \\
138 & e\_W1mag & mag & ? uncertainty on W1mag \\
139 & q\_W1mag & -- & quality of W1mag \\
140 & W2mag & mag & ? W2 magnitude \\
141 & e\_W2mag & mag & ? uncertainty on W2mag \\
142 & q\_W2mag & -- & quality of W2mag \\
143 & W3mag & mag & ? W3 magnitude \\
144 & e\_W3mag & mag & ? uncertainty on W3mag \\
145 & q\_W3mag & -- & quality of W3mag \\
146 & W4mag & mag & ? W4 magnitude \\
147 & e\_W4mag & mag & ? uncertainty on W4mag \\
148 & q\_W4mag & -- & Quality of W4mag \\
149 & refWISE & -- & Reference for WISE photometry \\
150 & jointWISE & -- & [ J] WISE joint photometry flag (4) \\
151 & photvar & mag & ? photometric variability \\
152 & f\_photvar & -- & filter of photometric variability \\
153 & Xray-PosErr & arcsec & ? Offset between J2000 E1991 position and X-ray source \\
154 & Xcnts & ct/s & ? ROSAT X-ray counts \\
155 & e\_Xcnts & ct/s & ? uncertainty on Xcnts \\
156 & HR1 & -- & ? ROSAT HR1 hardness ratio \\
157 & e\_HR1 & -- & ? uncertainty on HR1 \\
158 & HR2 & -- & ? ROSAT HR2 hardness ratio \\
159 & e\_HR2 & -- & ? uncertainty on HR2 \\
160 & r\_Xcnts & -- & Reference for ROSAT X-ray detection \\
161 & n\_Xcnts & -- & [ JD] ROSAT joint photometry flag (4) \\
162 & VMag & mag & ? Absolute V magnitude (calculated) \\
163 & B-V & mag & ? B-V color (calculated) \\
164 & V-K & mag & ? V-Ks color (calculated) \\
165 & V-I & mag & ? V-I color (calculated) \\
166 & J-K & mag & ? J-Ks color (calculated) \\
167 & EB-V & mag & ? Reddening \\
168 & r\_EB-V & -- & reference for EB-V \\
169 & BolMag & mag & ? Bolometric magnitude \\
170 & e\_BolMag & mag & ? uncertainty on BolMag \\
171 & r\_BolMag & -- & reference for BolMag \\
172 & Lx & [J-7/s] & ? Log of X-ray flux \\
173 & e\_Lx & [J-7/s] & ? uncertainty on Lx \\
174 & LxLbol & -- & ? Log(Lx/Lbol) \\
175 & f\_LxLbol & -- & limit flag on LxLbol \\
176 & SpType & -- & Spectral Type \\
177 & r\_SpType & -- & Reference for SpType \\
178 & n\_SpType & -- & [ JD] Spectral Type joint flag (4) \\
179 & Teff1 & K & ? Effective Temperature 1 \\
180 & e\_Teff1 & K & ? uncertainty on Teff1 \\
181 & r\_Teff1 & -- & reference for Teff1 \\
182 & Teff2 & K & ? Effective Temperature 2 \\
183 & e\_Teff2 & K & ? uncertainty on Teff2 \\
184 & r\_Teff2 & -- & reference for Teff2 \\
185 & Teff3 & K & ? Effective Temperature 3 \\
186 & e\_Teff3 & K & ? uncertainty on Teff3 \\
187 & r\_Teff3 & -- & reference for Teff3 \\
188 & Teff4 & K & ? Effective Temperature 4 \\
189 & e\_Teff4 & K & ? uncertainty on Teff4 \\
190 & r\_Teff4 & -- & reference for Teff4 \\
191 & vsini1 & km/s & ? Rotational Velocity 1 \\
192 & e\_vsini1 & km/s & ? uncertainty on vsini1 \\
193 & f\_vsini1 & -- & [ ule] limit flag on vsini1 (5) \\
194 & r\_vsini1 & -- & reference for vsini1 \\
195 & vsini2 & km/s & ? Rotational Velocity 2 \\
196 & e\_vsini2 & km/s & ? uncertainty on vsini2 \\
197 & f\_vsini2 & -- & [ ule] limit flag on vsini2 (5) \\
198 & r\_vsini2 & -- & reference for vsini2 \\
199 & vsini3 & km/s & ? Rotational Velocity 3 \\
200 & e\_vsini3 & km/s & ? uncertainty on vsini3 \\
201 & f\_vsini3 & -- & [ ule] limit flag on vsini3 (5) \\
202 & r\_vsini3 & km/s & reference for vsini3 \\
203 & CaHIndex & -- & ? CaH 697.5 nm 3nm absorption band Index \\
204 & e\_CaHIndex & -- & ? uncertainty on CaHIndex \\
205 & r\_CaHIndex & -- & reference for CaHIndex \\
206 & CaHNarrowIndex & -- & ? CaH 697.5 nm .5nm absorption band Index \\
207 & e\_CaHNarrowIndex & -- & ? uncertainty on CaHNarrowIndex \\
208 & r\_CaHNarrowIndex & -- & reference for CaHNarrowIndex \\
209 & EWHa1 & 0.1nm & ? H-alpha (656.3 nm) equivalent width 1 \\
210 & e\_EWHa1 & 0.1nm & ? uncertainty on EWHa1 \\
211 & r\_EWHa1 & -- & reference for EWHa1 \\
212 & EWHa2 & 0.1nm & ? H-alpha (656.3 nm) equivalent width 2 \\
213 & e\_EWHa2 & 0.1nm & ? uncertainty on EWHa2 \\
214 & r\_EWHa2 & -- & reference for EWHa2 \\
215 & EWHa3 & 0.1nm & ? H-alpha (656.3 nm) equivalent width 3 \\
216 & r\_EWHa3 & -- & reference for EWHa3 \\
217 & EWLi1 & 0.1pm & ? Lithium 670.8 nm doublet equivalent width 1 \\
218 & e\_EWLi1 & 0.1pm & ? uncertainty on EWLi1 \\
219 & f\_EWLi1 & -- & [ ule] limit flag on EWLi1 (5) \\
220 & r\_EWLi1 & -- & reference for EWLi1 \\
221 & EWLi2 & 0.1pm & ? Lithium 670.8 nm doublet equivalent width 2 \\
222 & e\_EWLi2 & 0.1pm & ? uncertainty on EWLi2 \\
223 & f\_EWLi2 & -- & [ ule] limit flag on EWLi2 (5) \\
224 & r\_EWLi2 & -- & reference for EWLi2 \\
225 & EWLi3 & 0.1pm & ? Lithium 670.8 nm doublet equivalent width 3 \\
226 & e\_EWLi3 & 0.1pm & ? uncertainty on EWLi3 \\
227 & f\_EWLi3 & -- & [ ule] limit flag on EWLi3 (5) \\
228 & r\_EWLi3 & -- & reference for EWLi3 \\
229 & EWLi4 & 0.1pm & ? Lithium 670.8 nm doublet equivalent width 4 \\
230 & e\_EWLi4 & 0.1pm & ? uncertainty on EWLi4 \\
231 & r\_EWLi4 & -- & reference for EWLi4 \\
232 & ALi1 & -- & ? Lithium abundance 1 \\
233 & r\_ALi1 & -- & reference for ALi1 \\
234 & ALi2 & -- & ? Lithium abundance 2 \\
235 & r\_ALi2 & -- & reference for ALi2 \\
236 & EWK7699 & 0.1nm & ? Potassium 769.9 nm equivalent width \\
237 & e\_EWK7699 & 0.1nm & ? uncertainty on EWK7699 \\
238 & r\_EWK7699 & -- & reference for EW7699 \\
239 & EWNa8200 & 0.1nm & ? Sodium 820.0 nm equivalent width \\
240 & e\_EWNa8200 & 0.1nm & ? uncertainty on EWNa8200 \\
241 & Na8200Index & -- & ? Sodium 820.0 nm gravity index \\
242 & e\_Na8200Index & -- & ? uncertainty on Na8200Index \\
243 & r\_Na8200Index & -- & reference on Na8200Index \\
244 & FeH1 & [Sun] & ? Metallicity 1 \\
245 & e\_FeH1 & [Sun] & ? uncertainty on FeH1 \\
246 & r\_FeH1 & -- & reference for FeH1 \\
247 & FeH2 & [Sun] & ? Metallicity 2 \\
248 & e\_FeH2 & [Sun] & ? uncertainty on FeH2 \\
249 & r\_FeH2 & -- & reference for FeH2 \\
250 & FeH3 & [Sun] & ? Metallicity 3 \\
251 & r\_FeH3 & -- & reference for FeH3 \\
252 & BaH & [Sun] & ? Barium abundance \\
253 & e\_BaH & [Sun] & ? uncertainty on BaH \\
254 & r\_BaH & -- & reference for BaH \\
255 & RHK1 & -- & ? Mount Wilson Activity Index 1 \\
256 & r\_RHK1 & -- & reference for RHK1 \\
257 & RHK2 & -- & ? Mount Wilson Activity Index 2 \\
258 & r\_RHK2 & -- & reference for RHK2 \\
259 & logg1 & -- & ? Surface Gravity 1 \\
260 & e\_logg1 & -- & ? uncertainty on logg1 \\
261 & r\_logg1 & -- & reference for logg1 \\
262 & logg2 & -- & ? Surface Gravity 2 \\
263 & e\_logg2 & -- & ? uncertainty on logg2 \\
264 & r\_logg2 & -- & reference for logg2 \\
265 & GROUP & -- & Final Membership \\
266 & q\_GROUP & -- & quality of GROUP \\
267 & Age & Myr & ? Stellar Age \\
268 & r\_Age & -- & reference for Age \\
269 & GROUP1 & -- & Membership 1 \\
270 & q\_GROUP1 & -- & quality of GROUP1 (6) \\
271 & r\_GROUP1 & -- & reference for GROUP1 \\
272 & GROUP2 & -- & Membership 2 \\
273 & q\_GROUP2 & -- & quality of GROUP2 (6) \\
274 & r\_GROUP2 & -- & reference for GROUP2 \\
275 & GROUP3 & -- & Membership 3 \\
276 & q\_GROUP3 & -- & quality of GROUP3 (6) \\
277 & r\_GROUP3 & -- & reference for GROUP3 \\
278 & GROUP4 & -- & Membership 4 \\
279 & q\_GROUP4 & -- & quality of GROUP4 (6) \\
280 & r\_GROUP4 & -- & reference for GROUP4 \\
281 & GROUP5 & -- & Membership 5 \\
282 & q\_GROUP5 & -- & quality of GROUP5 (6) \\
283 & r\_GROUP5 & -- & reference for GROUP5 \\
284 & GROUP6 & -- & Membership 6 \\
285 & q\_GROUP6 & -- & quality of GROUP6 (6) \\
286 & r\_GROUP6 & -- & reference for GROUP6 \\
287 & GROUP7 & -- & Membership 7 \\
288 & q\_GROUP7 & -- & quality of GROUP7 (6) \\
289 & r\_GROUP7 & -- & reference for GROUP7 \\
290 & GROUP8 & -- & Membership 8 \\
291 & q\_GROUP8 & -- & quality of GROUP8 (6) \\
292 & r\_GROUP8 & -- & reference for GROUP8 \\
293 & GROUP9 & -- & Membership 9 \\
294 & q\_GROUP9 & -- & quality of GROUP9 (6) \\
295 & r\_GROUP9 & -- & reference for GROUP9 \\
296 & GROUP10 & -- & Membership 10 \\
297 & q\_GROUP10 & -- & quality of GROUP10 (6) \\
298 & r\_GROUP10 & -- & reference for GROUP10 \\
299 & GROUP11 & -- & Membership 11 \\
300 & q\_GROUP11 & -- & quality of GROUP11 (6) \\
301 & r\_GROUP11 & -- & reference for GROUP11 \\
302 & LACEwING-F & -- & LACEwING Identification (Field Star Mode) \\
303 & q\_LACEwING-F & \% & ? quality of LACEwING-F \\
304 & LACEwING-Y & -- & LACEwING Identification (Young Star Mode) \\
305 & q\_LACEwING-Y & \% & ? quality of LACEwING-Y \\
306 & HRV1 & km/s & ? Radial Velocity 1 (7) \\
307 & e\_HRV1 & km/s & ? uncertainty on HRV1 \\
308 & q\_HRV1 & -- & [ e] quality of HRV1 \\
309 & r\_HRV1 & -- & reference for HRV1 \\
310 & HRV2 & km/s & ? Radial Velocity 2 (7) \\
311 & e\_HRV2 & km/s & ? uncertainty on HRV2 \\
312 & q\_HRV2 & -- & [ e] quality of HRV2 \\
313 & r\_HRV2 & -- & reference for HRV2 \\
314 & HRV3 & km/s & ? Radial Velocity 3 (7) \\
315 & e\_HRV3 & km/s & ? uncertainty on HRV3 \\
316 & q\_HRV3 & -- & [ e] quality of HRV3 \\
317 & r\_HRV3 & -- & reference for HRV3 \\
318 & HRV4 & km/s & ? Radial Velocity 4 (7) \\
319 & e\_HRV4 & km/s & ? uncertainty on HRV4 \\
320 & q\_HRV4 & -- & [ e] quality of HRV4 \\
321 & r\_HRV4 & -- & reference for HRV4 \\
322 & HRV5 & km/s & ? Radial Velocity 5 (7) \\
323 & e\_HRV5 & km/s & ? uncertainty on HRV5 \\
324 & q\_HRV5 & -- & [ e] quality of HRV5 \\
325 & r\_HRV5 & -- & reference for HRV5 \\
326 & HRV6 & km/s & ? Radial Velocity 6 (7) \\
327 & e\_HRV6 & km/s & ? uncertainty on HRV6 \\
328 & q\_HRV6 & -- & [ e] quality of HRV6 \\
329 & r\_HRV6 & -- & reference for HRV6 \\
330 & HRV7 & km/s & ? Radial Velocity 7 (7) \\
331 & e\_HRV7 & km/s & ? uncertainty on HRV7 \\
332 & q\_HRV7 & -- & [ e] quality of HRV7 \\
333 & r\_HRV7 & -- & reference for HRV7 \\
334 & HRV8 & km/s & ? Radial Velocity 8 (7) \\
335 & e\_HRV8 & km/s & ? uncertainty on HRV8 \\
336 & r\_HRV8 & -- & reference for HRV8 \\
337 & plx1 & mas & ? Parallax 1 (7) \\
338 & e\_plx1 & mas & ? uncertainty on plx1 \\
339 & r\_plx1 & -- & reference for plx1 \\
340 & plx2 & mas & ? Parallax 2 (7) \\
341 & e\_plx2 & mas & ? uncertainty on plx2 \\
342 & r\_plx2 & -- & reference for plx2 \\
343 & plx3 & mas & ? Parallax 3 (7) \\
344 & e\_plx3 & mas & ? uncertainty on plx3 \\
345 & r\_plx3 & -- & reference for plx3 \\
346 & plx4 & mas & ? Parallax 4 (7) \\
347 & e\_plx4 & mas & ? uncertainty on plx4 \\
348 & r\_plx4 & -- & reference for plx4 \\
349 & plx5 & mas & ? Parallax 5 (7) \\
350 & e\_plx5 & mas & ? uncertainty on plx5 \\
351 & r\_plx5 & -- & reference for plx5 \\
352 & plx6 & mas & ? Parallax 6 (7) \\
353 & e\_plx6 & mas & ? uncertainty on plx6 \\
354 & r\_plx6 & -- & reference for plx6 \\
355 & plx7 & mas & ? Parallax 7 (7) \\
356 & e\_plx7 & mas & ? uncertainty on plx7 \\
357 & r\_plx7 & -- & reference for plx7 \\
358 & plx8 & mas & ? Parallax 8 (7) \\
359 & e\_plx8 & mas & ? uncertainty on plx8 \\
360 & r\_plx8 & -- & reference for plx8 \\
\enddata
\tablecomments{Guide to the Contents of the Catalog of Suspected Nearby Young Stars, which is available as an online-only machine-readable table. Note that multiple column blocks are present for certain quantities ($T_{eff}$, EWHa, EWLi, etc.) in the interest of preserving every published data value. These should not inherently be preferred over each other, except that the weighted mean parallax columns (\#47-\#48) and weighted mean RV columns (\#49-\#50) should be preferred over the individual measurements that were combined to produce them (\#306-\#360). (1) LiSample flags are: L = star has a measured lithium EW greater than 10\AA {\it and} 2-$\sigma$ of the published error or 50 m\AA. F = star has lithium, but is a member of a more distant group beyond 120 pc. A = star is in the same star system as a lithium-sample star. Lower-case letters l,f,a indicate the objects are not system primaries. (2) Bonafide flags are: B = in final Bona-fide sample. R = rejected from initial high-confidence sample. Lower-case letters b,r indicate the objects are not system primaries. (3) Multiplicity types are AB for astrometric binaries, IB for interferometric binaries, EB for eclipsing binaries, OB for occultation binaries, VB for visual binaries; SB for spectroscopic binaries; S for spectroscopic single stars; SB1, SB2, SB3 for single-, double-, and triple-lined spectroscopic binaries. (4) Joint flags: J = quoted quantity actually refers to an unresolved detection that includes this object. D = quoted quantity has been deblended into approximate individual measurements. (5) Limit flags: l = lower limit. u = upper limit. e = uncertainty set to typical value. (6) Quality strings have been reproduced from the source papers and are a mix of different scales, with the exception of ``BF'', which always denotes an identification as a bona-fide member. (7) Source RVs and parallaxes for the weighted means in columns 47-50.}
\end{deluxetable}

\end{document}